\documentclass[aps,letter,prx,reprint,superscriptaddress,notitlepage,floatfix,nolongbibliography]{revtex4-2}
\usepackage{amsmath}
\usepackage{amssymb}
\usepackage{graphicx}
\usepackage{bm}
\usepackage{xcolor}
\usepackage{csquotes}
\usepackage[T1]{fontenc}
\usepackage[utf8]{inputenc}
\usepackage{standalone}
\usepackage{mathrsfs}
\usepackage{xpatch}
\usepackage{makecell}
\makeatletter
\xpatchcmd{\@ssect@ltx}{\@xsect}{\protected@edef\@currentlabelname{#8}\@xsect}{}{}
\xpatchcmd{\@sect@ltx}{\@xsect}{\protected@edef\@currentlabelname{#8}\@xsect}{}{}
\makeatother
\usepackage[colorlinks=false, hidelinks]{hyperref}
\usepackage{bookmark}
\def\ii{\text{i}}

\graphicspath{{media/}}

\begin{document}


\title{Chirality across scales in tissue dynamics}

\author{Sihan Chen}
\thanks{Both authors contributed equally to this work.}
\affiliation{Kadanoff Center for Theoretical Physics, The University of Chicago, Chicago, IL 60637, USA}
\affiliation{James Franck Institute, The University of Chicago, Chicago, IL 60637, USA}
\author{Doruk Efe G{\"o}kmen}
\thanks{Both authors contributed equally to this work.}
\affiliation{NSF-Simons National Institute for Theory and Mathematics in Biology, 186 E Chestnut St, Chicago, IL 60611, USA}
\affiliation{James Franck Institute, The University of Chicago, Chicago, IL 60637, USA}
\author{Michel Fruchart}
\affiliation{Gulliver, ESPCI Paris, Université PSL, CNRS, 75005 Paris, France}
\author{Miriam Krumbein}
\affiliation{The Swiss Institute for Dryland Environmental and Energy Research, The Jacob Blaustein Institutes for Desert Research, Ben-Gurion University of the Negev, Sede Boqer Campus, Midreshet Ben-Gurion, Israel}
\author{Pascal Silberzan}
\affiliation{Institut Curie, 26 Rue d'Ulm, 75005 Paris, France}
\author{Victor Yashunsky}
\email{yashunsk@bgu.ac.il}
\affiliation{The Swiss Institute for Dryland Environmental and Energy Research, The Jacob Blaustein Institutes for Desert Research, Ben-Gurion University of the Negev, Sede Boqer Campus, Midreshet Ben-Gurion, Israel}
\affiliation{Laboratoire Physique des Cellules et Cancer UMR168, Institut Curie, Paris Sciences et Lettres, Centre National de la Recherche Scientifique, Sorbonne Université, 75248 Paris, France - Equipe Labellisée Ligue Contre le Cancer}
\author{Vincenzo Vitelli}
\email{vitelli@uchicago.edu}
\affiliation{Kadanoff Center for Theoretical Physics, The University of Chicago, Chicago, IL 60637, USA}
\affiliation{James Franck Institute, The University of Chicago, Chicago, IL 60637, USA}
\date{\today}

\begin{abstract}
Chiral processes that lack mirror symmetry pervade nature from enantioselective molecular interactions to the asymmetric development of organisms. An outstanding challenge at the interface between physics and biology consists in bridging the multiple scales between microscopic and macroscopic chirality. Here, we combine theory, experiments and modern inference algorithms to study a paradigmatic example of dynamic chirality transfer across scales: the generation of tissue-scale flows from subcellular forces. The distinctive properties of our microscopic graph model and the corresponding coarse-grained viscoelasticity are that (i) net cell proliferation is spatially inhomogeneous and (ii) cellular dynamics cannot be expressed as an energy gradient. To overcome the general challenge of inferring microscopic model parameters from noisy high-dimensional data, we develop a nudged automatic differentiation algorithm (NADA) that can handle large fluctuations in cell positions observed in single tissue snapshots. This data-calibrated microscopic model quantitatively captures proliferation-driven tissue flows observed at large scales in our experiments on fibroblastoma cell cultures. Beyond chirality, our inference algorithm can be used to extract interpretable graph models from limited amounts of noisy data of living and inanimate cellular systems such as networks of convection cells and flowing foams.
\end{abstract}
\maketitle

\begin{figure*}[]
\centering
\hspace{-2em}\includegraphics[width = 1.9\columnwidth]{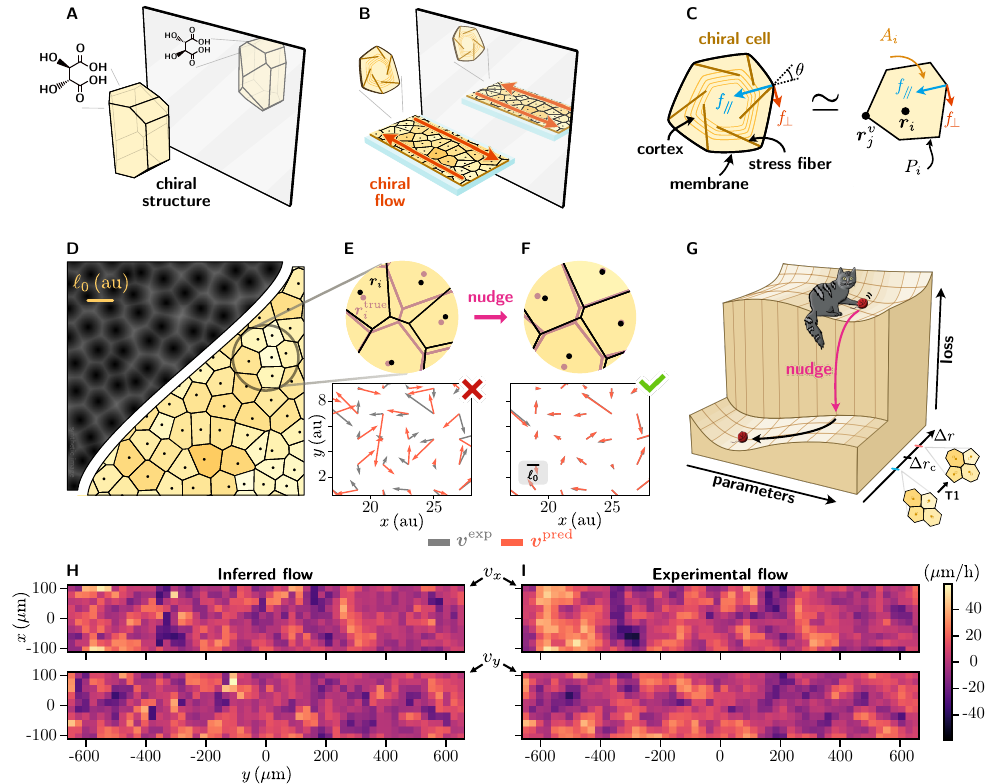}
\caption{{\bf Inferring dynamic cellular models of chiral tissues from data.} 
(A) The macroscopic chirality in the equilibrium structure of tartaric acid crystals has been famously traced back to molecular-level chirality by Louis Pasteur.
(B) Bridging macroscopic chiral flows of living tissues to the microscopic chiral forces between cells involves a far more challenging task of coarse graining a multitude of non-equilibrium processes intrinsic to life. Example of chiral flows in a cell monolayer is shown in red arrows. 
(C) Sketch of a chiral cell.
The stress fibers (in brown) make an angle $\theta$ with respect to the radial direction. 
When they contract, the cell membrane (thick black contour) rotates with respect to the actomyosin cortex (thin brown lines) because of the transverse component $f_\perp$ of the forces (in red). 
In the graph model, each cell is approximated by a polygon. 
The degrees of freedom are the positions $\bm{r}_i$ of the cells (black dots).
An instantaneous Voronoi tessellation associates a polygon to each cell, which represents its membrane. 
In the model, the area $A_i$ and perimeter $P_i$ of cell $i$ tend to relax to preferred values $A_0$ and $P_0$.
(D-G) To infer the parameters of the microscopic model from experimental data, we use a nudged version of automatic differentiation:
In addition to optimizing model parameters to match the experimental flow data $\bm{v}^{\rm exp}$ by gradient descent (black arrow in panel G), we also very slightly nudge the input position data $\bm{r}^{\rm exp}$ to improve the fitting of the model flow $\bm{v}^{\rm pred}$ (pink arrow in panel G).
The reasoning for this is explained in panels E-F: small errors on the segmented cell centers lead to an incorrect network topology when performing Voronoi tessellation. 
With the wrong topology, the inferred velocity $\bm{v}^{\rm pred}$ for the optimum model parameters is visibly incorrect (panel E). Nudging the position data by a very small amount (see Extended Data Fig.~\ref{fig.m0}F-H) fixes the topology and leads to excellent predictions (panel F).
This is because the change of topology (T1 transitions) leads to an abrupt change in the loss landscape (panel G).
Panel D illustrates the conversion between microscopy data (here mock data) on the left side and a Voronoi tesselation of cells centers on the right side.
(H-I) AD, combined with noise regularisation through nudging, can fit microscopic model parameters and reconstruct experimental tissue flow in fibrosarcoma cell cultures, achieving a normalised mean-squared error of approximately $30\%$. 
    }
	\label{fig.1}
\end{figure*}

Chiral processes, from enantioselective molecular interactions to asymmetric organ development, occur at all scales.
By studying tartaric acid crystallizing in wine barrels, Pasteur 
realized that the macroscopic chirality of these crystals could be traced all the way down to the chiral microstructure of the molecules they were made of (Fig.~\ref{fig.1}A)~\cite{pasteur1897researches}. 
Biology exploits chirality as a design principle across all length scales, from the very molecules of life 
to cells, tissues and organs that do not have mirror symmetry (Fig.~\ref{fig.1}B)~\cite{Phillips2013,shadkhoo2019role,Mani2023,Tan2024,naganathan2014active,Shankar2022, Casademunt2025,maroudassacks2021topological,Tan2022,Taniguchi2011,sato2015cell,Inaki2016,Wan2016,tee2015cellular,Inaki2018,Ray2018,yamamoto2020collective,Middelkoop2021,yamamoto2023epithelial,Badih2024,rahman2024biomechanical,Ma2023}.
Yet, how chirality propagates across scales in the presence of the non-equilibrium processes associated with life, from metabolic cycles to cell proliferation, has so far eluded quantitative coarse-graining approaches. 



Here, we combine theory, experiments and modern inference algorithms to study a paradigmatic example of \textit{dynamic} chirality transfer across scales: the generation of tissue-scale flows from subcellular forces. A cell exerts forces on its neighbors through its cytoskeleton, a network of biopolymers such as actin and microtubules within which molecular motor proteins (e.g. kinesin and myosin) generate stress through nonequilibrium metabolic cycles~\cite{Alberts,Fletcher2010485,chen2020motor}.
As shown in Fig.~\ref{fig.1}C, the cytoskeleton can spontaneously develop a chiral structure in which stress fibers (actin or microtubule) resemble a vortex \cite{tee2015cellular,Wan2016,yamamoto2023epithelial,rahman2024biomechanical}. 
Documented examples of a chiral cytoskeleton include both fibroblast cells~\cite{tee2015cellular}, that play a crucial role in tissue inflammation and repair, as well as Caco2 cells, a typical epithelial cell line derived
from colorectal adenocarcinoma~\cite{yamamoto2023epithelial}, see Extended Data Fig.~\ref{fig.cell} for a visual summary of these experimental findings.

\noindent{\bf Chiral cells.} As shown in Fig.~\ref{fig.1}C, a chiral cytoskeleton can generate forces with a transverse component (along the cell membrane) $f_\perp$ (in red) in addition to the usual normal component $f_\parallel$ (in blue) pulling inwards. 
This can be grasped in purely geometric terms: the total active force $\bm{f}$ generated by each contractile stress fiber is always pulling along the fiber, so when the fiber makes an angle $\theta$ with the cell boundary, then  $f_\parallel \simeq f \cos\theta$ while $f_\perp \simeq f \sin\theta$. The two force components are thus related via
\definecolor{force_blue}{HTML}{00A1E6}
\definecolor{force_red}{HTML}{E64500}
\begin{equation}
\begin{aligned}
\bm{f_\perp}=
-
\alpha_{0} \hat{\bm e}_z \times  \bm{f_\parallel}\,,
\label{chiral_vertex_model0}
\end{aligned}
\end{equation}
where $\alpha^{\rm o} \equiv \tan\theta$ is a cortical chirality parameter that measures the breaking of chiral symmetry at the single cell level. 
Note that $\bm f_\perp$  has a non-zero circulation (i.e. curl) around the cell membrane (Fig~\ref{fig.1}C) that tends to rotate isolated cells and prevents writing it as the gradient of a potential energy.
Within a tissue, cells can be tightly packed, hindering rotational motion, but these transverse  forces remain active and perform path-dependent work \cite{scheibner2020odd,fruchart2023odd}.

In order to bridge the gap between subcellular active forces and tissue-scale dynamics, we seek inspiration from a class of discrete dynamical equations originally devised to model foams ~\cite{Alt2017,Vedula2012,Fletcher2014,Bi2016,Xi2018,Vedula2012,Fletcher2014,sussman2018no,farhadifar2007influence,staple2010mechanics,bi2015density,Weaire1984}, but with a crucial twist. The chiral dynamics of our microscopic (i.e. discrete, cell-level) model will be non-variational, i.e. not expressible as an energy gradient, manifesting the active \textit{interactions} between live cells (which is distinct from self-propulsion of individual cells).  This chiral microscopic model and the coarse-grained active elasticity we derive from it account for changes in cell-shape (e.g. strains), offering a complementary perspective to descriptions based on liquid crystals \cite{maroudassacks2021topological,balasubramaniam2022active,alert2022active,yashunsky2022chiral} that focus on orientational order (e.g. a nematic director), see SI. 


\medskip
\noindent{\bf Microscopic graph model of chiral tissues.} 
In our graph model, each cell within the tissue, labelled by $i$, is represented as a polygon (Fig.~\ref{fig.1}C) with area $A_i$ and perimeter $P_i$ located around its center $\bm{r}_i$.
We propose (and later validate with experimental data) that the time evolution of cell positions in chiral tissues is described by 
\begin{equation}
\begin{aligned}
\partial_t \bm r_i = 
-\frac{\partial E}{\partial \bm r_i}  
+ 
\alpha^{\rm o} 
\left(\hat{\bm e}_z \times  \frac{\partial E_C}{\partial \bm r^v_j} \right)
\cdot \frac{\partial \bm r^v_j}{\partial \bm r_i}
\label{chiral_vertex_model}
\end{aligned}
\end{equation}
in which $\bm{r}^v_j$ are the positions of the vertices of the polygon $i$, implicitly defined from the cell-centers $\bm{r}_i$ by a standard protocol called instantaneous Voronoi tessellation~\cite{voronoi1908a,voronoi1908b}. 

The first term on the right-hand side of  Eq.~\eqref{chiral_vertex_model} is a typical 
gradient descent in an energy landscape that tends to enforce $P_i \simeq P_0$ and $A_i \simeq A_0$, where $A_0$ and $P_0$ are the parameters representing the optimal cellular area and perimeter respectively.
More precisely, $E = E_A + E_B + E_C$ can be seen as a cost function in which $E_A = \sum_i ({K_A}/{2}) (A_i - A_0)^2$ accounts for the 3D incompressibility of cells within the tissue, $E_B = -\sum_i K_P P_0 P_i$ accounts for cell-cell adhesion, and $E_C = \sum_i ({K_P}/{2}) P_i^2$ accounts for the active contractility of the cytoskeleton \cite{Bi2016,Farhadifar2007,staple2010mechanics,chen2020motor}.

The second term in Eq.~\eqref{chiral_vertex_model} is what we add to account for the transverse non-variational force $\bm f_\perp$
in Eq.~(\ref{chiral_vertex_model0}) (red arrow in Fig.~\ref{fig.1}C), with $\bm f_{\parallel}=-\partial E_c/\partial {\bm r}^v_j$.
It only involves $E_C$ because it is generated by the active contraction of the stress fibers within the cell, see Fig.~\ref{fig.1}C. 
Crucially, the point of application of $\bm f_\perp$ is on the cell boundary (parameterized by the vertices $\bm{r}^v_j$), not its center $\bm{r}_i$. 
This subtle but important difference, captured mathematically by the Jacobian $\partial \bm r^v_j/\partial \bm r_i$ in Eq. (\ref{chiral_vertex_model}), prevents us from simplifying the derivatives using the chain rule and makes the dynamics non-variational (see Method).
In addition, the cross product in Eq.~\eqref{chiral_vertex_model} makes the dynamics chiral.
Equation \eqref{chiral_vertex_model} illustrates with the forces $\bm f_\perp$ and $\bm f_\parallel$ the more general point that graph models of tissues can embed biological information about cell states both on the points (here, the cell centers) and on the links (cells edges).

In order to validate a discrete cellular model like Eq. \eqref{chiral_vertex_model} from experimental data, we need to infer its parameters.
However, this requires fitting  noisy, high dimensional experimental data to a graph model of fluctuating connectivity that, without an appropriate regularization, would be very sensitive to noise --- a general issue 
not limited to biology. 
\medskip

\noindent{\bf Inference algorithm (NADA).} To circumvent this 
problem, we have developed a Nudged Automatic Differentiation Algorithm (NADA). The input of our algorithm are snapshots of the cell boundaries from which the instantaneous cell centers $\bm{r}_i^{\rm exp}$ are obtained by segmentation (Fig.~\ref{fig.1}D). Using optical flow velocimetry (OFV) one can also obtain the experimental velocity field $\bm v^{\rm exp}$.
Our model \eqref{chiral_vertex_model} effectively predicts a velocity field $\bm v^{\rm pred} = F_p(\bm{r}_i^{\rm exp})$ (by coarse-graining $\partial_t \bm{r}_i$ to match the scale of OFV data). A naive inference strategy consists in trying to adjust the parameters $p$
of the model $F_p$ to minimize the loss function $\mathscr{L} = \lVert \bm{v}^{\rm pred} - \bm{v}^{\rm exp} \rVert$. 
In practice, this can be performed using gradient descent with the aid of a technique called automatic differentiation applied to  Eq.~\eqref{chiral_vertex_model}, see Methods.
Unfortunately, this strategy does not quite work as expected: Fig.~\ref{fig.1}E (bottom) shows the large mismatch between $\bm v^{\rm exp}$ (gray) and $\bm v^{\rm pred}$ (orange) when using this naive strategy.
The reason for the mismatch is that, even if the experimentally measured cell centers $\bm{r}_i^{\rm exp}$ are ever so slightly noisy, such tiny amounts can still have large impacts on the model because they produce drastic changes (known as T1 transitions) in the topology of the Voronoi tessellation representing the reconstructed tissue (compare the black and purple cell boundaries in Fig.~\ref{fig.1}E).

Our way out of this generic problem is to \enquote{nudge} the data in the "right" direction to revert the topology changes due to noise (Fig.~\ref{fig.1}F). 
But finding the "right" direction looks like a daunting task: in what direction should one nudge the data, and by how much? 
Our idea is simple: do not just fit the model to the data, also fit the data to the model. 
More precisely, the noisy data points $\bm{r}_i^{\rm exp}$ are nudged in a direction chosen to best reduce the loss $\mathscr{L}$, by an amount constrained to be small (e.g. of the order of the experimental uncertainty).
The details of this algorithm are described in the Methods.
To see why it works, recall that noise produces topological changes in the reconstructed tissue (Fig.~\ref{fig.1}E-F). 
These topological changes lead to discontinuous changes in the loss landscape $\mathscr{L}(p,\bm{r}^{\rm exp})$ (Fig.~\ref{fig.1}G and Extended Data Figure \ref{fig.m0}), so a tiny change in $\bm{r}^{\rm exp}$ can lead to a drastic improvement in the accuracy of the method. 

In panels H and I, we show the results of applying the NADA method on tissue flow in \textit{in vitro} fibrosarcoma HT1080 cell culture (Methods). 
Comparing panels H and I shows that our microscopic model with parameters fitted with NADA reproduces most features of the experimental velocity field with a normalised mean-squared error of approximately $30\%$, if the parameters are constrained to be simply constant over space. 
We emphasize that our method requires a very moderate amount of data and can even be applied to single snapshots in the case of rapidly evolving tissues. 
More quantitative comparisons between the inferred and measured flows are given in the Methods (Extended Data Fig.~\ref{fig.m0}).

\begin{figure*}[t]
\centering
\hspace{-2em}\includegraphics[width = 1.75\columnwidth]{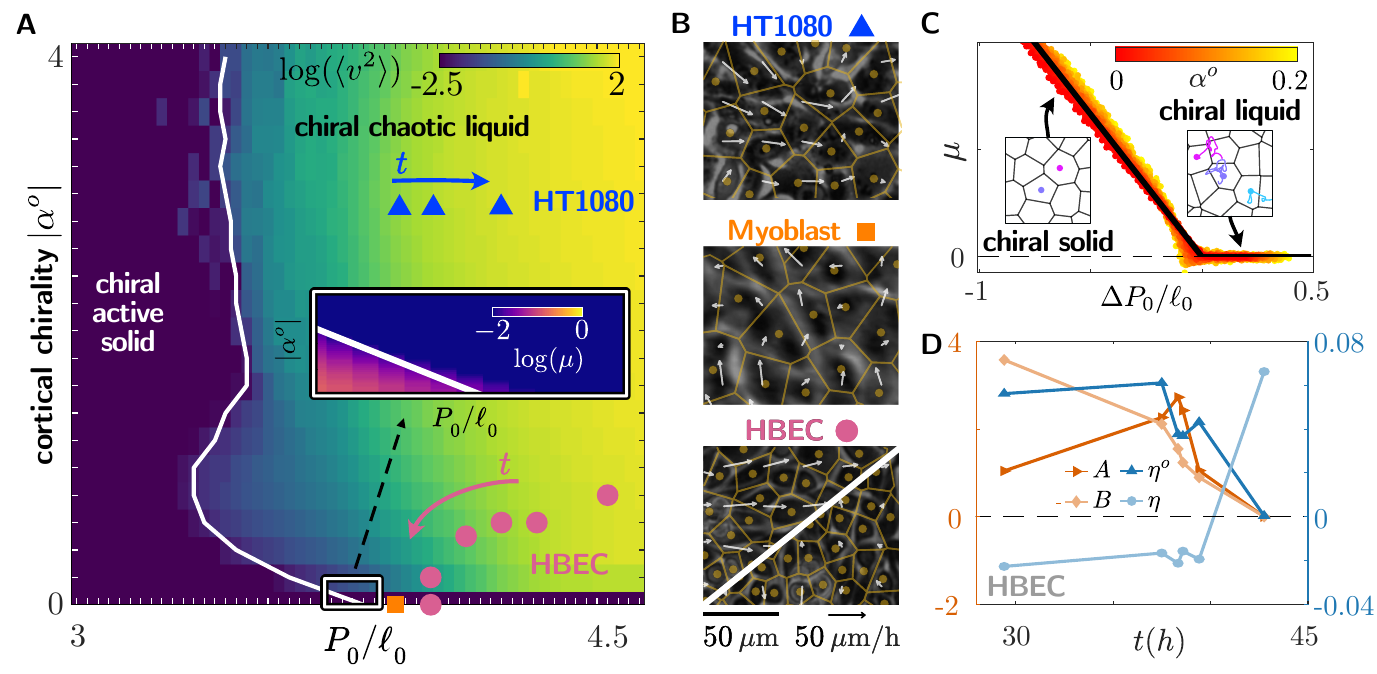}
\caption{{\bf Active viscoelasticity, phase diagram and dynamical rigidity transitions.}
(A) Numerical phase diagram defined by the steady-state cell velocity. Inferred experimental parameters are shown in symbols, see panel B for labeling. See Methods for a table of parameters. When $\alpha^o=0$ the tissues reduce to achiral solid/liquid with vanishing steady-state velocity. Inset: shear modulus $\mu$ for small $|\alpha^o|$ values (zooming in the region under the rectangular magnifier), see Extended Data Fig.~\ref{fig.m1}B for details. 
(B) Typical Voronoi tessellation and velocity fields in experiments. While HT1080 cells and C2C12 myoblasts maintain relatively constant cell density and velocity statistics over time, HBEC exhibit a marked increase in cell density and a decrease in velocity (the velocity field changes from top left to bottom right). 
(C) Universal behavior of the elastic shear modulus $\mu$ as function of $\Delta P_0$, the distance to the critical point.
The shear modulus $\mu$ decreases as $\Delta P_0$ increases until vanishing at the critical point $\Delta P_0 =0$. Inset: cells remain stationary in the chiral solid phase and exhibit chaotic motion in the chiral liquid phase. 
(D) Temporal evolution of the ratios of the viscoelastic moduli of HBEC.
 We plot the trajectory of the inferred parameters on the numerical phase diagram obtained from cell velocities, see panel B, and find distinct behaviors among the three cell types (see Methods for detailed parameters). HT1080 exhibits persistent strong chirality $\alpha^o=2.9$ and sits deeply in the chiral liquid phase. We find no observable chirality for C2C12 myoblasts, whose parameters are close to the normal solid-liquid transition. The parameters of HBEC  evolve significantly over time. At the start point of the trajectory ($t=29.4 \rm h$),  $\alpha^o=-0.8$ (opposite chirality to HT1080) and the tissue is in the chiral liquid phase. Both $|\alpha^o|$ and $P_0/\ell_0$ gradually decrease over time, and the tissue enters the normal liquid phase in the end of the trajectory ($t=42.8 \rm h$). 
Here, $\ell_0^2$ is the mean cell area. In (D) $A$ and $B$ are in unit of $\gamma K_P$, and  $\eta$ and $\eta^o$ are in unit of $\gamma$, with $\gamma$ being the frictional coefficient between cells and the substrate. In the simulations of (A) and (C) we have set $\gamma=1$, $K_P=1$, $A_0=\ell_0^2=1$ and $K_A=0.1$. 
    }
	\label{fig.2}
\end{figure*}

\begin{figure*}[t]
\centering
\hspace{-2em}\includegraphics[width = 1.8\columnwidth]{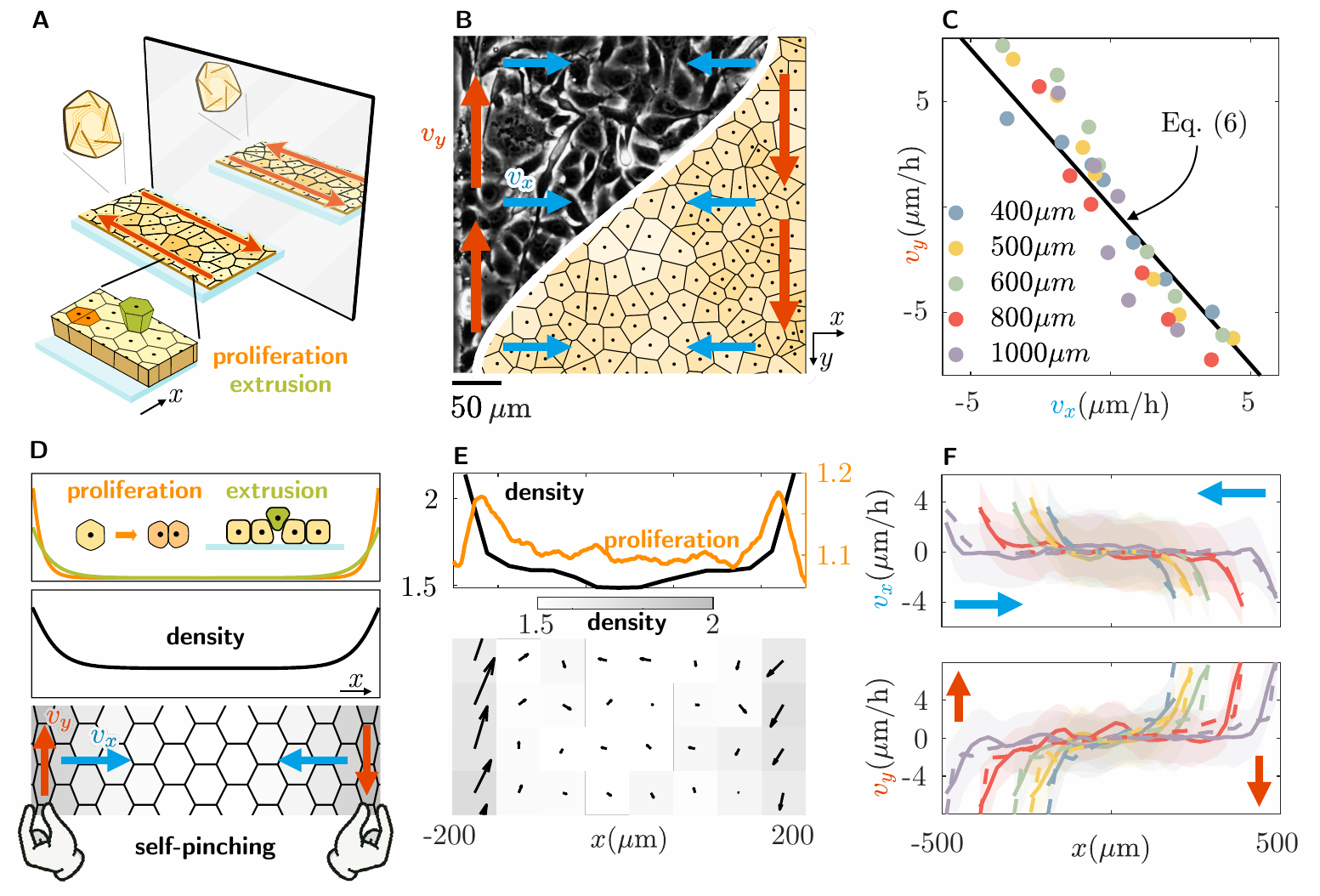}
\caption{{\bf From micro to macro: coarse-grained theory captures mechanics of chiral tissues} (A) Illustration of the experimental sample and its broken mirror symmetry. Proliferation and extrusion dynamically regulate cell density within the sample. (B) Snapshot of HT1080 tissue. Chiral and converging flows are indicated by red and blue arrows, respectively. (C) $v_y$ vs $v_x$ of edge currents for various sample widths, see panel F for errorbars. Solid lines indicate theoretical prediction in Eq.~(\ref{continuum}), with $\alpha^o=2.9$ and $P_0/\ell_0=4.0$ extracted from the inference (no adjustable parameter used for the fitting).  (D) Inhomogeneous tissue growth effectively ``pinches'' the cells near the boundaries, which further leads an inhomogeneous density field $\rho$, resulting in the macroscopic flows. (E) Top: the density profile (in unit of $10^{-3} {\rm \mu m}^{-2}$) and the proliferation rate (renormalized by its value in the bulk) measured in experiment. Bottom: heatmap of the average density field on a selected region, together with the velocity field indicated by the arrows. (F) $v_x$ and $v_y$ profile for various sample widths are shown in solid lines. Simulations results (with proliferation/extrusion) are shown in dashed lines. Blue and red arrows indicate the corresponding flow directions in panel B. In the simulations all parameters of the microscopic model are extracted from the inference: $\alpha^o=2.9$,  $P_0/\ell_0=4.0$, $K_A=0$ and $K_P=0.43 {\rm h}^{-1}$ (no adjustable parameter used for the fitting). See Extended Data Fig.~\ref{fig.m5} for details of the simulations with proliferation/extrusion. 
    }
\label{fig.3}
\end{figure*}

\medskip
\noindent{\bf Macroscopic rheology and phase diagram.} 
The rheological properties of an ordinary  material describe how it deforms when submitted to externally imposed forces. 
For instance, its overall rigidity is characterized by the shear modulus, which is very high for a chunk of steel and vanishes in fluids.
But tissues are living materials: their rheology also describes how they react to internally generated forces due to mechanisms ranging from stress fibers in the cytoskeleton to cell proliferation~\cite{Marchetti2013,Hallatschek2023}.
Recent works have suggested that abrupt changes in tissue rheology, akin to solid-liquid phase transitions, may play a key part in tissue morphogenesis, both in development and diseases~\cite{Petridou2019,Mongera2018,Wu2023,Lenne2022,Petridou2021}.
Intuitively, this suggests that a tissue may be represented as a point moving in a phase-diagram describing its possible phases as a function of parameters, very much like a material. However, a tissue is no ordinary material, its phases generically describe far from equilibrium steady states. 

Figure~\ref{fig.2} shows such a non-equilibrium phase diagram for the chiral dynamical system defined by the microscopic Equation~\eqref{chiral_vertex_model}, as a function of the optimal cell perimeter $P_0$ and of the cortical chirality $\alpha^{\rm o}$ ($\ell_0^2$ being the average cell area).
We observe in Fig.~\ref{fig.2}A that a phase-transition line, drawn in white, separates a region where the tissue acts as a chiral solid (dark blue) from a region where it acts as a chiral liquid (yellow/green). 
Here, liquids and solids are distinguished both by the average velocity of the cell centers and by the shear modulus $\mu$ which vanishes in the liquid phase (inset of Fig.~\ref{fig.2}A and Fig.~\ref{fig.2}C).
The phase transition shown in Fig.~\ref{fig.2}A extends to chiral tissues the rigidity transition observed in non-chiral graph models~\cite{bi2015density,Bi2016}, that we recover when $\alpha^{\rm o} = 0$.
On top of the phase diagram of Fig.~\ref{fig.2}A, we have superimposed the best fit parameters obtained using NADA for three \emph{in vitro} tissues: the HT1080 fibrosarcoma discussed above (blue triangle), C2C12 myoblasts (orange square), and human bronchial epithelial cells (HBEC; pink circles), see Methods for more details and Fig.~\ref{fig.2}B for snapshots of the tissues, OFV velocity fields, and cell segmentations.

The chiral HT1080 culture exhibits a finite and roughly time-independent $\alpha^{\rm o}$, and in the experiment we analyzed, it is far away from any phase transition.
In contrast, we infer that the C2C12 myoblast culture is in a (non-chiral) liquid phase, and is very close to the solid/liquid phase transition.
Most interestingly, the HBEC culture changes as a function of time and we observe that (i) it starts with a finite cortical chirality, that vanishes as time goes on and (ii) it approaches the liquid/solid phase transition, although it always stays in the liquid phase.
The cases of C2C12 myoblasts and HBEC are intriguing in view of the hypothesis that biological systems may function close to criticality~\cite{Mora2011,Hidalgo2014,Krotov2014,Munoz2018,Petridou2019,Lenne2022}.


\medskip
\noindent{\bf Odd viscoelasticity.} 
A precise description of the rheology of a tissue entails determining quantities known as elastic moduli and viscosities (an introduction in the context of embryonic organization is given in Ref.~\cite{Petridou2019}), or more generally frequency-dependent viscoelastic coefficients.
In our case, the chirality and activity of the tissue require a suitably chosen non-variational extension of viscoelasticity, known as odd viscoelasticity~\cite{scheibner2020odd,fruchart2023odd}.
For small deformations of the tissue, and neglecting inertia, we can describe the evolution of the displacement $\bm{u}$ as (Methods)
\begin{equation}
\label{simplified_viscoelastic}
\begin{split}
    \zeta \partial_t \bm{u} = &-\bm{\nabla} P +
    [(\mu + \eta \partial_t) + (K^{\rm o} + \eta^{\rm o} \partial_t)\bm{\epsilon}] \Delta \bm{u} \\
    &+
    [(B + \eta^B \partial_t) - (A + \eta^{A} \partial_t)\bm{\epsilon}] \bm{\nabla}(\bm{\nabla} \cdot \bm{u}).
\end{split}
\end{equation}
In this equation, the deformation field $\bm{u}$ describes the displacements of arbitrary points in the tissue with respect to a reference configuration, $\zeta$ represents friction with the substrate, $P$ is the pressure, $\mu$ and $B$ are the shear and bulk elastic moduli that describe the elasticity of usual solids, $\eta$ and $\eta^{\text{B}}$ are the shear and bulk viscosity of usual fluids, while $K^{\rm o}$, $A$, $\eta^{\rm o}$ and $\eta^{\text{A}}$ are active versions of these rheological parameters that are peculiar to nonequilibrium chiral tissues, see Methods for more details.

In Methods, we perform an analytic coarse-graining of a mean-field version of Eq.~\eqref{chiral_vertex_model} on a hexagonal lattice to estimate the viscoelastic moduli as well as numerical simulations of the full model that are in remarkable agreement with our mean-field calculations (Extended Data Fig.~\ref{fig.moduli}).
Based on these results, we can predict the mechanical response of tissues; for instance, Fig.~\ref{fig.2}D shows the time evolution of the predicted viscoelastic moduli in the HBEC tissue. 
We observe that (i) the tissue is initially chiral ($A, \eta^{\rm o} \neq 0$) and this chirality vanishes as time goes on; (ii) the viscosity of the tissue changes sign, in a similar way as in active bacteria where zero and negative viscosity has been experimentally observed~\cite{lopez2015turning}; (iii) both the passive bulk elastic modulus $B$ and the active modulus $A$ (coupling compression and torque) tend to vanish at long times, when the tissue approaches the critical point of a phase transition, in a similar way as what happens in passive critical solids~\cite{Lubensky2015,Liu2010}. Note that the shear moduli $K^o$ and $\mu$ naturally vanish in the liquid phase. 
The scaling of the moduli at the transition, illustrated in Fig.~\ref{fig.2}C, is further discussed in the Methods and Extended Data Fig.~\ref{fig.m1}.

\medskip
{\bf Living mechanics of tissue growth and flow.}
Biological tissues are made of cells that can proliferate and die, allowing the tissue to grow and change shape. 
At the level of continuum mechanics, this is captured by the continuity equation
\begin{equation}
    \partial_t \rho + \nabla \cdot [\rho  \bm{v}] = 
    [k_{\text{prolif}}(\bm{r}) - k_{\text{extr}}(\bm{r})] \rho
    \label{continuity_equation}
\end{equation}
for the density field $\rho$, in which $\bm{v} \simeq \partial_t \bm{u}$ is the velocity field that couples Eq.~(\ref{continuity_equation}) to Eq.~(\ref{simplified_viscoelastic}), while $k_{\text{prolif}}(\bm{r})$ and $k_{\text{extr}}(\bm{r})$ are respectively the rates of proliferation and extrusion (i.e. removal of cells). 
When these rates are spatially inhomogeneous, they tend to induce inhomogeneous density fields that can drive morphogenetic flows
and regulate biological function. 
We refer to the coupled system of Eqs.~(\ref{simplified_viscoelastic}-\ref{continuity_equation}) as living mechanics to stress the interplay of population dynamics and active rheology necessary to understand the response of biological tissues~\cite{Hallatschek2023}.

In chiral tissues, induced flows perpendicular to the underlying density gradients can be sustained in a steady-state.
Figure~\ref{fig.3}A-B shows chiral HT1080 cells flowing within an adhesive channel. 
In this case, we observe that cells are denser at the edge of the channel (Fig.~\ref{fig.3}D-E), suggesting higher net proliferation rate ($k_{\rm prolif}-k_{\rm extr}$) around the boundaries. This pattern may arise from either elevated proliferation at the edges or increased extrusion in the bulk, or both. 
Although the extrusion rate is unknown, we observe an inhomogeneous proliferation rate that peaks near the boundaries (Fig.~\ref{fig.3}E). Similar patterns of elevated proliferation near confinement boundaries have been observed in other cell monolayers  \cite{nelson2005emergent,Carpenter2024}.

This inhomogeneous proliferation can be interpreted as a “self-pinching”, i.e., the tissue compresses itself at the edges and increases the local density, see Fig.~\ref{fig.3}D. Such a density field can be maintained through a dynamical balance between proliferation and extrusion of cells. 
Intuitively, the self-pinching mechanism can trigger transverse forces and hence flow near the sample edge.
This macroscopic mechanism mimics at the continuum level the nature of microscopic intracellular forces: recall that $f_\perp \approx \alpha_0 f_\parallel$, where $\alpha_0$ is the microscopic cortical chirality parameter we inferred using NADA. 
We stress that the edge flows exist in the absence of self-propulsion of individual cells and stop if there is no self-pinching, i.e. no inhomogeneous net proliferation. 

The active viscoelastic theory in Eq. \eqref{simplified_viscoelastic} coupled with Eq. (\ref{continuity_equation}) allow us to predict the corresponding tissue flows: it gives the tissue velocity field $\bm{v} \simeq \partial_t \bm{u}$ as a function of gradients of the displacement $\bm{u}$, that we link to the density field using the geometric relation $\rho-\rho_0 = \rho_0 \bm{\nabla}\cdot \bm{u}$, in which $\rho_0$ is the density of the undeformed tissue. 
In the channel geometry of our experiments confined in the $x$ direction where $\partial_y \equiv 0$ (Fig.~\ref{fig.3}A), this leads to $\partial_x \rho = \rho_0 \partial_x^2 u_x$. Neglecting viscous terms in \eqref{simplified_viscoelastic}, we end up with
\begin{equation}
\begin{aligned}
\begin{pmatrix}
v_x \\
v_y
\end{pmatrix}
=
\frac{1}{\rho_0 \zeta}
\begin{pmatrix}
B+\mu\\
A-K^o
\end{pmatrix}
\partial_x \rho
\end{aligned}
\label{flow}
\end{equation}
where we recall that the active elastic moduli $A$ and $K^{\rm o}$ are only present in chiral tissues.
Hence, we expect persistent converging flows directed toward the bulk at the edge of the channel (blue arrows in Fig.~\ref{fig.3}B), in addition to chiral flows along the edge (red arrows).
This is indeed what is observed in our experiments, see Fig.~\ref{fig.3}C,F and Ref.~\cite{yashunsky2022chiral}. 


To go further, we estimate the elastic moduli using mean-field theory in the solid phase (Methods), and extrapolate the values to the liquid phase, leading to
\begin{equation}
    \frac{v_y}{v_x}  \simeq -\frac{\alpha^o}{3 - a P_0/\ell_0}\,.
    \label{continuum}
\end{equation}
in which $a = 1/[2 \sqrt{2\sqrt{3}}]$ and $\ell_0^2$ is the average area of cells in the microscopic model.
Next we substitute in Eq.~(\ref{continuum}) the values of the microsopic parameters $\alpha^o$ and $P_0/\ell_0$ inferred from bulk experimental data using NADA.
Strikingly, the ratio $v_y/v_x$ we predict without fitting any macroscopic moduli (black line in Fig.~\ref{fig.3}C) captures measurements of the experimental boundary flows for different widths (colored dots in Fig.~\ref{fig.3}C). 

These predictions are possible because our coarse-graining procedure quantitatively tracks chirality across scales by relating analytically  the macroscopic elastic moduli in the continuum theory (describing the behavior of the system near the boundary) to the microscopic parameters inferred by NADA from bulk measurements. 
The relation between macroscopic measurements of $v_y$ and $v_x$ offers an alternative method to extract the microscopic cortical chirality parameter $\alpha^o$ (Methods). This can, in turn, be used to test hypotheses about how individual proteins give rise to cell chirality \cite{tee2015cellular, yamamoto2023epithelial, lu2024polarity}. 

We can further confirm the self-pinching mechanism summarized in Fig.~\ref{fig.3}D independently of continuum theory through direct simulations of the chiral microscopic model \eqref{chiral_vertex_model} in which inhomogeneous cell proliferation and extrusion are included, see Methods for details of the implementation.
In the presence of chiral interactions between cells, particular care has to be taken to reduce discontinuities in the flow field when the number of cells changes (Extended Data Fig.~\ref{fig.m5}).
The results of our simulations are shown in Fig.~\ref{fig.3}F (dashed lines) and compared with the experiments (continuous lines).
The qualitative features of the flow are well captured by the chiral microscopic model, confirming our picture. 

Circling back to the broad quest for microscopic origins of macroscopic chirality, we envision the bioinspired approach developed in this work to shed new light on outstanding questions in classical physics such as the seemingly erratic motion of inanimate convection cells whose number fluctuates in heated fluid layers (see Extended Fig.~\ref{fig_rb}).
Beyond chirality, our noise-regularized inference algorithm (embedded in an automated model-discovery pipeline~\cite{Schmitt2024}) 
could help test hypotheses about biochemical mechanisms acting at the subcellular level from rheological measurements of a whole tissue. 

\medskip
\noindent\textbf{Acknowledgements.}
We thank Daniel Seara for critical feedback on the manuscript. The BiPMS group is a member of the Institut Pierre-Gilles de Gennes and has benefited from the technical contributions of the joint service unit Unité d'Appui et de Recherche 3750 of the French National Centre for Scientific Research. The BiPMS group is a member of the LabEx Cell(n) Scale (Grant Nos. ANR-11-LABX-0038 and ANR-10-IDEX-0001-02). We gratefully acknowledge funding from the Canceropôle Ile-de-France and the French National Cancer Institute. The research was supported by The Israel Science Foundation (grants no. 838/23, 2044/23). S.C. and M.F. acknowledge a Kadanoff–Rice fellowship funded by the National Science Foundation under award no. DMR-2011854. D.E.G acknowledges support by the NSF-Simons National Institute for Theory and Mathematics in Biology (NITMB) Fellowship supported via grants from the NSF (DMS-2235451) and Simons Foundation (MPS-NITMB-00005320). M.F. acknowledges partial support from the National Science Foundation under grant DMR-2118415 and the Simons Foundation. V.V. acknowledges partial support from the Army Research Office under grant W911NF-22-2-0109 and W911NF-23-1-0212. M.F. and V.V acknowledge partial support from the France Chicago center through a FACCTS grant. This research was partly supported from the National Science Foundation through the Center for Living Systems (grant no. 2317138), the National Institute for Theory and Mathematics in Biology, the Simons Foundation and the Chan Zuckerberg Foundation. This work was completed in part with resources provided by the University of Chicago’s Research Computing Center.

\clearpage
\makeatletter
\renewcommand{\fnum@figure}[1]{{Extended Data Fig.~\thefigure}.~}
\makeatother
\setcounter{figure}{0}
\renewcommand{\thefigure}{\arabic{figure}}
\begin{center}
    \bf METHODS
\end{center}
\section{Microscopic model}
\subsection{Chiral cells}
\begin{figure}
    \centering
    \includegraphics[width = 1.\columnwidth]{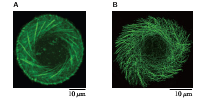}
    \caption{{\bf Examples of chiral cytoskeleton.} (A) Chiral cytoskeleton in fibroblasts, adapted from Ref.~\cite{tee2015cellular}. (B) Chiral cytoskeleton in Caco2, adapted from Ref.~\cite{yamamoto2023epithelial}. 
    }
    \label{fig.cell}
\end{figure}

In Extended Data Fig.~\ref{fig.cell} we provide two examples of chiral cytoskeletal structures in epithelial cells observed from single cell imaging. The exact origin of the chirality in the cytoskeletal structures of the three cell types studied in the present work is not known. Other chiral mechanisms such as chirally distributed myosin motors have also been reported~\cite{lu2024polarity}. Although our chiral microscopic model is inspired by the chiral cytoskeletal structures, it is also a minimal extension of traditional graph models to include the chiral intercellular force. Hence, the model is not limited to a unique origin of cellular chirality. 

\subsection{Chiral microscopic model}
Consider a two-dimensional (2D) monolayer of cells representing a confluent biological tissue (i.e. with no gap between cells).
A cell performs active mechanical action on its neighbors through its cytoskeleton -- a network formed by biopolymers such as actin and microtubules within which molecular motors generate stress through out-of-equilibrium metabolic cycles~\cite{Alberts,Fletcher2010485,chen2020motor}.
To describe the combination of these active forces with the passive mechanics of the cell's materials, we start with a conventional (achiral) graph model, known as the vertex model~\cite{farhadifar2007influence,staple2010mechanics,bi2015density,Bi2016,sussman2018no}. Cells are modeled as polygons with vertex positions $\bm r^v_j$, from which the perimeter $P_i$ and the area $A_i$ of each cell are calculated. The graph model assumes an energy
\begin{equation}
E = \sum_i \left[\frac{K_A}{2} (A_i - A_0)^2 
-K_P P_0 P_i
+ \frac{K_P}{2} P_i^2 \right]
\label{e2.1}
\end{equation}
where $P_0$ and $A_0$ are the preferred perimeter and area, respectively.  
We label by $E=E_A + E_B + E_C$ the three terms in \eqref{e2.1}.
The effective energy $E_A$ accounts for the 3D incompressibility of cells within the tissue, while $E_B$ accounts for cell-cell adhesion and $E_C$ accounts for the active contractility of the cytoskeleton \cite{Bi2016,Farhadifar2007,staple2010mechanics,chen2020motor}.
It is common to combine $E_B+E_C=\sum_iK_P[(P_i-P_0)^2-P_0^2]/2$ where it becomes apparent that these terms tend to enforce $P_i \simeq P_0$. 

There are two common methods to describe the dynamics of the graph model. One is to track the positions of all vertices ($\bm r^v_j)$, the other is to track only the positions of cell centers ($\bm r_i$), and derive vertex positions at every time step using Voronoi tessellation. Both methods yield qualitatively similar dynamics and we use the later (Voronoi model) in this paper. We assume overdamped dynamics
\begin{equation}
\begin{aligned}
&\gamma \partial_t \bm r_i = -\frac{\partial E}{\partial \bm r_i}=\sum_j\bm f^v_j \cdot \frac{\partial \bm r^v_j}{\partial \bm r_i}
\quad
\label{e0.1}
\end{aligned}
\end{equation}
in which $\gamma$ represents the friction between the cells and the substrate. In the last equality, we decompose the force acting on cell centers $-\partial E/\partial \bm r_i$ using the chain rule, in which $\bm f^v_j = -\partial E/\partial \bm r^v_j$ measures the force acting on vertices. 
The instantaneous Voronoi tessellation automatically performs T1 transitions (exchanges of neighboring cells), which modify the topology of the system and enable it to fluidize. We then eliminate $\gamma$ by absorbing it in $K_A$ and $K_P$, i.e., we let $K_A/\gamma \to K_A$ and $ K_P/\gamma \to K_P $, reducing the number of parameters.  

As shown in the main text, when chirality is present, part of the contractile forces acting on the membranes is converted to chiral forces. In our graph model, the membranes, or cell borders, are the vertices. Hence, chirality modifies the force acting on vertices to $\bm f^v_j = -\partial E/\partial \bm r^v_j+ \alpha^o \hat{\bm e}_z\times \partial E_C/\partial \bm r^v_j$, where the contractile force $-\partial E_C/\partial \bm r^v_j$ is converted to chiral forces via $\alpha^o$. Substitute this in Eq.~(\ref{e0.1}) gives Eq.~(\ref{chiral_vertex_model}) of the main text. Note that Eq.~(\ref{chiral_vertex_model}) preserves momentum. Importantly, the chiral force is non-variational, i.e., it cannot be written as a derivative of a potential energy. One direct consequence of this is the failure of the chain rule,
\begin{equation}
\begin{aligned}
\hat{\bm e}_z \times  \frac{\partial E}{\partial \bm r_i}=\hat{\bm e}_z \times  \left(\bm f^v_j \cdot \frac{\partial \bm r^v_j}{\partial \bm r_i}\right)\neq
\left(\hat{\bm e}_z \times  \bm f^v_j\right) \cdot \frac{\partial \bm r^v_j}{\partial \bm r_i}\,.
\end{aligned}
\label{chain}
\end{equation}
Equation (\ref{chain}) suggests that applying the chiral tilting on cell centers is not equivalent to applying the chiral tilting on vertices. Other consequences of the non-variational forces include the odd viscoelastic moduli which can not be derived from a potential energy or an action. Although chirality has been introduced to graph models in prior works in the forms of external torques and anisotropic line tensions~\cite{sato2015cell,yamamoto2020collective}, the non-variational forces associated with the chirality have not been studied previously. 

Our graph model builds a relation between the cell velocity $\bm v_i$ and the cell position $\bm r_i$ with four parameters $K_A$, $K_P$, $P_0$ and $\alpha^o$. Note that the parameter $A_0$ does not contribute to the velocity because it only produces a homogeneous pressure. For any mechanical prediction such as viscoelastic moduli, an additional parameter $\gamma$ (friction) will be required, as the frictional force $f$ and the velocity $v$ are related via $f=\gamma v$. To extract $\gamma$, the velocity data is not sufficient and a mechanical measurement would be required. 

\subsection{T1 transitions and singularities}
In our graph model, a vertex is typically defined by the circumcenters of three neighboring cell centers, leading to an analytical function $\bm r_v(\bm r_1,\bm r_2,\bm r_3)$, where $\bm r_v$ is the vertex position and $\bm r_i$ is the cell position (see SI for details). The set formed by all neighboring triplets defines the topology of the model. For a given topology, Eq.~(\ref{chiral_vertex_model}) is analytical: $E$ is an analytical function of $\bm r_v$ and $\bm r_v$ is an analytical function of $\bm r_i$. 

The topology changes when the cells switch neighbors through T1 transitions. Due to the different topologies before and after T1 transitions, the force exerted on cells can be discontinuous. Hence, each T1 transition is a singularity. Forces are ill-defined exactly at the T1 transitions, because four or more cell centers share the same distance to the vertex, and there is no analytical function to describe $\bm r^v(\bm r)$. These singularities also make it impossible to track a particular linear mode, because the mode changes completely after T1 transitions. 

Despite the mathematically ill definition, these singularities can be ignored in any statistical measurements. This is because the singularities are sparsely distributed points in the phase space, and the probability that a particular snapshot sits on one singularity is zero.

\section{NADA}

\subsection{Algorithm and implementation}

\begin{figure*}
    \centering
    \includegraphics[]{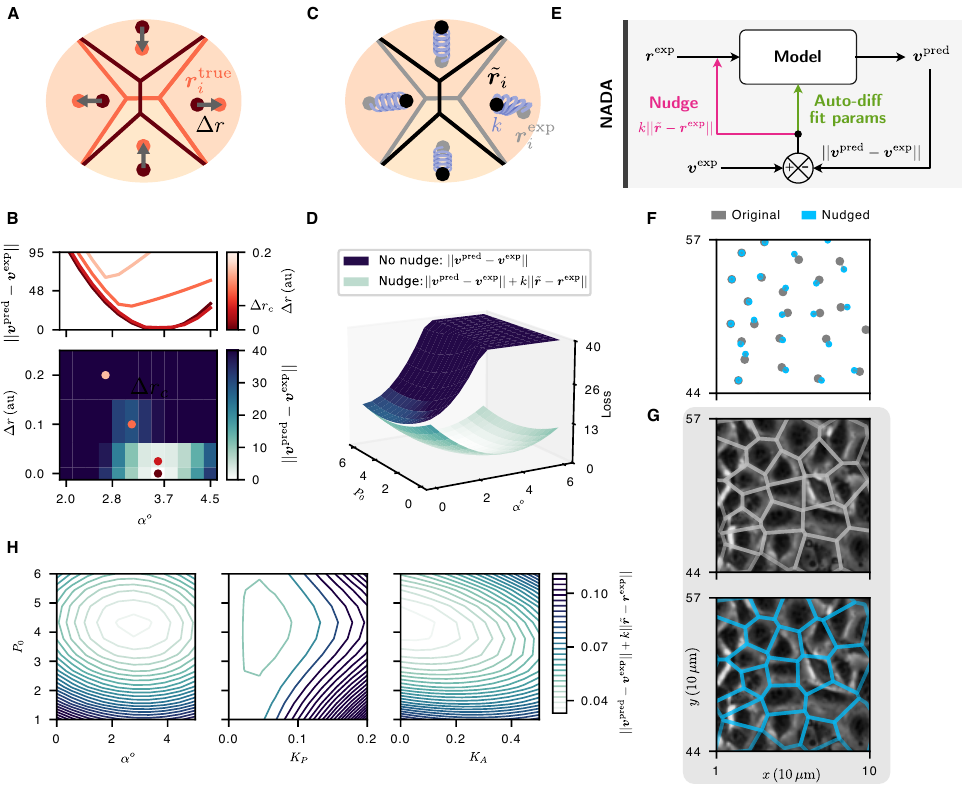}
    \caption{{\bf Nudged Automatic Differentiation Algorithm (NADA) and the Optimisation Landscape.} 
    (A) A small parametrized displacement $\Delta r$ to nuclei positions, such that $\Delta r > \Delta r_{\rm c}$ leads to a T1 transition.
    (B) The loss landscape $||\bm{v}^{\rm pred}-\bm{v}^{\rm exp}||$ in a synthetic system undergoing an abrupt change across this T1 transition, where the optimum jumps to an incorrect parameter value with a small displacement $\Delta r_{\rm c} \sim 1/30$ of typical cell size.
    Such alterations of the network topology can be inadvertently introduced by noise, for example during segmentation of experimental data.
    (C) Corrupted topology can be repaired if we nudge the positions $\tilde{\bm{r}}$ during optimisation of parameters, while enforcing proximity to the original segmented positions $\bm{r}^{\rm exp}$ by attaching springs with stiffness $k$.
    (D) Despite adding the non-negative harmonic potential, the global minimum of the new optimisation loss function $||\bm{v}^{\rm pred}-\bm{v}^{\rm exp}||+ k ||\tilde{\bm{r}}-\bm{r}^{\rm exp}|| $ is much lower than the original loss $||\bm{v}^{\rm pred}-\bm{v}^{\rm exp}||$ (for which the noisy positions $\bm{r}^{\rm exp}$ are quenched).
    The correct model parameters of the synthetic system can then be inferred through gradient descent on the nudged landscape.
    (E) Therefore, to infer the parameters of the vertex from experimental data, we use a nudged version of automatic differentiation. In addition to fitting the model parameters (green arrow), the segmented input position data $\bm{r}^{\rm exp}$ is also modified slightly (pink arrow) to improve the fit of the model flow $\bm{v}^{\rm pred}$ to the experimental flow $\bm{v}^{\rm exp}$ obtained by OFV.
    (F) Comparison of the original cell positions in gray $\bm{r}^{\rm exp}$ (obtained by segmenting experimental images) and the nudged positions $\tilde{\bm{r}}$ after inference optimisation.
    (G) The Voronoi tessellation with original positions $\bm{r}^{\rm exp}$ and nudged positions $\tilde{\bm{r}}$, overlayed with an image of HT1080 tissue.
    (H) Slices of the nudged loss landscape for the HT1080 tissue, indicating the optimal model parameters inferred from the experiment.
    Moreover, sloppy and stiff model parameters can be identified by inspecting the contours.
    }
    \label{fig.m0}
\end{figure*}
We detail here the implementation of NADA. 

The input of NADA is the positions of cell centers $\bm r_i^{\rm exp}$ and the velocity field $\bm v^{\rm exp}$. The spacing of the velocity field is coarse-grained to $25\rm \mu m$, close to the size of a single cell. In the examples we show in the main text, these data are obtained from videos of epithelial monolayers. $\bm r_i^{\rm exp}$ at a particular time $t$ is extracted from the segmentation of that particular snapshot, and the center position of cell $i$ is defined as the average position of all pixels belonging to cell $i$. $\bm v^{\rm exp}$ is calculated by optical flow velocimetry (OFV) using two adjacent snapshots. 

While the OFV provides relatively reliable velocity data, the segmentation results are usually less accurate. Mistakenly recognized cell borders result in errors $\Delta \bm r_i =\bm r_i^{\rm exp}-\bm r_i^{\rm true}$ between measured positions and true positions $\bm r_i^{\rm true}$. Our graph model is very sensitive to these errors because they can trigger topological changes, see Extended Data Fig.~\ref{fig.m0}A for example. We study the effect of these errors in a synthetic system where we introduce artificial noises on cell positions with magnitude $\Delta r$, and try to infer the parameters from the contaminated data. Our goal is to identify the best set of four parameters, $K_P$, $P_0$, $K_A$ and $\alpha^o$ ($A_0$ does not affect the velocity field; it only contributes to a constant pressure). To do so we define a loss function $\mathscr{L}=||\bm v^{\rm pred}-\bm v^{\rm exp}||$, i.e., the distance ($L^2$ norm) between the predicted velocity field $\bm v^{\rm pred}$ and the measured velocity field. $\bm v^{\rm pred}=F_p(\bm r^{\rm exp})$ is calculated from Eq.~(\ref{chiral_vertex_model}) using $\bm r_i^{\rm exp}$ and a set of model parameters $\bm p$. In this numerical experiment, we fix the rest of the parameters and infer only $\alpha^o$, with the ground truth $\alpha^o=3.4$. This is realized through an automatic differentiation algorithm that is developed based on Ref.~\cite{shumilin2023method}. In Extended Data Fig.~\ref{fig.m0}B we plot the loss landscape as function of both $\alpha^o$ and $\Delta r$. We find that as long as the noise does not trigger a topological change, i.e., $\Delta r<\Delta r_c$, the loss landscape stays nearly the same as the noise-free landscape at $\Delta r=0$, and we have an accurate inference of $\alpha^o$. However, when $\Delta r$ exceeds $\Delta r_c$, the loss landscape changes drastically with a discontinuous jump, and the inference result is no longer reliable. 

In NADA, the noises on cell centers are regularized by nudging the centers to restore the true model topology. Although we do not know the true positions $\bm r_i^{\rm true}$ of the experimental data, the topological information is hidden in $\bm v^{\rm exp}$: Only the correct topology can reproduce the observed velocity field. Specifically, we nudge the cell centers to new positions $\tilde{\bm r}_i=\bm r_i^{\rm exp}+\bm \epsilon_i$, with $\bm \epsilon_i$ being trainable parameters. For a given set of model parameters $\bm p$, we calculate the predicted cell velocities using Eq.~(\ref{chiral_vertex_model}) and $\tilde{\bm r}_i$, and coarse grain it to the predicted velocity field $\bm v^{\rm pred}=F_p( \tilde{\bm r})$. We then minimize the loss function $\mathscr{L}=||\bm v^{\rm pred}-\bm v^{\rm exp}|| + k|| \bm \epsilon||$. In the loss function, the first term measures the distance between $\bm v^{\rm pred}$ and $\bm v^{\rm exp}$. The second term attaches effective springs between the nudged positions and the original positions, see Extended Data Fig.~\ref{fig.m0}C. This prevents cells from being nudged too far from their original positions. 

In the algorithm, we first minimize the loss function with respect to both $\bm p$ and $\Delta \bm r$ using stochastic gradient descent (SGD), which performs nudging. A key challenge is finding an appropriate spring constant $k$: if $k$ is too large, the nudging is effectively inhibited; if it is too small, cells can move arbitrarily, which destroys all information contained in the original positions. We choose $k$ based on the following principle: we let $||\bm v^{\rm pred}-\bm v^{\rm exp}||$ and 
$k||\Delta \bm r||$ contribute similarly to loss function in the final state of SGD. This ensures that the springs and the velocity difference have similar constraining effects on the parameters. We test NADA using the synthetic data and show the loss landscape in Extended Data Fig.~\ref{fig.m0}D. We see that NADA yields much smaller loss than the standard inference without nudging, despite the additional non-negative term in the loss function. It also reaches a more accurate inference result. The algorithm is graphyically summarized in Extended Data Fig.~\ref{fig.m0}E.

We then perform nudging on experimental data of HT1080. In Extended Data Fig.~\ref{fig.m0}FG we compare the cell positions and the Voronoi tessellations before and after nudging. We find that the nudging does not significantly modify cell positions, yet it greatly improves the inference results. 
To further increase the accuracy of the inferred parameters, once the nudging is done, we fix $\tilde{\bm r}$ and calculate the loss landscape for varying $\bm p$. We then find the minimum of the loss landscape, and use $\bm p$ there as the final parameters. Examples of the loss landscapes of HT1080 are shown in Extendend Data Fig.~\ref{fig.m0}H. 

\subsection{Results}
\begin{table*}
\begin{tabular}{||c c c c c c c c c c||} 
 \hline
cell & phase & $P_0/\ell_0$ & $\alpha^o$ & $B$ & $A$ & $\mu$ & $K^o$ & $\eta$ & $\eta^o$\\ [0.5ex] 
 \hline\hline
 HT1080 & chiral liquid & $4.0$ & $2.9$ & 2.2& -7.1 &0  &0 & -0.1 &-0.09 \\ 
 \hline
 \makecell{HBEC (start)} & chiral liquid & 4.5 & -0.8 & 3.6 &1.0 &0 &0 & -0.02 & 0.05 \\
 \hline
\makecell{HBEC (end)} & liquid & 4.0 & 0.0 & 0.0 & 0.0 &0 &0 & 0.07 & 0.0\\
 \hline
myoblast & liquid & 3.9 & 0.0 & 0.0 & 0.0 &0 &0 & 0.07 & 0.0\\
 \hline
\end{tabular}
\caption{Inference results and predicted viscoelastic moduli. Elastic moduli are in unit of $\gamma K_P$. Viscous moduli are in unit of $\gamma$. }
\label{table.infer}
\end{table*}
Here we detail the infer results shown in Fig.~\ref{fig.2}A. 

For HT1080 we show infer results at $t=12.5 \rm h$, $t=17.5 \rm h$ and $t= 22.5 \rm h$. All results exhibit strong chirality $\alpha^o=2.9$.  $P_0/\ell_0$ has a tendency to increase with time, which is consistent with the observed increase in average cell density: the average cell size $\ell_0$ decreases $20\%$ within $24$ hours. The increase of $P_0/\ell_0$ is slow though and we approximate the tissue by a steady-state with $P_0/\ell_0\approx 4.0 \pm 0.2$. The inferred parameters predict that the tissue is deeply in the chiral chaotic liquid phase, which is consistent with the observed drastic velocity field. For myoblasts we obtain vanishing chirality, indicating a non-chiral liquid phase, in alignment with the observed nearly frozen tissue. For HBEC we show time evolution of the parameters from $t=29.4 \rm h$ to $t=42.8 \rm h$, see Fig.~\ref{fig.2}D for the exact time points. The inferred parameters predict that the tissue switches from the chiral chaotic liquid phase to the non-chiral liquid phase, in agreement with the observed slowing down of the velocity field. Interestingly, while the average cell density for HBEC also increases with time, the inferred $P_0/\ell_0$ value decreases with time. This suggests that the absolute value of $P_0$ must also decrease with time, i.e., cells become less adhesive to each other as the density increases.

The inferred model parameters are used in numerical simulations to calculate viscoelastic moduli. Interestingly, we infer $K_A=0$ for all cells. While $K_A$ should be positive due to the 3D incompressibility, the vanishing inferred $K_A$ means that $K_A\ll K_P \ell_0^2$ for all cells, such that the effect of $K_A$ on the velocity field is negligible. The vanishing $K_A$ also leads to vanishing bulk modulus $B$ in the non-chiral case. Because the inference uses only velocity data without mechanical measurements, the absolute magnitude of the moduli can be rescaled by an unknown parameter $\gamma$ (friction). However, the relative magnitude of the moduli can still be obtained. 

The results are summarized in Table.~\ref{table.infer}. 

\section{Experimental methods}

\subsection{Cell culture}
Human bronchial epithelial cells (HBECs), generously provided by J. Minna’s laboratory (Dallas, TX), were cultured in a supplemented keratinocyte serum-free medium with L-glutamine (Keratinocyte-SFM with L-glutamine; Gibco). HT1080 cells (ATCC CCL-121) were maintained in Dulbecco’s modified Eagle’s medium/ high glucose, L-glutamine and sodium pyruvate (Hyclone, Cytiva, Cat no. SH30243.01). C2C12 myoblast cells (ATCC CRL-1772) were cultured in Dulbecco’s modified Eagle’s medium (ATCC). All cell media were supplemented with $10~\%$ fetal bovine serum (Gibco, Thermo Fisher Scientific, Cat no.A5256701), and penicillin (100 U/mL)-streptomycin (0.1 mg/mL) (Hyclone, Cytiva, Cat no. SV30010). 
Cell cultures were  passaged once a week and incubated at $37~^{\circ}C$ under $5~\%$ $CO_2$ partial pressure and $95~\%$ relative humidity. For time-lapse experiments, cells were seeded in polystyrene 6- or 12-well tissue culture plates, and imaging was started once the cultures reached confluence. 
\subsection{Live cell microscopy}
Time-lapse microscopy experiments were performed on automated IX71 (Olympus) and Axio observer Z1 (Zeiss) inverted microscopes. 
Both microscopes were equipped with temperature, humidity, and $CO_2$ regulations (Life Imaging Services GmbH; Switzerland) maintain cell conditions as in culture.
Phase contrast images were acquired using a $5\times$ and $10\times$ objectives. The intervals between images were set to 5 or 15 minutes.
\subsection{Image processing}
Image analysis was conducted using custom Python code. We utilized the Gunnar Farneback method from the OpenCV library to derive the velocity field from microscopic images, generating displacement vectors for every pixel across successive images of the same field, rather than relying on a sparse set of points. Flow fields were computed with an averaging window size of $11.1 \times 11.1 {\rm \mu m}^2$. To obtain flow profiles, the flow fields were averaged over time, space, and across stripes of the same width.

Cell segmentation was performed from phase contrast images using a pre-trained deep learning network (Cellpose 3.0) using the "cyto3" model \cite{stringer2025cellpose3}. The cell diameter of $20 {\rm \mu m}$ was set to reflect the typical diameter of individual cells in the image.

The proliferation rate of HT1080 cells was estimated using a Random Forest classifier. Phase-contrast images were used for classification, leveraging the characteristic round morphology and bright appearance of dividing cells. The model was trained on a dataset of images that had been manually annotated to identify dividing cells. The number of pixels predicted as dividing cells was then averaged temporally and spatially, and across stripes of equal width, and normalized to the corresponding value in unconfined cultures.

\section{Continuum description of tissues}


Biological tissues are complex materials that are neither solid or fluid. Their mechanical properties are summarized by a complex frequency-dependent dynamical moduli tensor ${\cal A}_{ijk\ell}(\omega)$ that encodes the viscoelastic response of the material: a small oscillatory strain $\epsilon_{kl}(\omega) e^{i\omega t}$ induces an oscillatory stress $\sigma_{ij}(\omega) e^{i\omega t}$ (on top of the prestress $\sigma^0$), related by $\sigma_{ij}(\omega) ={\cal A}_{ijk\ell}(\omega) \epsilon_{k\ell}(\omega)$. 

In this section, we review the continuum description of biological tissues adopted in the main text.
Our approach closely follows Refs.~\cite{Alert2020,CochetEscartin2014,Recho2016,Ranft2010,marchetti2013hydrodynamics,Julicher2007,Prost2015}, in which we incorporate the effect of chirality through odd viscoelasticity~\cite{fruchart2023odd,avron1998odd,scheibner2020odd,banerjee2021active,banerjee2017odd,chen2021realization}.
In short, the chiral and non-equilibrium character of the system is encoded at the level of the continuum theory in the fact that one can have ${\cal A}_{ijk\ell}\neq {\cal A}_{k\ell ij}$.

\subsection{Balance equations}

The fluid is described by balance equations for mass and linear momentum. We assume that angular momentum quickly relaxes, and we also ignore cell polarity or nematic alignment (see discussion in SI). In addition, a dynamic constitutive equation for the stress has to be considered.

The mass balance equation takes the form
\begin{equation}
    \partial_t \rho + \partial_i [\rho v_i] =  k(\rho) \rho
\end{equation}
in which $\rho(t,\bm{r})$ is the cell density, $\bm{v}(t,\bm{r})$ is the tissue velocity, while $k(\rho)$ is a net cell proliferation rate that combines cell divisions and deaths.
In principle, $k$ can also depend on $\bm{v}$, on the stress $\bm{\sigma}$, or be explicitly spatially dependent.

The linear momentum balance equation takes the form
\begin{equation}
    \rho D_t v_i = - \zeta v_i + \partial_j \sigma_{ij} + f^{\text{ext}}_i 
\end{equation}
in which $D_t = \partial_t + \bm{v} \cdot \nabla$ is the convective derivative, $\zeta$ represents cell–substrate friction, $\sigma_{ij} = \sigma_{ij}^{\text{h}} + \sigma_{ij}^{\text{v}}$ is the stress tensor (decomposed into a hydrostatic part $\sigma_{ij}^{\text{h}}$ present in the fluid at rest plus a viscoelastic part $\sigma_{ij}^{\text{v}}$), and $\bm{f}^{\text{ext}}$ are other external forces.
The hydrostatic stress is given by an equation of state, which we take to be $\sigma_{ij}^{\text{h}} = -p \delta_{ij}$ where $p = c_{\text{s}}^2 \rho$ and $c_{\text{s}}$ is the speed of sound.
One could also consider an odd hydrostatic stress accounting for torques in the tissue at rest.
It is often assumed that inertial effects can be neglected, so that the tissue is always force-balanced, namely
\begin{equation}
    \zeta v_i \simeq \partial_j \sigma_{ij} + f^{\text{ext}}_i 
\label{overdamped}
\end{equation}

\subsection{Chiral active viscoelasticity}
\label{chiral_active_viscoelasticity}

Finally, one typically considers a constitutive equation for the viscoelastic stress of the form~\cite{banerjee2021active,Banerjee2021err,Floyd2024}
\begin{equation}
\mathscr{D}^{(a)}_t \sigma_{ij} = \kappa_{ijk\ell} \partial_\ell v_k - R_{ijk\ell} \sigma_{k\ell} 
    \label{JSM}
\end{equation}
Equation \eqref{JSM} describes a viscoelastic medium, which exhibits both viscous and elastic behaviors.
These are encoded in the parameters $\kappa_{ijk\ell}$ (akin to an elastic tensor) and $R_{ijk\ell}$ (which is a tensor of relaxation rates).
The choice and physical meaning of these parameters is discussed below.
In addition, $\mathscr{D}^{(a)}_t$ is the Gordon–Schowalter derivative defined as~\cite{Beris1994,Bodnar2022,Gordon1972}
\begin{equation}
    \mathscr{D}^{(a)}_t X = D_t X + \Omega X - X \Omega + a(E X + X E)
    \label{GSD}
\end{equation}
for a rank-two tensor field $X_{ij}$, 
in which $\Omega_{ij} = (\partial_i v_j - \partial_j v_i)/2$ is the vorticity tensor while $E_{ij} = (\partial_i v_j + \partial_j v_i)/2$, and matrix products are implied (i.e. $\mathscr{D}^{(a)}_t X_{ij} = D_t X_{ij} + \Omega_{ik} X_{kj} + \cdots$).
Equation \eqref{JSM} is known as the Johnson-Segalman model~\cite{Johnson1977} (here generalized to odd viscosity and elasticity following Refs.~\cite{banerjee2021active,Banerjee2021err,Floyd2024}).
The parameter $a$ is known as the slip parameter, and interpolates between various convected derivatives ($a = 1$ corresponds to the upper-convected derivative; $a = 0$ to the corotational derivative; $a = -1$ to the lower-convected derivative) that entail different rheological models (see Refs.~\cite{Beris1994,Bodnar2022,Gordon1972,Johnson1977,Hinch2021,Castillo2022,Eggers2023,Stone2023} for details, discussions, and controversies).  
Here, we focus on the linear response where the choice of the convected derivative is immaterial. 

Viscoelastic media can be described as a combination of elasticity and viscosity in different ways (in the same way as an electrical impedance can be decomposed into resistance, inductance, and capacitance in different ways).
Equation~\eqref{JSM} corresponds to a Maxwell material, in which viscosity and elasticity are added in series. Another description, known as a Kelvin-Voigt material, consists in adding these in parallel.
As in electronics, these descriptions are mathematically equivalent but do not involve the same viscosity and elasticity tensors. 
To illustrate this, we linearize Eq.~\eqref{JSM} and go to Fourier space (in frequency/time), leading to an equation of the form
\begin{equation}
    \dot{e}_{ij} = \mathcal{C}_{ijk\ell}(\omega) \sigma_{k\ell}
    \label{compliance}
\end{equation}
in which $\dot{e}_{k\ell} \equiv \partial_\ell v_k$, and in which $\mathcal{C}(\omega)$ is known as the compliance tensor, which can be decomposed, at lowest orders in $\omega$, as  
\begin{equation}
    \label{CMaxwell}
    \mathcal{C} = \ii \omega \kappa_{\text{M}}^{-1} + \eta_{\text{M}}^{-1}
\end{equation}
where $\kappa_{\text{M}}^{-1}$ has the dimension of the inverse of an elastic stiffness, while $\eta_{\text{M}}^{-1}$ has the dimension of the inverse of a viscosity.
In Eq.~\eqref{CMaxwell}, the quantities $\mathcal{C}$, $\kappa_{\text{M}}^{-1}$, and $\eta_{\text{M}}^{-1}$ are rank-4 tensors. They can alternatively be interpreted as linear operators (matrices), as described in Sec.~\ref{matrix_repr} and Ref.~\cite{fruchart2023odd}. 
Equation \eqref{CMaxwell} corresponds to the Maxwell representation.

Equivalently, one can write Eq.~\eqref{compliance} as
\begin{equation}
    \sigma_{ij} = \mathcal{A}_{ijk\ell}(\omega) \dot{e}_{k\ell}
\end{equation}
in which $\mathcal{A}(\omega)$ is a generalized elasticity tensor which can be decomposed as
\begin{equation}
    \mathcal{A} = \frac{1}{\ii \omega} \kappa_{\text{KV}} +\eta_{\text{KV}}
\end{equation}
in which $\kappa_{\text{KV}}$ has the dimension of an elastic stiffness, while $\eta_{\text{KV}}$ has the dimension of a viscosity
This corresponds to the Kelvin-Voigt representation.
Both representations are related through $\mathcal{C} = \kappa^{-1} [\ii \omega + R]$ and $\mathcal{A} = \mathcal{C}^{-1}$ (in which the inverse is a matrix inverse in the matrix representation of Sec.~\ref{matrix_repr}).

The tensor $\mathcal{A}$ is related to the storage and loss moduli $G'$ and $G''$ through $\ii \omega \mathcal{A}(\omega) = G'(\omega) + \ii G''(\omega)$, where $G'(\omega) \equiv C(\omega)$ can be seen as a frequency-dependent elasticity tensor and $\eta(\omega) \equiv G''(\omega)/\omega$ as a frequency-dependent viscosity tensor.
For passive systems, the symmetry ${\cal A}_{ijkl}={\cal A}_{klij}$ must hold. 
However, in chiral active systems the symmetry is in general broken, i.e., ${\cal A}_{ijkl}\neq {\cal A}_{klij}$~\cite{banerjee2017odd,scheibner2020odd,banerjee2021active,chen2021realization,fruchart2023odd}.

Here we focus on the long-time response of the tissues, i.e., the limit $\omega \to 0$. As long as $C_{ijk\ell} \equiv C_{ijkl}(\omega \to 0) \neq 0$, we have ${\cal A}_{ijkl}(\omega \to 0) = C_{ijk\ell}$ and the corresponding long-time response is elastic, while it is viscous otherwise. All viscoelastic moduli outside of this section denote their values in the $\omega \to 0$ limit unless otherwise mentioned.




\subsection{Matrix representation of the stress-strain relation}
\label{matrix_repr}

In the paper the stress and the strain are decomposed into four base modes. For arbitrary displacement field $\bm u(\bm x)$, its components in the four base modes are: $\raisebox{-0.2ex}{\includegraphics[height=1.6ex]{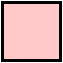}}=\partial_x u_x+\partial_y u_y$ (isotropic expansion), $\raisebox{-0.3ex}{\includegraphics[height=2.1ex]{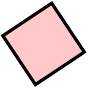}}=\partial_x u_y-\partial_y u_x$ (rotation), $\raisebox{-0.0ex}{\includegraphics[height=1.2ex]{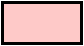}}=\partial_x u_x-\partial_y u_y$ (pure shear 1) and $\raisebox{-0.2ex}{\includegraphics[height=1.8ex]{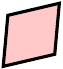}}=\partial_x u_y+\partial_y u_x$ (pure shear 2). The corresponding stress components of a stress tensor 
$\sigma$ are $\raisebox{-0.8ex}{\includegraphics[height=3ex]{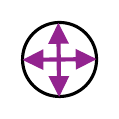}}=(\sigma_{xx}+\sigma_{yy})/2$ (isotropic expansion), $\raisebox{-0.4ex}{\includegraphics[height=2.2ex]{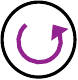}}=(\sigma_{yx}-\sigma_{xy})/2$ (rotation), $\raisebox{-0.6ex}{\includegraphics[height=2.2ex]{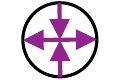}}=(\sigma_{xx}-\sigma_{yy})/2$ (pure shear 1) and $\raisebox{-0.7ex}{\includegraphics[height=2.8ex]{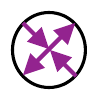}}=(\sigma_{yx}+\sigma_{xy})/2$ (pure shear 2). 

For elastic materials, the stress-strain relation of isotropic and rotational-invariant systems follows a generic form~\cite{scheibner2020odd,fruchart2023odd}. The two shear components are related by
\begin{equation}
  \includegraphics[width=0.25\textwidth]
  {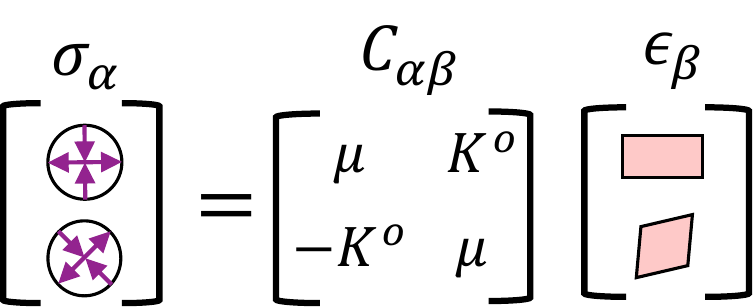}  
  \label{odd_1}
\end{equation}
The isotropic expansion and the rotation are related by
\begin{equation}
  \includegraphics[width=0.3\textwidth]
  {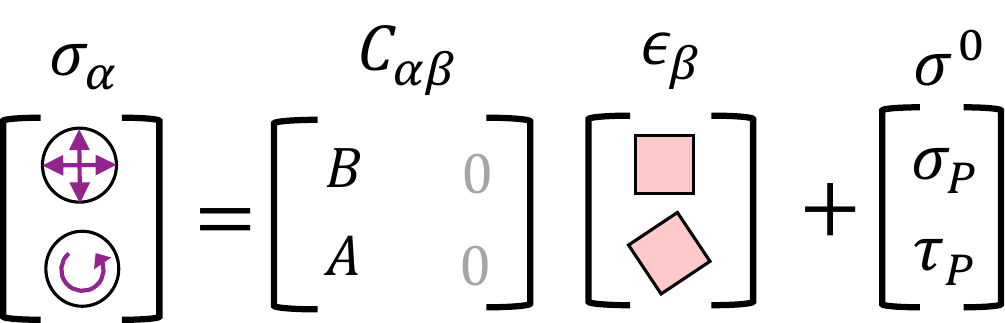}  
  \label{odd_2}
\end{equation}
with $\sigma_P$ and $\tau_P$ being the pre-isotropic-stress and the pre-torque.

In the viscous case, the coupling between the shear stress and the shear strain rate becomes
\begin{equation}
  \includegraphics[width=0.23\textwidth]{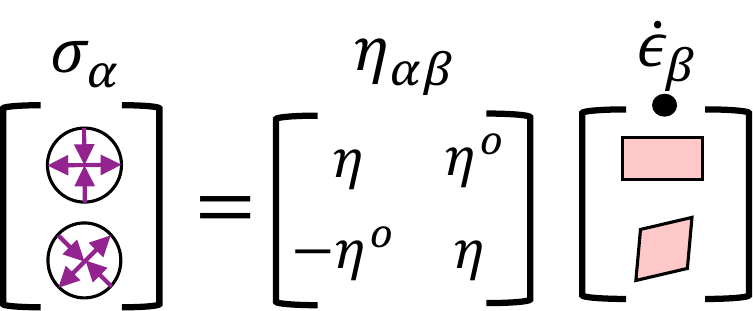}  
  \label{odd_3}
\end{equation}
and the relation between the isotropic expansion and the rotation is
\begin{equation}
  \includegraphics[width=0.3\textwidth]{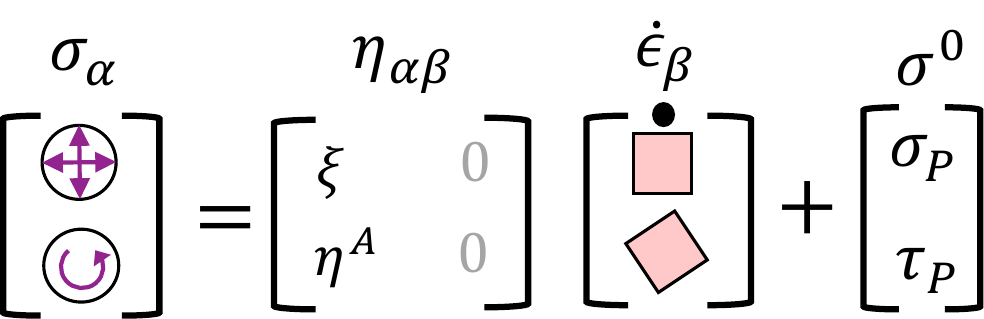}  
  \label{odd_4}
\end{equation}

\section{Mean-field theory}
\begin{figure*}[t]
\centering
\hspace{-2em}\includegraphics[width = 2.\columnwidth]{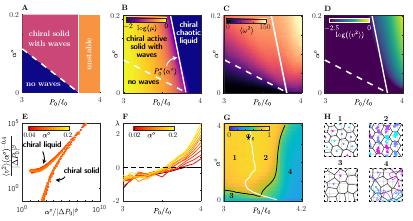}
\caption{{\bf Phase diagram and dynamic rigidity transition.  } (A) Phase diagram predicted by the MFT based on a hexagonal lattice. Solid white line separates the diagram to a stable solid phase and an unstable phase (corresponding to liquid). Dashed white line is a line of exceptional points, and also the onset of odd elastic waves. (B) Numerical results of the shear modulus $\mu$. $\mu$ vanishes at the solid white line $P_0^*(\alpha^o)=3.81-|\alpha^o|$, indicating the dynamical phase transition from chiral active solid to chiral chaotic liquid. Dashed white line indicates the onset of odd elastic waves, see (C). In (B) the data of $\mu$ is smoothed to reduce the effect of noise: The data point at $(P_0,\alpha^o)$ represents the averaged $\mu$ in the range $P_0\pm0.1$ and $\alpha^o\pm 0.04$. (C) Numerical results of the oscillation frequency $\omega$ (imaginary parts of the eigenvalues of the Jacobian matrix).  Dashed white line indicates the onset of odd elastic waves defined by $\langle \omega^2\rangle = 40$. (D) Numerical results of the steady-state cell velocity. (E) Widom-like scaling function of the cell velocity $\langle v^2 \rangle$. Scaling exponents are $\phi=4$ and $ \beta = 3$.   (F) Numerical results of the largest Lyapunov exponent $\lambda$. $\lambda$ for $\alpha^o=0$ is shown by the black line. (G)  Numerical results of the hexatic order parameter $\Psi_6$. Black solid line indicates $\Psi_6=0.7$, and the white solid line is the solid-liquid phase transition obtained from Fig.~\ref{fig.2}A. (H) Tissue configuration and cell trajectories for four phases in (G). 1: hexatic solid. 2: hexatic liquid. 3: disordered solid. 4: disordered liquid. We have set $A_0=\ell_0^2=1$, $K_P=1$ and $K_A=0.1$ in the simulations.
    }
    \label{fig.m1}
\end{figure*}

\begin{figure*}[t]
\centering
\hspace{-2em}\includegraphics[width = 1.8\columnwidth]{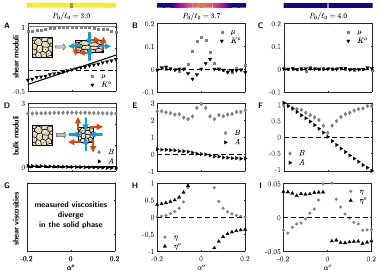}
\caption{{\bf Active viscoelasticity.} Numerical results of viscoelasticitic moduli (symbols). MFT results are shown in solid lines for comparison. For each set of parameters used we show the corresponding shear modulus $\mu$ on top, using the same color code in Extended Data Fig.~\ref{fig.m1}B, where the squares indicate $\alpha^o=0$ where the model reduces to the non-chiral graph model.  (A-C) Normal shear modulus $\mu$ and odd shear modulus $K^o$. Inset of (A): under a pure shear deformation, the normal forces on the boundaries (blue) produce a shear stress in the same direction as the strain.The chiral forces (red) produce a shear stress in a different direction, corresponding to the odd shear moduli $K^o$.  (D-F) Normal bulk modulus $B$ and odd bulk modulus $A$. Inset of (D): under an isotropic expansion, the normal forces on the boundaries (blue) produce an pressure. The chiral forces (red) produce a torque, corresponding to the odd bulk modulus $K^o$.   (G-I) Normal shear viscosity $\eta$ and odd shear viscosity $\eta^o$. We have set $\gamma=1$, $A_0=\ell_0^2=1$, $K_P=1$ and $K_A=0.1$ in the simulations. 
    }
	\label{fig.moduli}
\end{figure*}
Here we detail the derivation of the mean-field theory (MFT) for the chiral graph model. In the MFT, we assume a ground state of uniformly distributed cells which forms a honeycomb lattice-like Voronoi tessellation (with side length $d$). All cell centers form a triangular lattice with discrete translational symmetry with three nearest neighbor bond vectors
\begin{equation}
\begin{aligned}
\bm a^{(1)}&=\sqrt 3 d
\begin{pmatrix}
\sqrt 3 / 2 \\
1/2
\end{pmatrix},\\
 \bm a^{(2)}&=\sqrt 3 d
\begin{pmatrix}
0 \\
1
\end{pmatrix},\\
\bm a^{(3)}&=\sqrt 3 d
\begin{pmatrix}
-\sqrt 3 / 2 \\
1/2
\end{pmatrix}
\label{m3.1}
\end{aligned}
\end{equation}
Each cell center has six nearest neighbors. The rest three neighbor bond vectors are $\bm a^{(4)}=-\bm a^{(1)}$, $\bm a^{(5)}=-\bm a^{(2)}$ and $\bm a^{(6)}=-\bm a^{(3)}$. Note that the mean cell area $\ell_0^2 = (3\sqrt{3}/2)d^2 = 1$, and we have $d /\ell_0= 2^{1/2} /3^{3/4}\approx 0.62$. Each cell center has 6 vertices with vectors ${\bm b}^{(n)}=({\bm a}^{(n+1)}+{\bm a}^{(n+2)})/3$. For simplicity we also assume $K_A=0$ because it does not qualitatively affect the fluid-solid phase transition in graph models~\cite{Bi2016}. 

In the reference state, all cells feel zero net force. In SI, we analytically calculate how the cells respond to small displacement field $\bm u(\bm x)$ of the cell centers. 
 This allows us to extract a Fourier-space dynamical matrix $D(\bm q)$, which governs  all linear waves $\bm u(\bm x) \sim\exp(i\bm q\cdot \bm x-i\omega t)$  via 
\begin{equation}
\begin{aligned}
i\omega &\begin{pmatrix}
&u_\parallel\\
&u_\perp 
\end{pmatrix}= -D(\bm q)\begin{pmatrix}
& u_\parallel\\
&u_\perp 
\end{pmatrix}
\label{s6}
\end{aligned}
\end{equation}
where $u_\parallel$ and $u_\perp$ are the longitudinal and transverse components of $\bm u$ with respect to $\bm q$. The value of $\omega$ determines the stability and the oscillation of the linear modes, which is discussed in the main text. 

Due to the lattice structure $D(\bm q)$ is anisotropic with discrete rotational symmetry. We are interested in its behavior in the 12 directions $\bm a^{(n)}$ and $\bm b^{(n)}$. These directions are special because the graph model is made up of the nearest-neighbor and the next-nearest-neighbor interactions, and all nearest neighbors and next-nearest neighbors are in the directions of $\bm a^{(n)}$ or $\bm b^{(n)}$. For wave vector in $\hat{\bm e}_x$ (parallel to $\bm b^{(2)}$), we have
\begin{equation}
\begin{aligned}
&D(q\hat{\bm e}_x) = \frac{4K_P\sin^2(3dq/4)}{d} \cdot\\ &\begin{pmatrix}
& [4+2\cos(3dq/2)]d-P_0/3   & 2\alpha^od\\
&[-4+2\cos(3dq/2)]\alpha^od & 6d-P_0
\end{pmatrix},
\label{s6}
\end{aligned}
\end{equation}

We find that the most unstable mode is located at $q=2\pi/(3d)$. This mode determines the stability of the model, because as long as $q=2\pi/(3d)$ is stable, all other modes must also be stable, see SI. The dynamical matrix of this mode is 
\begin{equation}
\begin{aligned}
&D(q\hat{\bm e}_x) = \frac{4K_P}{d}  &\begin{pmatrix}
& 6d-P_0/3   & 2\alpha^od\\
&-2\alpha^od & 6d-P_0
\end{pmatrix},
\label{s6}
\end{aligned}
\end{equation}
with characteristic frequencies
\begin{equation}
\begin{aligned}
\omega(q)=2i{K_Ps}\left[2\pm\sqrt{1-3\epsilon^2}\right]
\label{s8}
\end{aligned}
\end{equation}
where $P_0^{\rm MF}=6d\approx 3.72 \ell_0$, $s = 1-P_0/P_0^{\rm MF}$ and $\epsilon = \alpha^o/s$. It is stable for $P_0<P_0^{\rm MF}$ and unstable for $P_0>P_0^{\rm MF}$.  In addition, when $|\epsilon|>1/\sqrt 3$ we expect oscillation of the linear wave, which is also called odd elastic wave~\cite{scheibner2020odd}. $|\epsilon|=1/\sqrt 3$ is thus an exceptional point where the two eigenvectors coalesce. Note that the exceptional point can vary for different $q$. $\bm q=q\hat{\bm e}_y$ (parallel to $\bm a^{(2)}$) gives similar results, see SI. 

In the small $|\bm q|$ limit, the dynamical matrix reduces to
\begin{equation}
\begin{aligned}
D(\bm q) = \frac{3K_Pdq^2}{4}\begin{pmatrix}
& (18d-P_0)   & -18\alpha^od\\
&18\alpha^od & 3 (6d-P_0) 
\end{pmatrix}.
\label{s5}
\end{aligned}
\end{equation}

This dynamical matrix corresponds to elastic moduli
\begin{subequations}
\begin{align}
    B &= 0.31 \gamma K_P(P_0/\ell_0) \\
    \mu&= 1.4 \gamma K_P (3.72-P_0/\ell_0)\label{MFmu} \\
    A &= 0 \\
    K^o &= 1.7 \gamma K_P\alpha^o
\end{align}
\label{MF} 
\end{subequations}
see~\cite{fruchart2023odd} for coarse-graining procedures. 
Notably, while $A$ is always zero, $K^o$ is nonzero for nonzero $\alpha^o$. Such a property is not possible in lattices with only nearest-neighbor interactions, but can emerge in lattices with next-nearest-neighbor interactions~\cite{scheibner2020odd}. The structure of the graph model indeed generates next-nearest-neighbor interactions, see SI. The property $A=0$ is not limited to the lattice structure, but can be generalized to any structure under affine deformation. This is because $A$ measures the change of torque due to bulk expansion. An affine bulk expansion  increases the perimeter of each cell with a fixed ratio. Because $E_C \sim \sum P_i^2$, both the chiral force and the boundary size increases with the same ratio. As a result, the torque measured at the boundary remains constant and $A=0$. This means that nonzero $A$ must be associated with non-affine displacements. 


In addition to the dynamical matrix, we also calculate the prestress $\sigma^0$ of the MFT, see SI:
\begin{equation}
\begin{aligned}
\sigma^0&=(K_P/2\bar A)(P_0^{\rm MF})^2\begin{pmatrix}
& s   & -\alpha^o\\
&\alpha^o & s
\end{pmatrix}
\label{s12}
\end{aligned}
\end{equation}
which is composed of an isotropic stress $\sigma_P$ plus a torque $\tau_P\sim\alpha^o$. 

\section{Phase diagram and dynamical phase transition}
We detail here the definition and the theoretical and numerical results of the phase diagram. 

Traditional (achiral) graph models are known to have two phases: a solid phase with finite shear modulus $\mu$ and a liquid phase with $\mu=0$. The same definition also applies to the chiral graph model. When $\alpha^o\neq 0$, the two phases become the chiral solid phase and the chiral liquid phase with odd viscoelastic moduli. 

We show the MFT phase diagram in Extended Data Fig.~\ref{fig.m1}A. The MFT predicts a solid-liquid phase transition at $P_0^{\rm MF}/\ell_0=3.72$ (solid white line), which is independent of $\alpha^o$, see Eq.~(\ref{MFmu}). The chiral active solid phase is further divided by the dashed white line based on the presence of odd elastic waves (oscillations of overdamped modes), $|a^o| = (P_0^{\rm MF}-P_0)/(\sqrt 3 P_0^{\rm MF})$. The dashed line is also a line of exceptional points. 

The MFT phase diagram is qualitatively verified by numerical results, see Extended Data Fig.~\ref{fig.m1}B-D. In Extended Data Fig.~\ref{fig.m1}B we show the shear modulus $\mu$ for $|\alpha^o|\leq 0.2$ and identify phase transitions at $P_0^*(\alpha^o)/\ell_0=3.81-|\alpha^o|$. For $\alpha^o=0$, this transition reduces to the rigidity transition of the conventional graph models located at $P_0^*=3.81$. 
The critical preferred perimeter $P_0^*$ reduces for increasing $|\alpha^o|$, suggesting that activity tends to fluidize the system. 
A similar trend was observed in the self-propelled Voronoi model~\cite{Bi2016}. However, this trend does not hold for larger $|\alpha^o|$, see the velocity data in Fig.~\ref{fig.2}A. The reason is discussed later in Extended Data Fig.~\ref{fig.m1}G. 

We further analyze the eigenvalues of the Jacobian matrix, $\lambda^J$. The calculation is done using automatic differentiation, see SI for details. In the chiral solid phase, $\omega={\rm Im}(\lambda^J)$ denotes the oscillation frequency of the linear modes. In Extended Data Fig.~\ref{fig.m1}C we find a transition from small $\omega^2$ to large $\omega^2$ as $\alpha^o$ increases, suggesting a crossover from a non-odd-elastic-wave phase to an odd-elastic-wave phase. From this we define a transition line that qualitatively agrees with the theoretical prediction of the MFT. Note that we are unbale to track down a particular elastic mode but can only study the average behavior of all modes, because each T1 transition is a singularity that completely changes the elastic modes. Interestingly, we also find large $\omega^2$ values in the odd chaotic fluid phase, although their physical meanings are unclear because the system is not in a static state. 

In the chiral liquid phase, cells move continuously in a chaotic manner. We find that except for small $\alpha^o$, the steady-state cell velocity can also serve as an indicator for the dynamical phase transition, see Extended Data Fig.~\ref{fig.m1}D. Hence, in Fig.~\ref{fig.2}A of the main text we use $\langle v^2\rangle$ as an indicator of the phase transition instead of $\mu$, because calculating $\mu$ is numerically challenging for large $\alpha^o$ due to the strong chaos. In Extendede Data Fig.~\ref{fig.m1}E we plot $\langle v^2\rangle$  as functions of $\Delta P_0=P_0-P_0^*(\alpha^o)$, the distance to the critical point. We find that a Widom-like scaling
\begin{equation}
\langle v^2 \rangle/|\alpha^o|^{0.4} = |\Delta P_0|^\beta{\cal V}_{\pm}(|\alpha^o|/|\Delta P_0|^\phi)
\end{equation}
can collapse all data onto ${\cal V}_{\pm}$, the scaling functions in the odd solid phase and the odd liquid phase, respectively. We find two scaling exponents $\beta=3$ and $\phi=4$. This scaling relation holds for $|\alpha^o|>0.04$ and fails for small $|\alpha^o|$, because the velocity naturally vanishes when $\alpha^o\to 0$.

In Extended Data Fig.~\ref{fig.m1}F we plot the largest lyapunov exponent $\lambda$. For nonzero $\alpha^o$, $\lambda$ changes from negative values for small $P_0$ to positive values for large $P_0$, indicating a phase transition from a non-chaotic phase (chiral active solid) to a chaotic phase (chiral chaotic liquid). When $\alpha^o=0$, we find that $\lambda$ approaches zero for large $P_0$, corresponding to a non-chaotic liquid phase. The transition point at which $\lambda=0$ reduces for increasing $\alpha^o$, which qualitatively agrees with $P_0^*(\alpha^o)$. 

We find in Fig.~\ref{fig.2}A that while small $|\alpha^o|$ tends to fluidize the system ($P_0^*$ reduces with increasing $|\alpha^o|$), for larger $|\alpha^o|$ there is a reentrant to the solid phase. To explain this, we measure the local hexatic order parameter $\Psi_6=\langle \sum_{j} \exp(i6\theta_{ij})\rangle_i$, where $\theta_{ij}$ is the angle of the bond between the cell $i$ and the cell $j$. We find that the large $|\alpha^o|$ enforces a hexatic order on the tissue, although local disorders still exist (Extended Data Fig.~\ref{fig.m1}G). Similar hexatic order has been found to be produced by cell division/motility~\cite{tang2024cell}. The hexatic order explains the reentrant to the solid phase: as we find in MFT, in a perfect hexatic lattice $\alpha^o$ does not affect $P_0^*$ at all (Extended Data Fig.~\ref{fig.m1}A). We define four phases based on the geometric properties: hexatic solid, hexatic liquid, disordered solid and disordered liquid (Extended Data Fig.~\ref{fig.m1}G), and show the corresponding trajectories in Extended Data Fig.~\ref{fig.m1}H. Interestingly, in the hexatic liquid phase (phase 2), cells chaotically move in a glassy manner: they spend most of the time moving within confined spaces without changing the topology, but topological rearrangements still appear at relatively low frequency. The hexatic phase relies on the assumption that all cells share a uniform $\alpha^o$. In reality, this assumption is probably not true as fluctuations always exist between individual cells. Hence, it is unlikely for the hexatic phase to be observed in experiments. 

We then numerically study how the viscoelastic moduli vary within the phase diagram. We show the two shear moduli in  Extended Data Fig.~\ref{fig.moduli}A-C. For $P_0/\ell_0=3.0$ (chiral active solid), $K^o$ shows a linear dependence on $\alpha^o$ and $\mu$ remains nearly constant (Extended Data Fig.~\ref{fig.moduli} A). Despite the disordered structure observed in the simulations, both $\mu$ and $K^o$ are quantitatively captured by the MFT prediction Eq.~(\ref{MF}). For $P_0/\ell_0=3.7$, we find that both shear moduli vanish at large $|\alpha^o|$ values (Extended Data Fig.~\ref{fig.moduli}B), indicating of the solid-fluid phase transition. For $P_0/\ell_0=4.0$ (chiral chaotic liquid), the shear moduli completely vanish for any $|\alpha^o|$ (Extended Data Fig.~\ref{fig.moduli}C)~\footnote{For Voronoi model with $\alpha^o=0$ it has been argued that $\mu$ can have a small but finite value in the liquid phase if $K_A>0$~\cite{sussman2018no}.}.

Finite bulk moduli $A$ and $B$ are observed for all non-zero $\alpha^o$ values (Extended Data Fig. \ref{fig.moduli}G-I). Both $A$ and $B$ for $P_0/\ell_0=3.0$ are in quantitative alignment with the MFT results (solid lines), see Extended Data Fig.~\ref{fig.moduli}D. Note that the MFT result $A=0$ holds for any affine version of the graph model (Method). Hence, the observed large values of $A$ in the chiral fluid phase (Extended Data Fig. \ref{fig.moduli}E, F) indicate strong non-affinity. 

The vanishing shear moduli in the chiral fluid phase allow us to measure the shear viscosities, i.e., the change of stresses due to shear flows. In the chiral solid phase, both the normal and odd shear viscosities, $\eta$ and $\eta^o$, diverge in the measurement due to the yield stress~\cite{hertaeg2024discontinuous}, see Extended Data Fig. \ref{fig.moduli}G, H. $\eta$ becomes negative for large $|a^o|$ (Extended Data Fig. \ref{fig.moduli}H, I), indicating that instead of dissipating energy, it begins to generate energy. While $\eta$ must remain positive in equilibrium systems, the energy injection driven by the non-variational forces allows for a negative viscosity. Similar negative viscosity has been experimentally observed in active bacteria~\cite{lopez2015turning}. The value of $\eta^o$ saturates to constant values for large $|\alpha^o|$. Interestingly, the values of $\eta^o$ are opposite to each other when approaching the solid phase from the positive/negative sides of $\alpha^o$ (Extended Data Fig.~\ref{fig.moduli}I), indicating a qualitative difference between the left-hand odd fluid to the right-hand odd fluid.

\begin{figure}[h]
\centering
\hspace{-2em}\includegraphics[width = 1.\columnwidth]{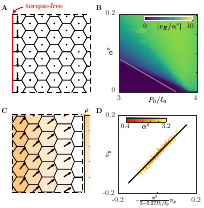}
\caption{{\bf Two mechanisms of macroscopic chiral flow} (A) Chiral edge current at torque-free boundaries. Arrows indicate flow direction for positive $\alpha^o$ values. (B) Numerical results of the chiral edge current $v_E$ at torque-free boundaries, with the gray line being the MFT prediction of the onset of chiral edge currents. (C) Chiral flow due to density gradient. Arrows indicate flow direction for positive $\alpha^o$ values. (D) Numerical results of the flow field generated by artificial ``pinching'' in $x$-direction. A universal relation is found for $0.4\leq \alpha^o\leq 3.2$ and $3\leq P_0/\ell_0\leq 4$, validating Eq.~(\ref{continuum}). In the simulation we have set $\ell_0^2=A_0=1$, $K_P=1$ and $K_A=0.1$. 
    }
	\label{fig.m4}
\end{figure}

\section{Macroscopic chiral flows}

\begin{figure}[t]
\centering
\hspace{-2em}\includegraphics[width = 1.\columnwidth]{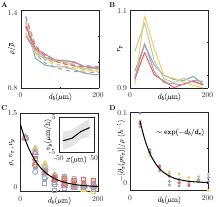}
\caption{{\bf Density, proliferation rate and flow in experiments of HT1080} (A) Experimental density fields (normalized by the average value) as function of $d_b$, the distance to the boundaries, are shown in solid lines. Numerical results are shown in dashed lines for comparison. (B) Experimental proliferation rate profile (normalized by the average value). (C) Experimental data of the density $\rho$ (circles), the converging flow $v_x$ (squares), and the chiral flow $v_y$ (triangles). Data renormalized by $Y\to [Y-{\rm min}(Y)]/[{\rm max}(Y)-{\rm min}(Y)]$. Solid black line is exponential fit with decay length $d_\rho=44 \rm\mu m$. Note that its value at $d_b=0$ is slightly above $1$, because the data closest to the boundaries are at $d_b = 12.5 {\rm \mu m}$. Inset: Experimental chiral flow for sample width $100 {\rm \mu m}$. (D) Estimation of the net proliferation rate from velocity profile, $k_{\rm prolif}-k_{\rm extr}=\partial_x(\rho v_x)/\rho$, based on mass conservation. Solid black line is exponential fit with $d_e=35 {\rm \mu m}$. 
    }
	\label{fig.m3}
\end{figure}

We discuss here the emergence of the macroscopic chiral flows, including the details of the experimental observations, the microscopic origin and the attempt to reproduce them in numerical simulations.

\subsubsection{Two mechanisms for macroscopic chiral flows}
Chiral edge current is a prevalent phenomenon in chiral active systems. A steady-state cellular flow indicates a persistent cellular force, which corresponds to a gradient of stress in the continuum limit. There are two possible mechanisms for stress gradients: the first one is based on the spatial inhomogeneities, for example the inhomogeneous proliferation/extrusion (self-pinching) discussed in the main text. The second one is based on boundary conditions, i.e., the average stress in the bulk is different from the average stress at the boundaries, hence producing a stress gradient around the boundaries. 

We start with the boundary conditions. As we discuss in Sec.~\ref{chiral_active_viscoelasticity}, a non-zero $\alpha^o$ naturally produces a pre-torque $\tau_P$ in the tissue. If the torque at the boundaries $\tau_B$ is different from $\tau_P$, a net force that is tangential to the boundaries is exerted on the cells that are in direct contact with the boundaries, see Extended Data Fig.~\ref{fig.m4}A. Here we consider a non-chiral and passive boundary, for example a boundary with frictional interaction with the tissue. In this case the boundary does not actively produce a torque, thus $\tau_B$ is in the same direction as $\tau_P$ with $|\tau_B|\leq |\tau_P|$, and the direction of the chiral flow is determined by the direction of $\tau_P$ or the sign of $\alpha^o$. We study an extreme case in which $\tau_B=0$ (torque-free boundary condition). We theoretically study the resulting tissue deformation using the MFT and find that such forces can lead to chiral edge currents, see SI. In the chiral solid phase, because of the finite elasticity, chiral edge currents only appear for $|\alpha^o|>{\sqrt 3(P_0^{\rm MF}-P_0)}/(4P_0^{\rm MF})$. In the unstable solid phase (odd liquid phase), on the other hand, any non-zero $\alpha^o$ generates edge currents. These predictions are qualitatively verified by numerical simulations, see Extended Data Fig.~\ref{fig.m4}B. When $|\alpha^o|$ is above the threshold of edge currents (gray line), we find $v_E\sim\alpha^o$. Hence, the edge currents can also be used to estimate $\alpha^o$. However, the observed chiral flows in HT1080 are unlikely to be driven by this mechanism: The inference of the bulk data suggests a positive $\alpha^o$ value, which would generate chiral edge currents at the torque-free boundaries in the opposite direction to the observed flows, see Fig.~\ref{fig.3}B and Extended Data Fig.~\ref{fig.m4}A. 

The inhomogeneous density field, on the other hand, produces chiral flows in the correct direction, see Extended Data Fig.~\ref{fig.m4}C. In this mechanism, cell proliferation/extrusion exerts `self-pinching', leading to an inhomogeneous density field, which further generates macroscopic flows. To illustrate the mechanism analytically, we calculate the steady-state velocity field in the continuum limit following Eq.~(\ref{simplified_viscoelastic})
with $\zeta=\gamma/\ell_0^2$. In the chiral active solid phase, we neglect viscous stresses and obtain 
\begin{equation}
\begin{aligned}
(\gamma/\ell_0^2)  	\begin{pmatrix}
v_x \\
v_y
\end{pmatrix}=
\begin{pmatrix}
(B+\mu)\partial^2_x u_x + K^o\partial^2_x u_y \\
(A-K^o)\partial^2_xu_x+ \mu \partial^2_xu_y 
\end{pmatrix}
\end{aligned}
\end{equation}
with $\bm u$ being the displacement field. For simplicity we approximate a `self pinching' in $x$ direction to an inhomogeneous displacement field $u_x$, which results in a density field $\rho(x) = -\partial_x u_x$. In this case 
\begin{equation}
\begin{aligned}
(\gamma/\ell_0^2)  	\begin{pmatrix}
v_x \\
v_y
\end{pmatrix}=
-\partial_x \rho\begin{pmatrix}
B+\mu\\
A-K^o
\end{pmatrix}
\end{aligned}
\label{flow}
\end{equation}
Substituting the MFT results for elastic moduli yields Eq.~(\ref{continuum}). In this mechanism, chiral flows emerge in the transverse direction of the density gradient. Such a phenomenon is also known as the odd diffusivity or the odd transport coefficient~\cite{hargus2021odd}. 
1
We then test the theoretical prediction of Eq.~(\ref{continuum}) using numerical simulations with artificial pinching: we apply an inhomogeneous displacement field $u_x$ on random snapshots of the simulations and measure the instantaneous velocity field $\bm v$. We find that the prediction agrees quantitatively with the numerical results, see Extended Data Fig.~\ref{fig.m4}D. Strikingly, while the prediction relies on the MFT results which is only accurate for the hexatic lattice in the chiral solid phase, it captures the numerical results for both the chiral solid and chiral liquid phases with disordered structures. 

Equation~(\ref{continuum}) also serves as an independent method to estimate $\alpha^o$. We notice that the denominator of Eq.~(\ref{continuum}) is not very sensitive to $P_0/\ell_0$: for $3<P_0/\ell_0<5$, Eq.~(\ref{continuum}) yields $0.46<v_y/(\alpha^ov_x)<0.61$. Hence, $v_y/v_x\approx 0.5\alpha^o$ is a good estimator for a wide range of $P_0/\ell_0$. Together with a linear regression of Fig.~\ref{fig.3}E, we obtain an independent estimate $\alpha^o\approx 3.6$, which is in reasonable agreement with the inference result $\alpha^o=2.9$. 

\subsubsection{Experimental flow field}
In Extended Data Figs.~\ref{fig.m3} AB we plot the density and proliferation profiles for various sample widths as functions of $d_b$, the distance to the boundaries. Both profiles peak near the boundaries, aligning the inhomogeneous growth mechanism. Moreover, we find that $\rho$, $v_x$ and $v_y$ all follow the same exponential profile, see Extended Data Fig.~\ref{fig.m3}C. This is consistent Eq.~(\ref{flow}), which strongly supports the ``self-pinching'' mechanism.  In the inset of Extended Data Fig.~\ref{fig.m3}C we plot the chiral flow of a small sample width $100 \rm \mu m$ ($4$ cell widths), whose existence implies that the observed chirality has a single-cell origin. 

\begin{figure}[t]
\centering
\hspace{-2em}\includegraphics[width = 1.\columnwidth]{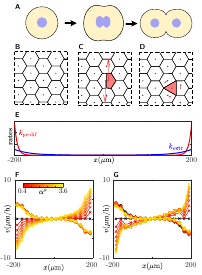}
\caption{{\bf  Simulation with proliferation/extrusion} (A) Illustration of cell division as a continuous-in-time process. (B-D) Illustration of discrete-in-time proliferation in graph models. (B) Graph model before proliferation.  (C) Division-like proliferation generates large local velocities (arrows) just after proliferation. New cell is marked in red. (D) Proliferation on a random vertex generates small  local velocities just after proliferation.  (E) Proliferation and extrusion rates used in the simulations. We use exponential decay with decay length $6.25 {\rm \mu m}$ (proliferation) and $35 \rm \mu m$ (extrusion). The maximal extrusion rate is set to $k_0=0.33 {\rm h}^{-1}$.  (F) Velocity profile for $P_0/\ell_0=3.0$ for various $\alpha^o$ values. Circles denote $v_x$ and crosses denote $v_y$. (G) Velocity profile for $P_0/\ell_0=4.0$. Colors indicate various $\alpha^o$ values in $0-3.6$.  
In the simulations we have set $A_0=\ell_0^2=625 {\rm \mu m}^2$, $K_A=0$ and  $K_P=0.43 {\rm h}^{-1}$.    }
    \label{fig.m5}
\end{figure}

In addition to the proliferation, the extrusion of cells also modifies the density field. While quantifying the extrusion rate directly from tissue videos is challenging, we can estimate the total rate ($k_{\rm prolif}-k_{\rm extr}$) from the continuity equation Eq.~(\ref{continuity_equation}). In the experiments, the average cell density gradually increases with time, but the magnitude of the increase is not large ($\ell_0$ decreases around $20\%$ within 24 hours).  Assuming the density profile is in a steady state for simplicity, we have $\partial_x(\rho v_x)/\rho =k_{\rm prolif}-k_{\rm extr}$.  
In Fig.~\ref{fig.m3}D we plot the profile of the cell flux $\partial_x(\rho v_x)/\rho$. We find that $\partial_x(\rho v_x)/\rho$ is always negative and follows a exponential decay with $d_b/d_e$, with $d_e\approx35\mu m$. This is in alignment with a hypothesis of the growth pattern: Most of the inhomogeneous proliferation occurs very close to the boundaries, while inhomogeneous extrusion also takes place near the boundaries but at a greater distance than proliferation. Far from the boundaries, both proliferation and extrusion are homogeneous, and the macroscopic flows are constrained near the boundaries.  We then use this observation to construct a numerical simulation with proliferation and extrusion, see discussion below. 

\subsubsection{Numerical simulation with proliferation and extrusion}
Can we reproduce the living mechanics of tissues and the resulting steady-state boundary flows in numerical simulations? Here, we present a simulation method incorporating both proliferation and extrusion, which qualitatively captures the dynamics of cell turnover and the associated flow fields.

Implementing proliferation and extrusion in graph models poses significant challenges. In biological systems, both processes are continuous in time. For instance, prior to cell division, a parent cell enlarges, then gradually constricts its mid-plane to generate two daughter cells (see Extended Data Fig.\ref{fig.m5}A). 
In Voronoi models, however, proliferation and extrusion have to be modeled by discrete-in-time events. The extrusion (or death) is typically  modeled by the removal of one cell. The proliferation is typically modeled by injecting a new cell close to one original cell which mimics cell division, see Extended Data Fig.~\ref{fig.m5}BC~\cite{tang2024cell}. In chiral graph models, however, the division-like injection creates strong local forces on the affected cells, see arrows in Extended Data Fig.~\ref{fig.m5}C. Such forces introduce strong temporal discontinuities in the velocity field which can lead to unphysical results. To reduce this effect, we adopt a different proliferation strategy by injecting new cells on a random vertex of the original Voronoi tessellation. By definition a vertex is on the circumcenter of three nearby cells, and injecting the new cell on a vertex can significantly reduce the influence on the neighboring cells, see Extended Data Fig.~\ref{fig.m5}D. 

We model the inhomogeneous proliferation/extrusion with rates $k_{\rm prolif}$ and $k_{\rm extr}$. Both rates are assumed to be exponential functions of $d_b$, the distance to the boundaries, see Extended Data Fig.~\ref{fig.m5}E. Specifically, we choose the two decay lengths to be $d_{\rm extr}=35 \rm \mu m$ and $d_{\rm prolif}=6.25 \rm \mu m$. Here $d_{\rm extr}$ is chosen according to Extended Data Fig.~\ref{fig.m3}D and  $d_{\rm prolif}$ is set to be much smaller than $d_{\rm extr}$, such that the total rate is in alignment with Extended Data Fig.~\ref{fig.m3}D. Specifically, we apply the following simulation procedure: In each timestep, every cell can be extruded at a rate $k_0 {\exp}(-d_b/d_{\rm extr})$. To ensure  constant total cell number, a new cell is proliferated immediately after an extrusion event. The location of the new cell is chosen to be on a  random vertex of the original Voronoi tessellation sampled with weight $k_{\rm prolif}\sim \exp(-d_b/d_{\rm prolif})$. We choose $k_0$ such that the steady-state density profile matches with the experiment, see Extended Data Fig.~\ref{fig.m3}A. 

We perform the simulation with parameters inferred from NADA with periodic boundary condition. The simulation results quantitatively reproduces the observed flow fields, see Fig. \ref{fig.3}E. In Extended Data Fig.~\ref{fig.m5}FG we show that $\alpha^o$ does not affect $v_x$ but dramatically changes $v_y$. We also find that changing $P_0/\ell_0$ does not qualitatively affect the flow profile. We observe wiggling in $v_y$ for $P_0/\ell_0=4.0$ due to the strong chaotic flow that rectifies the effect of discrete proliferation/extrusion.

\section{Chiral cellular dynamics in nonequilibrium patterns}

\begin{figure}[t]
\centering
\includegraphics[width = 1.05\columnwidth]{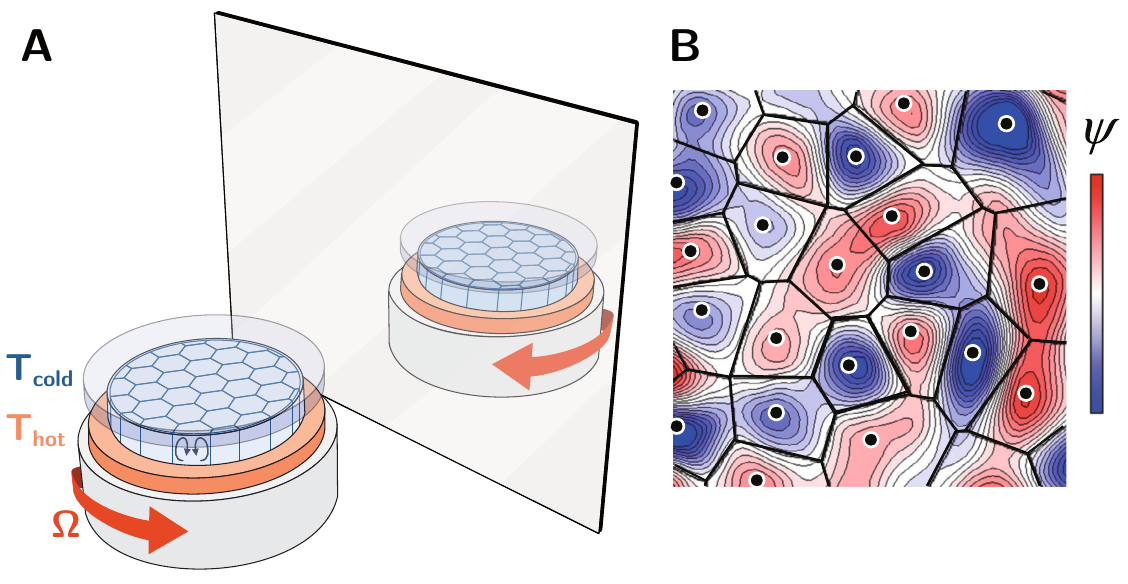}
\caption{{\bf Cellular dynamics in rotating Rayleigh-Bénard convection.}
(A) A cylindrical container heated from below is put under rotation at angular velocity $\Omega$. The temperature gradient $[T_{\text{hot}} - T_{\text{cold}}]/h$ (in which $h$ is the height of the container) leads to thermal convection.
This setup is known as rotating Rayleigh-Bénard (RRB) convection. In the appropriate regime, this can lead to more or less disordered hexagonal patterns of convection cells.
(B) A Voronoi tesselation algorithm can be applied to the streamfunction $\psi$ reconstructed from OFV measurements in experiments in order to identify convection cells~\cite{Fujita2020}.
Panel B is adapted from Ref.~\cite{Fujita2020}.
\label{fig_rb}
}
\end{figure}

\def\vb#1{\bm{#1}}

Cellular patterns reminiscent of chiral biological tissues arise in non-living systems ranging from crystal growth to fluid flows~\cite{Pismen2010,Cross1994,cross2009pattern}.
To illustrate the potential of our biologically inspired approach beyond cell biology, we consider a paradigmatic example:  Rayleigh-Bénard convection. This well known phenomenon occurs whenever a fluid layer is confined between two horizontal plates at different temperatures, with the hot plate below. When the temperature difference is large enough, an instability develops, leading to the appearance of spatial patterns that include disordered convection cells that occur in an experimental context ranging from a pan filled with boiling water in your kitchen to quasi-two-dimensional flows in the atmosphere. 

In convection problems, the building block is a convection cell (Fig.~\ref{fig_rb}), a non-living cell whose behavior and interactions between other convection cells is governed by Navier-Stokes equations.
In rotating Rayleigh-Bénard convection, the fluid is put under rotation, as shown in Fig.~\ref{fig_rb}A, see Refs.~\cite{Busse1980,Bodenschatz2000recent,Ahlers2006,Guarino2004,Ecke2023}. 
Taken together these ingredients make the dynamics chiral and non-variational, very much like in the biological tissues discussed in the main text.

Symmetry arguments suggest that the effective displacement field $\vb u (\vb x,t)$ describing the deformation of the pattern should follow the so-called phase diffusion equation~\cite{Echebarria1998,Echebarria2000,Echebarria2000b}
\begin{align}
\begin{split}
    \partial_t \vb u = &D_\parallel \nabla (\nabla \cdot \vb u) + D_\perp \Delta \vb u 
    \\
    + &D_\parallel^\times  \bm{\epsilon} \cdot \nabla (\nabla \cdot \vb u) + D_\perp^\times  \bm{\epsilon} \cdot \Delta \vb u. 
\end{split}
\label{eq:phase} 
\end{align}
where $D_\parallel$, $D_\perp$, $D_\parallel^\times$ and $D_\perp^\times$ are phenomenological coefficients known as phase diffusivities. While the phenomenological reasoning leading to Eq.~\eqref{eq:phase} is quite distinct from the coarse-grained derivation of   Eq.~\eqref{simplified_viscoelastic}, the phase diffusion equation is a special case of our chiral elastodynamic description. In the limit in which pressure and viscosities are ignored, Eq.~\eqref{simplified_viscoelastic}  and Eq.~\eqref{eq:phase} are isomorphic once the following dictionary is used to map phase diffusivities onto elastic moduli: $D_\parallel=B/{\zeta}$, $D_\perp=\mu/{\zeta}$, $D_\parallel^\times=A/\zeta$ and $D_\perp^\times=K^{\rm o}/\zeta$. This correspondence and especially the presence of the two cross diffusivities $D_\parallel^\times$ and $D_\perp^\times$ (i.e. the chiral active moduli $A$ and $K^{\rm o}$) suggests the existence of an effective cell-level description of the dynamics of patterns akin to our non-variational graph model of tissues. Very much like living cells, inanimate convection cells fleetingly form and die giving rise to T1 transitions analogous to the ones occurring in our biological tissues. 

The analogy does not stop at the mathematical level --- it extends to the still poorly understood phenomenology of rotating Rayleigh-Bernard convection. Robust traveling wall modes, akin to chiral edge flows in tissues, have been observed at the vertical boundary of the container in experiments and predicted using direct Navier Stokes simulations \cite{Rossby1969,Ecke1992,Kuo1993,Favier2020,Ecke2022}. It has been conjectured that boundary zonal flows, that persist in the presence of bulk convection, are related to these wall modes \cite{deWit2023,Wedi2022,Ecke2023,Ecke2022,Vasil2024}. However, the origin of these phenomena is still not fully understood. 

Traditional descriptions of convection cells in fluids are based on continuum amplitude or phase equations such as Eq.~(\ref{eq:phase}), rather than descriptions based on individual cells. 
A graph model such as Eq.~\eqref{chiral_vertex_model} in the main text would provide a complementary description of the dynamics of rotating Rayleigh-Bénard convection cells.
Such a microscopic model (inferred using NADA from experimental measurements like Fig.~\ref{fig_rb}B), would be consistent with Eq.~\eqref{eq:phase} upon coarse-graining, but remain valid even for narrow channels and discrete problems involving cells rearrangements and defects dynamics for which the continuum description typically breaks down.

\bibliography{citation}

\begin{thebibliography}{6}%
\makeatletter
\providecommand \@ifxundefined [1]{%
 \@ifx{#1\undefined}
}%
\providecommand \@ifnum [1]{%
 \ifnum #1\expandafter \@firstoftwo
 \else \expandafter \@secondoftwo
 \fi
}%
\providecommand \@ifx [1]{%
 \ifx #1\expandafter \@firstoftwo
 \else \expandafter \@secondoftwo
 \fi
}%
\providecommand \natexlab [1]{#1}%
\providecommand \enquote  [1]{``#1''}%
\providecommand \bibnamefont  [1]{#1}%
\providecommand \bibfnamefont [1]{#1}%
\providecommand \citenamefont [1]{#1}%
\providecommand \href@noop [0]{\@secondoftwo}%
\providecommand \href [0]{\begingroup \@sanitize@url \@href}%
\providecommand \@href[1]{\@@startlink{#1}\@@href}%
\providecommand \@@href[1]{\endgroup#1\@@endlink}%
\providecommand \@sanitize@url [0]{\catcode `\\12\catcode `\$12\catcode
  `\&12\catcode `\#12\catcode `\^12\catcode `\_12\catcode `\%12\relax}%
\providecommand \@@startlink[1]{}%
\providecommand \@@endlink[0]{}%
\providecommand \url  [0]{\begingroup\@sanitize@url \@url }%
\providecommand \@url [1]{\endgroup\@href {#1}{\urlprefix }}%
\providecommand \urlprefix  [0]{URL }%
\providecommand \Eprint [0]{\href }%
\providecommand \doibase [0]{https://doi.org/}%
\providecommand \selectlanguage [0]{\@gobble}%
\providecommand \bibinfo  [0]{\@secondoftwo}%
\providecommand \bibfield  [0]{\@secondoftwo}%
\providecommand \translation [1]{[#1]}%
\providecommand \BibitemOpen [0]{}%
\providecommand \bibitemStop [0]{}%
\providecommand \bibitemNoStop [0]{.\EOS\space}%
\providecommand \EOS [0]{\spacefactor3000\relax}%
\providecommand \BibitemShut  [1]{\csname bibitem#1\endcsname}%
\let\auto@bib@innerbib\@empty
\bibitem [{\citenamefont {Scheibner}\ \emph {et~al.}(2020)\citenamefont
  {Scheibner}, \citenamefont {Souslov}, \citenamefont {Banerjee}, \citenamefont
  {Sur{\'o}wka}, \citenamefont {Irvine},\ and\ \citenamefont
  {Vitelli}}]{scheibner2020odd}%
  \BibitemOpen
  \bibfield  {author} {\bibinfo {author} {\bibfnamefont {C.}~\bibnamefont
  {Scheibner}}, \bibinfo {author} {\bibfnamefont {A.}~\bibnamefont {Souslov}},
  \bibinfo {author} {\bibfnamefont {D.}~\bibnamefont {Banerjee}}, \bibinfo
  {author} {\bibfnamefont {P.}~\bibnamefont {Sur{\'o}wka}}, \bibinfo {author}
  {\bibfnamefont {W.~T.~M.}\ \bibnamefont {Irvine}},\ and\ \bibinfo {author}
  {\bibfnamefont {V.}~\bibnamefont {Vitelli}},\ }\bibfield  {title} {\bibinfo
  {title} {Odd elasticity},\ }\href {https://doi.org/10.1038/s41567-020-0795-y}
  {\bibfield  {journal} {\bibinfo  {journal} {Nature Physics}\ }\textbf
  {\bibinfo {volume} {16}},\ \bibinfo {pages} {475} (\bibinfo {year}
  {2020})}\BibitemShut {NoStop}%
\bibitem [{\citenamefont {Fruchart}\ \emph {et~al.}(2022)\citenamefont
  {Fruchart}, \citenamefont {Han}, \citenamefont {Scheibner},\ and\
  \citenamefont {Vitelli}}]{fruchart2022odd}%
  \BibitemOpen
  \bibfield  {author} {\bibinfo {author} {\bibfnamefont {M.}~\bibnamefont
  {Fruchart}}, \bibinfo {author} {\bibfnamefont {M.}~\bibnamefont {Han}},
  \bibinfo {author} {\bibfnamefont {C.}~\bibnamefont {Scheibner}},\ and\
  \bibinfo {author} {\bibfnamefont {V.}~\bibnamefont {Vitelli}},\ }\href@noop
  {} {\bibinfo {title} {The odd ideal gas: Hall viscosity and thermal
  conductivity from non-hermitian kinetic theory}} (\bibinfo {year} {2022}),\
  \Eprint {https://arxiv.org/abs/2202.02037} {arXiv:2202.02037} \BibitemShut
  {NoStop}%
\bibitem [{\citenamefont {Heinz}\ \emph {et~al.}(2005)\citenamefont {Heinz},
  \citenamefont {Paul},\ and\ \citenamefont {Binder}}]{heinz2005calculation}%
  \BibitemOpen
  \bibfield  {author} {\bibinfo {author} {\bibfnamefont {H.}~\bibnamefont
  {Heinz}}, \bibinfo {author} {\bibfnamefont {W.}~\bibnamefont {Paul}},\ and\
  \bibinfo {author} {\bibfnamefont {K.}~\bibnamefont {Binder}},\ }\bibfield
  {title} {\bibinfo {title} {Calculation of local pressure tensors in systems
  with many-body interactions},\ }\href@noop {} {\bibfield  {journal} {\bibinfo
   {journal} {Physical Review E—Statistical, Nonlinear, and Soft Matter
  Physics}\ }\textbf {\bibinfo {volume} {72}},\ \bibinfo {pages} {066704}
  (\bibinfo {year} {2005})}\BibitemShut {NoStop}%
\bibitem [{\citenamefont {Han}\ \emph {et~al.}(2021)\citenamefont {Han},
  \citenamefont {Fruchart}, \citenamefont {Scheibner}, \citenamefont
  {Vaikuntanathan}, \citenamefont {De~Pablo},\ and\ \citenamefont
  {Vitelli}}]{han2021fluctuating}%
  \BibitemOpen
  \bibfield  {author} {\bibinfo {author} {\bibfnamefont {M.}~\bibnamefont
  {Han}}, \bibinfo {author} {\bibfnamefont {M.}~\bibnamefont {Fruchart}},
  \bibinfo {author} {\bibfnamefont {C.}~\bibnamefont {Scheibner}}, \bibinfo
  {author} {\bibfnamefont {S.}~\bibnamefont {Vaikuntanathan}}, \bibinfo
  {author} {\bibfnamefont {J.~J.}\ \bibnamefont {De~Pablo}},\ and\ \bibinfo
  {author} {\bibfnamefont {V.}~\bibnamefont {Vitelli}},\ }\bibfield  {title}
  {\bibinfo {title} {Fluctuating hydrodynamics of chiral active fluids},\
  }\href {https://doi.org/10.1038/s41567-021-01360-7} {\bibfield  {journal}
  {\bibinfo  {journal} {Nature Physics}\ }\textbf {\bibinfo {volume} {17}},\
  \bibinfo {pages} {1260} (\bibinfo {year} {2021})}\BibitemShut {NoStop}%
\bibitem [{\citenamefont {Yashunsky}\ \emph {et~al.}(2022)\citenamefont
  {Yashunsky}, \citenamefont {Pearce}, \citenamefont {Blanch-Mercader},
  \citenamefont {Ascione}, \citenamefont {Silberzan},\ and\ \citenamefont
  {Giomi}}]{yashunsky2022chiral}%
  \BibitemOpen
  \bibfield  {author} {\bibinfo {author} {\bibfnamefont {V.}~\bibnamefont
  {Yashunsky}}, \bibinfo {author} {\bibfnamefont {D.~J.}\ \bibnamefont
  {Pearce}}, \bibinfo {author} {\bibfnamefont {C.}~\bibnamefont
  {Blanch-Mercader}}, \bibinfo {author} {\bibfnamefont {F.}~\bibnamefont
  {Ascione}}, \bibinfo {author} {\bibfnamefont {P.}~\bibnamefont {Silberzan}},\
  and\ \bibinfo {author} {\bibfnamefont {L.}~\bibnamefont {Giomi}},\ }\bibfield
   {title} {\bibinfo {title} {Chiral edge current in nematic cell monolayers},\
  }\href {https://doi.org/10.1103/PhysRevX.12.041017} {\bibfield  {journal}
  {\bibinfo  {journal} {Physical Review X}\ }\textbf {\bibinfo {volume} {12}},\
  \bibinfo {pages} {041017} (\bibinfo {year} {2022})}\BibitemShut {NoStop}%
\bibitem [{\citenamefont {Lin}\ \emph {et~al.}(2023)\citenamefont {Lin},
  \citenamefont {Merkel},\ and\ \citenamefont {Rupprecht}}]{lin2023structure}%
  \BibitemOpen
  \bibfield  {author} {\bibinfo {author} {\bibfnamefont {S.-Z.}\ \bibnamefont
  {Lin}}, \bibinfo {author} {\bibfnamefont {M.}~\bibnamefont {Merkel}},\ and\
  \bibinfo {author} {\bibfnamefont {J.-F.}\ \bibnamefont {Rupprecht}},\
  }\bibfield  {title} {\bibinfo {title} {Structure and rheology in vertex
  models under cell-shape-dependent active stresses},\ }\href@noop {}
  {\bibfield  {journal} {\bibinfo  {journal} {Physical Review Letters}\
  }\textbf {\bibinfo {volume} {130}},\ \bibinfo {pages} {058202} (\bibinfo
  {year} {2023})}\BibitemShut {NoStop}%
\end{thebibliography}%


\begin{thebibliography}{109}%
\makeatletter
\providecommand \@ifxundefined [1]{%
 \@ifx{#1\undefined}
}%
\providecommand \@ifnum [1]{%
 \ifnum #1\expandafter \@firstoftwo
 \else \expandafter \@secondoftwo
 \fi
}%
\providecommand \@ifx [1]{%
 \ifx #1\expandafter \@firstoftwo
 \else \expandafter \@secondoftwo
 \fi
}%
\providecommand \natexlab [1]{#1}%
\providecommand \enquote  [1]{``#1''}%
\providecommand \bibnamefont  [1]{#1}%
\providecommand \bibfnamefont [1]{#1}%
\providecommand \citenamefont [1]{#1}%
\providecommand \href@noop [0]{\@secondoftwo}%
\providecommand \href [0]{\begingroup \@sanitize@url \@href}%
\providecommand \@href[1]{\@@startlink{#1}\@@href}%
\providecommand \@@href[1]{\endgroup#1\@@endlink}%
\providecommand \@sanitize@url [0]{\catcode `\\12\catcode `\$12\catcode
  `\&12\catcode `\#12\catcode `\^12\catcode `\_12\catcode `\%12\relax}%
\providecommand \@@startlink[1]{}%
\providecommand \@@endlink[0]{}%
\providecommand \url  [0]{\begingroup\@sanitize@url \@url }%
\providecommand \@url [1]{\endgroup\@href {#1}{\urlprefix }}%
\providecommand \urlprefix  [0]{URL }%
\providecommand \Eprint [0]{\href }%
\providecommand \doibase [0]{https://doi.org/}%
\providecommand \selectlanguage [0]{\@gobble}%
\providecommand \bibinfo  [0]{\@secondoftwo}%
\providecommand \bibfield  [0]{\@secondoftwo}%
\providecommand \translation [1]{[#1]}%
\providecommand \BibitemOpen [0]{}%
\providecommand \bibitemStop [0]{}%
\providecommand \bibitemNoStop [0]{.\EOS\space}%
\providecommand \EOS [0]{\spacefactor3000\relax}%
\providecommand \BibitemShut  [1]{\csname bibitem#1\endcsname}%
\let\auto@bib@innerbib\@empty
\bibitem [{\citenamefont {Pasteur}(1897)}]{pasteur1897researches}%
  \BibitemOpen
  \bibfield  {author} {\bibinfo {author} {\bibfnamefont {L.}~\bibnamefont
  {Pasteur}},\ }\href@noop {} {\emph {\bibinfo {title} {Researches on the
  molecular asymmetry of natural organic products}}},\ \bibinfo {number} {14}\
  (\bibinfo  {publisher} {WF Clay},\ \bibinfo {year} {1897})\BibitemShut
  {NoStop}%
\bibitem [{\citenamefont {Phillips}\ \emph {et~al.}(2013)\citenamefont
  {Phillips}, \citenamefont {Kondev}, \citenamefont {Theriot},\ and\
  \citenamefont {Garcia}}]{Phillips2013}%
  \BibitemOpen
  \bibfield  {author} {\bibinfo {author} {\bibfnamefont {R.}~\bibnamefont
  {Phillips}}, \bibinfo {author} {\bibfnamefont {J.}~\bibnamefont {Kondev}},
  \bibinfo {author} {\bibfnamefont {J.}~\bibnamefont {Theriot}},\ and\ \bibinfo
  {author} {\bibfnamefont {H.}~\bibnamefont {Garcia}},\ }\href@noop {} {\emph
  {\bibinfo {title} {Physical Biology of the Cell}}}\ (\bibinfo  {publisher}
  {Garland Science},\ \bibinfo {year} {2013})\BibitemShut {NoStop}%
\bibitem [{\citenamefont {Shadkhoo}\ and\ \citenamefont
  {Mani}(2019)}]{shadkhoo2019role}%
  \BibitemOpen
  \bibfield  {author} {\bibinfo {author} {\bibfnamefont {S.}~\bibnamefont
  {Shadkhoo}}\ and\ \bibinfo {author} {\bibfnamefont {M.}~\bibnamefont
  {Mani}},\ }\href {https://doi.org/10.1371/journal.pcbi.1007454} {\bibfield
  {journal} {\bibinfo  {journal} {PLoS Computational Biology}\ }\textbf
  {\bibinfo {volume} {15}},\ \bibinfo {pages} {e1007454} (\bibinfo {year}
  {2019})}\BibitemShut {NoStop}%
\bibitem [{\citenamefont {Liu}\ \emph {et~al.}(2023)\citenamefont {Liu},
  \citenamefont {Lemaire}, \citenamefont {Munro},\ and\ \citenamefont
  {Mani}}]{Mani2023}%
  \BibitemOpen
  \bibfield  {author} {\bibinfo {author} {\bibfnamefont {S.}~\bibnamefont
  {Liu}}, \bibinfo {author} {\bibfnamefont {P.}~\bibnamefont {Lemaire}},
  \bibinfo {author} {\bibfnamefont {E.}~\bibnamefont {Munro}},\ and\ \bibinfo
  {author} {\bibfnamefont {M.}~\bibnamefont {Mani}},\ }\href
  {https://doi.org/10.1101/2022.11.05.515310} {\bibfield  {journal} {\bibinfo
  {journal} {bioRxiv}\ ,\ \bibinfo {pages} {515310}} (\bibinfo {year}
  {2023})}\BibitemShut {NoStop}%
\bibitem [{\citenamefont {Tan}\ \emph {et~al.}(2024)\citenamefont {Tan},
  \citenamefont {Amiri}, \citenamefont {Seijo-Barandiaran}, \citenamefont
  {Staddon}, \citenamefont {Materne}, \citenamefont {Tomas}, \citenamefont
  {Duclut}, \citenamefont {Popović}, \citenamefont {Grapin-Botton},\ and\
  \citenamefont {Jülicher}}]{Tan2024}%
  \BibitemOpen
  \bibfield  {author} {\bibinfo {author} {\bibfnamefont {T.~H.}\ \bibnamefont
  {Tan}}, \bibinfo {author} {\bibfnamefont {A.}~\bibnamefont {Amiri}}, \bibinfo
  {author} {\bibfnamefont {I.}~\bibnamefont {Seijo-Barandiaran}}, \bibinfo
  {author} {\bibfnamefont {M.~F.}\ \bibnamefont {Staddon}}, \bibinfo {author}
  {\bibfnamefont {A.}~\bibnamefont {Materne}}, \bibinfo {author} {\bibfnamefont
  {S.}~\bibnamefont {Tomas}}, \bibinfo {author} {\bibfnamefont
  {C.}~\bibnamefont {Duclut}}, \bibinfo {author} {\bibfnamefont
  {M.}~\bibnamefont {Popović}}, \bibinfo {author} {\bibfnamefont
  {A.}~\bibnamefont {Grapin-Botton}},\ and\ \bibinfo {author} {\bibfnamefont
  {F.}~\bibnamefont {Jülicher}},\ }\href
  {https://doi.org/10.1103/prxlife.2.033006} {\bibfield  {journal} {\bibinfo
  {journal} {PRX Life}\ }\textbf {\bibinfo {volume} {2}},\ \bibinfo {pages}
  {033006} (\bibinfo {year} {2024})}\BibitemShut {NoStop}%
\bibitem [{\citenamefont {Naganathan}\ \emph {et~al.}(2014)\citenamefont
  {Naganathan}, \citenamefont {F{\"u}rthauer}, \citenamefont {Nishikawa},
  \citenamefont {J{\"u}licher},\ and\ \citenamefont
  {Grill}}]{naganathan2014active}%
  \BibitemOpen
  \bibfield  {author} {\bibinfo {author} {\bibfnamefont {S.~R.}\ \bibnamefont
  {Naganathan}}, \bibinfo {author} {\bibfnamefont {S.}~\bibnamefont
  {F{\"u}rthauer}}, \bibinfo {author} {\bibfnamefont {M.}~\bibnamefont
  {Nishikawa}}, \bibinfo {author} {\bibfnamefont {F.}~\bibnamefont
  {J{\"u}licher}},\ and\ \bibinfo {author} {\bibfnamefont {S.~W.}\ \bibnamefont
  {Grill}},\ }\href {https://doi.org/10.7554/eLife.04165} {\bibfield  {journal}
  {\bibinfo  {journal} {elife}\ }\textbf {\bibinfo {volume} {3}},\ \bibinfo
  {pages} {e04165} (\bibinfo {year} {2014})}\BibitemShut {NoStop}%
\bibitem [{\citenamefont {Shankar}\ \emph {et~al.}(2022)\citenamefont
  {Shankar}, \citenamefont {Souslov}, \citenamefont {Bowick}, \citenamefont
  {Marchetti},\ and\ \citenamefont {Vitelli}}]{Shankar2022}%
  \BibitemOpen
  \bibfield  {author} {\bibinfo {author} {\bibfnamefont {S.}~\bibnamefont
  {Shankar}}, \bibinfo {author} {\bibfnamefont {A.}~\bibnamefont {Souslov}},
  \bibinfo {author} {\bibfnamefont {M.~J.}\ \bibnamefont {Bowick}}, \bibinfo
  {author} {\bibfnamefont {M.~C.}\ \bibnamefont {Marchetti}},\ and\ \bibinfo
  {author} {\bibfnamefont {V.}~\bibnamefont {Vitelli}},\ }\href
  {https://doi.org/10.1038/s42254-022-00445-3} {\bibfield  {journal} {\bibinfo
  {journal} {Nature Reviews Physics}\ }\textbf {\bibinfo {volume} {4}},\
  \bibinfo {pages} {380–398} (\bibinfo {year} {2022})}\BibitemShut {NoStop}%
\bibitem [{\citenamefont {Casademunt}\ and\ \citenamefont
  {Alert}(2025)}]{Casademunt2025}%
  \BibitemOpen
  \bibfield  {author} {\bibinfo {author} {\bibfnamefont {J.}~\bibnamefont
  {Casademunt}}\ and\ \bibinfo {author} {\bibfnamefont {R.}~\bibnamefont
  {Alert}},\ }\href@noop {} {\emph {\bibinfo {title} {Lectures on Cellular
  Biophysics: from Molecules to Tissues}}},\ G - Reference,Information and
  Interdisciplinary Subjects Series\ (\bibinfo  {publisher} {World Scientific
  Publishing Company Pte Limited},\ \bibinfo {year} {2025})\BibitemShut
  {NoStop}%
\bibitem [{\citenamefont {Maroudas‑Sacks}\ \emph {et~al.}(2021)\citenamefont
  {Maroudas‑Sacks}, \citenamefont {Garion}, \citenamefont {Shani‑Zerbib},
  \citenamefont {Livshits}, \citenamefont {Braun},\ and\ \citenamefont
  {Keren}}]{maroudassacks2021topological}%
  \BibitemOpen
  \bibfield  {author} {\bibinfo {author} {\bibfnamefont {Y.}~\bibnamefont
  {Maroudas‑Sacks}}, \bibinfo {author} {\bibfnamefont {L.}~\bibnamefont
  {Garion}}, \bibinfo {author} {\bibfnamefont {L.}~\bibnamefont
  {Shani‑Zerbib}}, \bibinfo {author} {\bibfnamefont {A.}~\bibnamefont
  {Livshits}}, \bibinfo {author} {\bibfnamefont {E.}~\bibnamefont {Braun}},\
  and\ \bibinfo {author} {\bibfnamefont {K.}~\bibnamefont {Keren}},\ }\href
  {https://doi.org/10.1038/s41567-020-01083-1} {\bibfield  {journal} {\bibinfo
  {journal} {Nature Physics}\ }\textbf {\bibinfo {volume} {17}},\ \bibinfo
  {pages} {251} (\bibinfo {year} {2021})}\BibitemShut {NoStop}%
\bibitem [{\citenamefont {Tan}\ \emph {et~al.}(2022)\citenamefont {Tan},
  \citenamefont {Mietke}, \citenamefont {Li}, \citenamefont {Chen},
  \citenamefont {Higinbotham}, \citenamefont {Foster}, \citenamefont {Gokhale},
  \citenamefont {Dunkel},\ and\ \citenamefont {Fakhri}}]{Tan2022}%
  \BibitemOpen
  \bibfield  {author} {\bibinfo {author} {\bibfnamefont {T.~H.}\ \bibnamefont
  {Tan}}, \bibinfo {author} {\bibfnamefont {A.}~\bibnamefont {Mietke}},
  \bibinfo {author} {\bibfnamefont {J.}~\bibnamefont {Li}}, \bibinfo {author}
  {\bibfnamefont {Y.}~\bibnamefont {Chen}}, \bibinfo {author} {\bibfnamefont
  {H.}~\bibnamefont {Higinbotham}}, \bibinfo {author} {\bibfnamefont {P.~J.}\
  \bibnamefont {Foster}}, \bibinfo {author} {\bibfnamefont {S.}~\bibnamefont
  {Gokhale}}, \bibinfo {author} {\bibfnamefont {J.}~\bibnamefont {Dunkel}},\
  and\ \bibinfo {author} {\bibfnamefont {N.}~\bibnamefont {Fakhri}},\ }\href
  {https://doi.org/10.1038/s41586-022-04889-6} {\bibfield  {journal} {\bibinfo
  {journal} {Nature}\ }\textbf {\bibinfo {volume} {607}},\ \bibinfo {pages}
  {287–293} (\bibinfo {year} {2022})}\BibitemShut {NoStop}%
\bibitem [{\citenamefont {Taniguchi}\ \emph {et~al.}(2011)\citenamefont
  {Taniguchi}, \citenamefont {Maeda}, \citenamefont {Ando}, \citenamefont
  {Okumura}, \citenamefont {Nakazawa}, \citenamefont {Hatori}, \citenamefont
  {Nakamura}, \citenamefont {Hozumi}, \citenamefont {Fujiwara},\ and\
  \citenamefont {Matsuno}}]{Taniguchi2011}%
  \BibitemOpen
  \bibfield  {author} {\bibinfo {author} {\bibfnamefont {K.}~\bibnamefont
  {Taniguchi}}, \bibinfo {author} {\bibfnamefont {R.}~\bibnamefont {Maeda}},
  \bibinfo {author} {\bibfnamefont {T.}~\bibnamefont {Ando}}, \bibinfo {author}
  {\bibfnamefont {T.}~\bibnamefont {Okumura}}, \bibinfo {author} {\bibfnamefont
  {N.}~\bibnamefont {Nakazawa}}, \bibinfo {author} {\bibfnamefont
  {R.}~\bibnamefont {Hatori}}, \bibinfo {author} {\bibfnamefont
  {M.}~\bibnamefont {Nakamura}}, \bibinfo {author} {\bibfnamefont
  {S.}~\bibnamefont {Hozumi}}, \bibinfo {author} {\bibfnamefont
  {H.}~\bibnamefont {Fujiwara}},\ and\ \bibinfo {author} {\bibfnamefont
  {K.}~\bibnamefont {Matsuno}},\ }\href
  {https://doi.org/10.1126/science.1200940} {\bibfield  {journal} {\bibinfo
  {journal} {Science}\ }\textbf {\bibinfo {volume} {333}},\ \bibinfo {pages}
  {339–341} (\bibinfo {year} {2011})}\BibitemShut {NoStop}%
\bibitem [{\citenamefont {Sato}\ \emph {et~al.}(2015)\citenamefont {Sato},
  \citenamefont {Hiraiwa},\ and\ \citenamefont {Shibata}}]{sato2015cell}%
  \BibitemOpen
  \bibfield  {author} {\bibinfo {author} {\bibfnamefont {K.}~\bibnamefont
  {Sato}}, \bibinfo {author} {\bibfnamefont {T.}~\bibnamefont {Hiraiwa}},\ and\
  \bibinfo {author} {\bibfnamefont {T.}~\bibnamefont {Shibata}},\ }\href
  {https://doi.org/10.1103/PhysRevLett.115.188102} {\bibfield  {journal}
  {\bibinfo  {journal} {Physical review letters}\ }\textbf {\bibinfo {volume}
  {115}},\ \bibinfo {pages} {188102} (\bibinfo {year} {2015})}\BibitemShut
  {NoStop}%
\bibitem [{\citenamefont {Inaki}\ \emph {et~al.}(2016)\citenamefont {Inaki},
  \citenamefont {Liu},\ and\ \citenamefont {Matsuno}}]{Inaki2016}%
  \BibitemOpen
  \bibfield  {author} {\bibinfo {author} {\bibfnamefont {M.}~\bibnamefont
  {Inaki}}, \bibinfo {author} {\bibfnamefont {J.}~\bibnamefont {Liu}},\ and\
  \bibinfo {author} {\bibfnamefont {K.}~\bibnamefont {Matsuno}},\ }\href
  {https://doi.org/10.1098/rstb.2015.0403} {\bibfield  {journal} {\bibinfo
  {journal} {Philosophical Transactions of the Royal Society B: Biological
  Sciences}\ }\textbf {\bibinfo {volume} {371}},\ \bibinfo {pages} {20150403}
  (\bibinfo {year} {2016})}\BibitemShut {NoStop}%
\bibitem [{\citenamefont {Wan}\ \emph {et~al.}(2016)\citenamefont {Wan},
  \citenamefont {Chin}, \citenamefont {Worley},\ and\ \citenamefont
  {Ray}}]{Wan2016}%
  \BibitemOpen
  \bibfield  {author} {\bibinfo {author} {\bibfnamefont {L.~Q.}\ \bibnamefont
  {Wan}}, \bibinfo {author} {\bibfnamefont {A.~S.}\ \bibnamefont {Chin}},
  \bibinfo {author} {\bibfnamefont {K.~E.}\ \bibnamefont {Worley}},\ and\
  \bibinfo {author} {\bibfnamefont {P.}~\bibnamefont {Ray}},\ }\href
  {https://doi.org/10.1098/rstb.2015.0413} {\bibfield  {journal} {\bibinfo
  {journal} {Philosophical Transactions of the Royal Society B: Biological
  Sciences}\ }\textbf {\bibinfo {volume} {371}},\ \bibinfo {pages} {20150413}
  (\bibinfo {year} {2016})}\BibitemShut {NoStop}%
\bibitem [{\citenamefont {Tee}\ \emph {et~al.}(2015)\citenamefont {Tee},
  \citenamefont {Shemesh}, \citenamefont {Thiagarajan}, \citenamefont
  {Hariadi}, \citenamefont {Anderson}, \citenamefont {Page}, \citenamefont
  {Volkmann}, \citenamefont {Hanein}, \citenamefont {Sivaramakrishnan},
  \citenamefont {Kozlov} \emph {et~al.}}]{tee2015cellular}%
  \BibitemOpen
  \bibfield  {author} {\bibinfo {author} {\bibfnamefont {Y.~H.}\ \bibnamefont
  {Tee}}, \bibinfo {author} {\bibfnamefont {T.}~\bibnamefont {Shemesh}},
  \bibinfo {author} {\bibfnamefont {V.}~\bibnamefont {Thiagarajan}}, \bibinfo
  {author} {\bibfnamefont {R.~F.}\ \bibnamefont {Hariadi}}, \bibinfo {author}
  {\bibfnamefont {K.~L.}\ \bibnamefont {Anderson}}, \bibinfo {author}
  {\bibfnamefont {C.}~\bibnamefont {Page}}, \bibinfo {author} {\bibfnamefont
  {N.}~\bibnamefont {Volkmann}}, \bibinfo {author} {\bibfnamefont
  {D.}~\bibnamefont {Hanein}}, \bibinfo {author} {\bibfnamefont
  {S.}~\bibnamefont {Sivaramakrishnan}}, \bibinfo {author} {\bibfnamefont
  {M.~M.}\ \bibnamefont {Kozlov}}, \emph {et~al.},\ }\href
  {https://doi.org//10.1038/ncb3137} {\bibfield  {journal} {\bibinfo  {journal}
  {Nature cell biology}\ }\textbf {\bibinfo {volume} {17}},\ \bibinfo {pages}
  {445} (\bibinfo {year} {2015})}\BibitemShut {NoStop}%
\bibitem [{\citenamefont {Inaki}\ \emph {et~al.}(2018)\citenamefont {Inaki},
  \citenamefont {Sasamura},\ and\ \citenamefont {Matsuno}}]{Inaki2018}%
  \BibitemOpen
  \bibfield  {author} {\bibinfo {author} {\bibfnamefont {M.}~\bibnamefont
  {Inaki}}, \bibinfo {author} {\bibfnamefont {T.}~\bibnamefont {Sasamura}},\
  and\ \bibinfo {author} {\bibfnamefont {K.}~\bibnamefont {Matsuno}},\ }\href
  {https://doi.org/10.3389/fcell.2018.00034} {\bibfield  {journal} {\bibinfo
  {journal} {Frontiers in Cell and Developmental Biology}\ }\textbf {\bibinfo
  {volume} {6}},\ \bibinfo {pages} {34} (\bibinfo {year} {2018})}\BibitemShut
  {NoStop}%
\bibitem [{\citenamefont {Ray}\ \emph {et~al.}(2018)\citenamefont {Ray},
  \citenamefont {Chin}, \citenamefont {Worley}, \citenamefont {Fan},
  \citenamefont {Kaur}, \citenamefont {Wu},\ and\ \citenamefont
  {Wan}}]{Ray2018}%
  \BibitemOpen
  \bibfield  {author} {\bibinfo {author} {\bibfnamefont {P.}~\bibnamefont
  {Ray}}, \bibinfo {author} {\bibfnamefont {A.~S.}\ \bibnamefont {Chin}},
  \bibinfo {author} {\bibfnamefont {K.~E.}\ \bibnamefont {Worley}}, \bibinfo
  {author} {\bibfnamefont {J.}~\bibnamefont {Fan}}, \bibinfo {author}
  {\bibfnamefont {G.}~\bibnamefont {Kaur}}, \bibinfo {author} {\bibfnamefont
  {M.}~\bibnamefont {Wu}},\ and\ \bibinfo {author} {\bibfnamefont {L.~Q.}\
  \bibnamefont {Wan}},\ }\href {https://doi.org/10.1073/pnas.1808052115}
  {\bibfield  {journal} {\bibinfo  {journal} {Proceedings of the National
  Academy of Sciences}\ }\textbf {\bibinfo {volume} {115}},\ \bibinfo {pages}
  {E11568} (\bibinfo {year} {2018})}\BibitemShut {NoStop}%
\bibitem [{\citenamefont {Yamamoto}\ \emph {et~al.}(2020)\citenamefont
  {Yamamoto}, \citenamefont {Hiraiwa},\ and\ \citenamefont
  {Shibata}}]{yamamoto2020collective}%
  \BibitemOpen
  \bibfield  {author} {\bibinfo {author} {\bibfnamefont {T.}~\bibnamefont
  {Yamamoto}}, \bibinfo {author} {\bibfnamefont {T.}~\bibnamefont {Hiraiwa}},\
  and\ \bibinfo {author} {\bibfnamefont {T.}~\bibnamefont {Shibata}},\
  }\href@noop {} {\bibfield  {journal} {\bibinfo  {journal} {Physical Review
  Research}\ }\textbf {\bibinfo {volume} {2}},\ \bibinfo {pages} {043326}
  (\bibinfo {year} {2020})}\BibitemShut {NoStop}%
\bibitem [{\citenamefont {Middelkoop}\ \emph {et~al.}(2021)\citenamefont
  {Middelkoop}, \citenamefont {Garcia-Baucells}, \citenamefont
  {Quintero-Cadena}, \citenamefont {Pimpale}, \citenamefont {Yazdi},
  \citenamefont {Sternberg}, \citenamefont {Gross},\ and\ \citenamefont
  {Grill}}]{Middelkoop2021}%
  \BibitemOpen
  \bibfield  {author} {\bibinfo {author} {\bibfnamefont {T.~C.}\ \bibnamefont
  {Middelkoop}}, \bibinfo {author} {\bibfnamefont {J.}~\bibnamefont
  {Garcia-Baucells}}, \bibinfo {author} {\bibfnamefont {P.}~\bibnamefont
  {Quintero-Cadena}}, \bibinfo {author} {\bibfnamefont {L.~G.}\ \bibnamefont
  {Pimpale}}, \bibinfo {author} {\bibfnamefont {S.}~\bibnamefont {Yazdi}},
  \bibinfo {author} {\bibfnamefont {P.~W.}\ \bibnamefont {Sternberg}}, \bibinfo
  {author} {\bibfnamefont {P.}~\bibnamefont {Gross}},\ and\ \bibinfo {author}
  {\bibfnamefont {S.~W.}\ \bibnamefont {Grill}},\ }\href
  {https://doi.org/10.1073/pnas.2021814118} {\bibfield  {journal} {\bibinfo
  {journal} {Proceedings of the National Academy of Sciences}\ }\textbf
  {\bibinfo {volume} {118}},\ \bibinfo {pages} {e2021814118} (\bibinfo {year}
  {2021})}\BibitemShut {NoStop}%
\bibitem [{\citenamefont {Yamamoto}\ \emph {et~al.}(2023)\citenamefont
  {Yamamoto}, \citenamefont {Ishibashi}, \citenamefont {Mimori-Kiyosue},
  \citenamefont {Hiver}, \citenamefont {Tokushige}, \citenamefont {Tarama},
  \citenamefont {Takeichi},\ and\ \citenamefont
  {Shibata}}]{yamamoto2023epithelial}%
  \BibitemOpen
  \bibfield  {author} {\bibinfo {author} {\bibfnamefont {T.}~\bibnamefont
  {Yamamoto}}, \bibinfo {author} {\bibfnamefont {T.}~\bibnamefont {Ishibashi}},
  \bibinfo {author} {\bibfnamefont {Y.}~\bibnamefont {Mimori-Kiyosue}},
  \bibinfo {author} {\bibfnamefont {S.}~\bibnamefont {Hiver}}, \bibinfo
  {author} {\bibfnamefont {N.}~\bibnamefont {Tokushige}}, \bibinfo {author}
  {\bibfnamefont {M.}~\bibnamefont {Tarama}}, \bibinfo {author} {\bibfnamefont
  {M.}~\bibnamefont {Takeichi}},\ and\ \bibinfo {author} {\bibfnamefont
  {T.}~\bibnamefont {Shibata}},\ }\href
  {https://doi.org/10.1101/2023.08.16.553476} {\bibfield  {journal} {\bibinfo
  {journal} {bioRxiv}\ ,\ \bibinfo {pages} {553476}} (\bibinfo {year}
  {2023})}\BibitemShut {NoStop}%
\bibitem [{\citenamefont {Badih}\ \emph {et~al.}(2024)\citenamefont {Badih},
  \citenamefont {Schaeffer}, \citenamefont {Vianay}, \citenamefont {Smilovici},
  \citenamefont {Blanchoin}, \citenamefont {Thery},\ and\ \citenamefont
  {Kurzawa}}]{Badih2024}%
  \BibitemOpen
  \bibfield  {author} {\bibinfo {author} {\bibfnamefont {G.}~\bibnamefont
  {Badih}}, \bibinfo {author} {\bibfnamefont {A.}~\bibnamefont {Schaeffer}},
  \bibinfo {author} {\bibfnamefont {B.}~\bibnamefont {Vianay}}, \bibinfo
  {author} {\bibfnamefont {P.}~\bibnamefont {Smilovici}}, \bibinfo {author}
  {\bibfnamefont {L.}~\bibnamefont {Blanchoin}}, \bibinfo {author}
  {\bibfnamefont {M.}~\bibnamefont {Thery}},\ and\ \bibinfo {author}
  {\bibfnamefont {L.}~\bibnamefont {Kurzawa}},\ }\href
  {https://doi.org/10.1101/2024.09.02.610752} {\bibfield  {journal} {\bibinfo
  {journal} {bioRxiv}\ ,\ \bibinfo {pages} {610752}} (\bibinfo {year}
  {2024})}\BibitemShut {NoStop}%
\bibitem [{\citenamefont {Rahman}\ \emph {et~al.}(2024)\citenamefont {Rahman},
  \citenamefont {Peters},\ and\ \citenamefont {Wan}}]{rahman2024biomechanical}%
  \BibitemOpen
  \bibfield  {author} {\bibinfo {author} {\bibfnamefont {T.}~\bibnamefont
  {Rahman}}, \bibinfo {author} {\bibfnamefont {F.}~\bibnamefont {Peters}},\
  and\ \bibinfo {author} {\bibfnamefont {L.~Q.}\ \bibnamefont {Wan}},\ }\href
  {https://doi.org/j.mbm.2024.100038} {\bibfield  {journal} {\bibinfo
  {journal} {Mechanobiology in medicine}\ }\textbf {\bibinfo {volume} {2}},\
  \bibinfo {pages} {100038} (\bibinfo {year} {2024})}\BibitemShut {NoStop}%
\bibitem [{\citenamefont {Ma}\ and\ \citenamefont {Bianco}(2023)}]{Ma2023}%
  \BibitemOpen
  \bibfield  {author} {\bibinfo {author} {\bibfnamefont {B.}~\bibnamefont
  {Ma}}\ and\ \bibinfo {author} {\bibfnamefont {A.}~\bibnamefont {Bianco}},\
  }\href {https://doi.org/10.1038/s41578-023-00561-1} {\bibfield  {journal}
  {\bibinfo  {journal} {Nature Reviews Materials}\ }\textbf {\bibinfo {volume}
  {8}},\ \bibinfo {pages} {403–413} (\bibinfo {year} {2023})}\BibitemShut
  {NoStop}%
\bibitem [{\citenamefont {Alberts}\ \emph {et~al.}(2017)\citenamefont
  {Alberts}, \citenamefont {Johnson}, \citenamefont {Lewis}, \citenamefont
  {Morgan}, \citenamefont {Raff}, \citenamefont {Roberts},\ and\ \citenamefont
  {Walter}}]{Alberts}%
  \BibitemOpen
  \bibfield  {author} {\bibinfo {author} {\bibfnamefont {B.}~\bibnamefont
  {Alberts}}, \bibinfo {author} {\bibfnamefont {A.~D.}\ \bibnamefont
  {Johnson}}, \bibinfo {author} {\bibfnamefont {J.}~\bibnamefont {Lewis}},
  \bibinfo {author} {\bibfnamefont {D.}~\bibnamefont {Morgan}}, \bibinfo
  {author} {\bibfnamefont {M.}~\bibnamefont {Raff}}, \bibinfo {author}
  {\bibfnamefont {K.}~\bibnamefont {Roberts}},\ and\ \bibinfo {author}
  {\bibfnamefont {P.}~\bibnamefont {Walter}},\ }\href@noop {} {\emph {\bibinfo
  {title} {Molecular Biology of the Cell}}},\ \bibinfo {edition} {6th}\ ed.\
  (\bibinfo  {publisher} {Garland Science},\ \bibinfo {address} {New York},\
  \bibinfo {year} {2017})\BibitemShut {NoStop}%
\bibitem [{\citenamefont {Fletcher}\ and\ \citenamefont
  {Mullins}(2010)}]{Fletcher2010485}%
  \BibitemOpen
  \bibfield  {author} {\bibinfo {author} {\bibfnamefont {D.~A.}\ \bibnamefont
  {Fletcher}}\ and\ \bibinfo {author} {\bibfnamefont {R.~D.}\ \bibnamefont
  {Mullins}},\ }\href {https://doi.org/10.1038/nature08908} {\bibfield
  {journal} {\bibinfo  {journal} {Nature}\ }\textbf {\bibinfo {volume} {463}},\
  \bibinfo {pages} {485} (\bibinfo {year} {2010})}\BibitemShut {NoStop}%
\bibitem [{\citenamefont {Chen}\ \emph {et~al.}(2020)\citenamefont {Chen},
  \citenamefont {Markovich},\ and\ \citenamefont {MacKintosh}}]{chen2020motor}%
  \BibitemOpen
  \bibfield  {author} {\bibinfo {author} {\bibfnamefont {S.}~\bibnamefont
  {Chen}}, \bibinfo {author} {\bibfnamefont {T.}~\bibnamefont {Markovich}},\
  and\ \bibinfo {author} {\bibfnamefont {F.~C.}\ \bibnamefont {MacKintosh}},\
  }\href {https://doi.org/10.1103/PhysRevLett.125.208101} {\bibfield  {journal}
  {\bibinfo  {journal} {Physical Review Letters}\ }\textbf {\bibinfo {volume}
  {125}},\ \bibinfo {pages} {208101} (\bibinfo {year} {2020})}\BibitemShut
  {NoStop}%
\bibitem [{\citenamefont {Scheibner}\ \emph {et~al.}(2020)\citenamefont
  {Scheibner}, \citenamefont {Souslov}, \citenamefont {Banerjee}, \citenamefont
  {Sur{\'o}wka}, \citenamefont {Irvine},\ and\ \citenamefont
  {Vitelli}}]{scheibner2020odd}%
  \BibitemOpen
  \bibfield  {author} {\bibinfo {author} {\bibfnamefont {C.}~\bibnamefont
  {Scheibner}}, \bibinfo {author} {\bibfnamefont {A.}~\bibnamefont {Souslov}},
  \bibinfo {author} {\bibfnamefont {D.}~\bibnamefont {Banerjee}}, \bibinfo
  {author} {\bibfnamefont {P.}~\bibnamefont {Sur{\'o}wka}}, \bibinfo {author}
  {\bibfnamefont {W.~T.~M.}\ \bibnamefont {Irvine}},\ and\ \bibinfo {author}
  {\bibfnamefont {V.}~\bibnamefont {Vitelli}},\ }\href
  {https://doi.org/10.1038/s41567-020-0795-y} {\bibfield  {journal} {\bibinfo
  {journal} {Nature Physics}\ }\textbf {\bibinfo {volume} {16}},\ \bibinfo
  {pages} {475} (\bibinfo {year} {2020})}\BibitemShut {NoStop}%
\bibitem [{\citenamefont {Fruchart}\ \emph {et~al.}(2023)\citenamefont
  {Fruchart}, \citenamefont {Scheibner},\ and\ \citenamefont
  {Vitelli}}]{fruchart2023odd}%
  \BibitemOpen
  \bibfield  {author} {\bibinfo {author} {\bibfnamefont {M.}~\bibnamefont
  {Fruchart}}, \bibinfo {author} {\bibfnamefont {C.}~\bibnamefont
  {Scheibner}},\ and\ \bibinfo {author} {\bibfnamefont {V.}~\bibnamefont
  {Vitelli}},\ }\href
  {https://doi.org/10.1146/annurev-conmatphys-040821-125506} {\bibfield
  {journal} {\bibinfo  {journal} {Annual Review of Condensed Matter Physics}\
  }\textbf {\bibinfo {volume} {14}},\ \bibinfo {pages} {471} (\bibinfo {year}
  {2023})}\BibitemShut {NoStop}%
\bibitem [{\citenamefont {Alt}\ \emph {et~al.}(2017)\citenamefont {Alt},
  \citenamefont {Ganguly},\ and\ \citenamefont {Salbreux}}]{Alt2017}%
  \BibitemOpen
  \bibfield  {author} {\bibinfo {author} {\bibfnamefont {S.}~\bibnamefont
  {Alt}}, \bibinfo {author} {\bibfnamefont {P.}~\bibnamefont {Ganguly}},\ and\
  \bibinfo {author} {\bibfnamefont {G.}~\bibnamefont {Salbreux}},\ }\href
  {https://doi.org/10.1098/rstb.2015.0520} {\bibfield  {journal} {\bibinfo
  {journal} {Philosophical Transactions of the Royal Society B: Biological
  Sciences}\ }\textbf {\bibinfo {volume} {372}},\ \bibinfo {pages} {20150520}
  (\bibinfo {year} {2017})}\BibitemShut {NoStop}%
\bibitem [{\citenamefont {Vedula}\ \emph {et~al.}(2012)\citenamefont {Vedula},
  \citenamefont {Leong}, \citenamefont {Lai}, \citenamefont {Hersen},
  \citenamefont {Kabla}, \citenamefont {Lim},\ and\ \citenamefont
  {Ladoux}}]{Vedula2012}%
  \BibitemOpen
  \bibfield  {author} {\bibinfo {author} {\bibfnamefont {S.~R.~K.}\
  \bibnamefont {Vedula}}, \bibinfo {author} {\bibfnamefont {M.~C.}\
  \bibnamefont {Leong}}, \bibinfo {author} {\bibfnamefont {T.~L.}\ \bibnamefont
  {Lai}}, \bibinfo {author} {\bibfnamefont {P.}~\bibnamefont {Hersen}},
  \bibinfo {author} {\bibfnamefont {A.~J.}\ \bibnamefont {Kabla}}, \bibinfo
  {author} {\bibfnamefont {C.~T.}\ \bibnamefont {Lim}},\ and\ \bibinfo {author}
  {\bibfnamefont {B.}~\bibnamefont {Ladoux}},\ }\href
  {https://doi.org/10.1073/pnas.1119313109} {\bibfield  {journal} {\bibinfo
  {journal} {Proceedings of the National Academy of Sciences}\ }\textbf
  {\bibinfo {volume} {109}},\ \bibinfo {pages} {12974–12979} (\bibinfo {year}
  {2012})}\BibitemShut {NoStop}%
\bibitem [{\citenamefont {Fletcher}\ \emph {et~al.}(2014)\citenamefont
  {Fletcher}, \citenamefont {Osterfield}, \citenamefont {Baker},\ and\
  \citenamefont {Shvartsman}}]{Fletcher2014}%
  \BibitemOpen
  \bibfield  {author} {\bibinfo {author} {\bibfnamefont {A.~G.}\ \bibnamefont
  {Fletcher}}, \bibinfo {author} {\bibfnamefont {M.}~\bibnamefont
  {Osterfield}}, \bibinfo {author} {\bibfnamefont {R.~E.}\ \bibnamefont
  {Baker}},\ and\ \bibinfo {author} {\bibfnamefont {S.~Y.}\ \bibnamefont
  {Shvartsman}},\ }\href {https://doi.org/10.1016/j.bpj.2013.11.4498}
  {\bibfield  {journal} {\bibinfo  {journal} {Biophysical Journal}\ }\textbf
  {\bibinfo {volume} {106}},\ \bibinfo {pages} {2291–2304} (\bibinfo {year}
  {2014})}\BibitemShut {NoStop}%
\bibitem [{\citenamefont {Bi}\ \emph {et~al.}(2016)\citenamefont {Bi},
  \citenamefont {Yang}, \citenamefont {Marchetti},\ and\ \citenamefont
  {Manning}}]{Bi2016}%
  \BibitemOpen
  \bibfield  {author} {\bibinfo {author} {\bibfnamefont {D.}~\bibnamefont
  {Bi}}, \bibinfo {author} {\bibfnamefont {X.}~\bibnamefont {Yang}}, \bibinfo
  {author} {\bibfnamefont {M.~C.}\ \bibnamefont {Marchetti}},\ and\ \bibinfo
  {author} {\bibfnamefont {M.~L.}\ \bibnamefont {Manning}},\ }\href
  {https://doi.org/10.1103/physrevx.6.021011} {\bibfield  {journal} {\bibinfo
  {journal} {Physical Review X}\ }\textbf {\bibinfo {volume} {6}},\ \bibinfo
  {pages} {021011} (\bibinfo {year} {2016})}\BibitemShut {NoStop}%
\bibitem [{\citenamefont {Xi}\ \emph {et~al.}(2018)\citenamefont {Xi},
  \citenamefont {Saw}, \citenamefont {Delacour}, \citenamefont {Lim},\ and\
  \citenamefont {Ladoux}}]{Xi2018}%
  \BibitemOpen
  \bibfield  {author} {\bibinfo {author} {\bibfnamefont {W.}~\bibnamefont
  {Xi}}, \bibinfo {author} {\bibfnamefont {T.~B.}\ \bibnamefont {Saw}},
  \bibinfo {author} {\bibfnamefont {D.}~\bibnamefont {Delacour}}, \bibinfo
  {author} {\bibfnamefont {C.~T.}\ \bibnamefont {Lim}},\ and\ \bibinfo {author}
  {\bibfnamefont {B.}~\bibnamefont {Ladoux}},\ }\href
  {https://doi.org/10.1038/s41578-018-0066-z} {\bibfield  {journal} {\bibinfo
  {journal} {Nature Reviews Materials}\ }\textbf {\bibinfo {volume} {4}},\
  \bibinfo {pages} {23–44} (\bibinfo {year} {2018})}\BibitemShut {NoStop}%
\bibitem [{\citenamefont {Sussman}\ and\ \citenamefont
  {Merkel}(2018)}]{sussman2018no}%
  \BibitemOpen
  \bibfield  {author} {\bibinfo {author} {\bibfnamefont {D.~M.}\ \bibnamefont
  {Sussman}}\ and\ \bibinfo {author} {\bibfnamefont {M.}~\bibnamefont
  {Merkel}},\ }\href {https://doi.org/10.1039/C7SM02127E} {\bibfield  {journal}
  {\bibinfo  {journal} {Soft Matter}\ }\textbf {\bibinfo {volume} {14}},\
  \bibinfo {pages} {3397} (\bibinfo {year} {2018})}\BibitemShut {NoStop}%
\bibitem [{\citenamefont {Farhadifar}\ \emph
  {et~al.}(2007{\natexlab{a}})\citenamefont {Farhadifar}, \citenamefont
  {R{\"o}per}, \citenamefont {Aigouy}, \citenamefont {Eaton},\ and\
  \citenamefont {J{\"u}licher}}]{farhadifar2007influence}%
  \BibitemOpen
  \bibfield  {author} {\bibinfo {author} {\bibfnamefont {R.}~\bibnamefont
  {Farhadifar}}, \bibinfo {author} {\bibfnamefont {J.-C.}\ \bibnamefont
  {R{\"o}per}}, \bibinfo {author} {\bibfnamefont {B.}~\bibnamefont {Aigouy}},
  \bibinfo {author} {\bibfnamefont {S.}~\bibnamefont {Eaton}},\ and\ \bibinfo
  {author} {\bibfnamefont {F.}~\bibnamefont {J{\"u}licher}},\ }\href
  {https://doi.org/10.1016/j.cub.2007.11.049} {\bibfield  {journal} {\bibinfo
  {journal} {Current Biology}\ }\textbf {\bibinfo {volume} {17}},\ \bibinfo
  {pages} {2095} (\bibinfo {year} {2007}{\natexlab{a}})}\BibitemShut {NoStop}%
\bibitem [{\citenamefont {Staple}\ \emph {et~al.}(2010)\citenamefont {Staple},
  \citenamefont {Farhadifar}, \citenamefont {R{\"o}per}, \citenamefont
  {Aigouy}, \citenamefont {Eaton},\ and\ \citenamefont
  {J{\"u}licher}}]{staple2010mechanics}%
  \BibitemOpen
  \bibfield  {author} {\bibinfo {author} {\bibfnamefont {D.~B.}\ \bibnamefont
  {Staple}}, \bibinfo {author} {\bibfnamefont {R.}~\bibnamefont {Farhadifar}},
  \bibinfo {author} {\bibfnamefont {J.-C.}\ \bibnamefont {R{\"o}per}}, \bibinfo
  {author} {\bibfnamefont {B.}~\bibnamefont {Aigouy}}, \bibinfo {author}
  {\bibfnamefont {S.}~\bibnamefont {Eaton}},\ and\ \bibinfo {author}
  {\bibfnamefont {F.}~\bibnamefont {J{\"u}licher}},\ }\href
  {https://doi.org/10.1140/epje/i2010-10677-0} {\bibfield  {journal} {\bibinfo
  {journal} {The European Physical Journal E}\ }\textbf {\bibinfo {volume}
  {33}},\ \bibinfo {pages} {117} (\bibinfo {year} {2010})}\BibitemShut
  {NoStop}%
\bibitem [{\citenamefont {Bi}\ \emph {et~al.}(2015)\citenamefont {Bi},
  \citenamefont {Lopez}, \citenamefont {Schwarz},\ and\ \citenamefont
  {Manning}}]{bi2015density}%
  \BibitemOpen
  \bibfield  {author} {\bibinfo {author} {\bibfnamefont {D.}~\bibnamefont
  {Bi}}, \bibinfo {author} {\bibfnamefont {J.~H.}\ \bibnamefont {Lopez}},
  \bibinfo {author} {\bibfnamefont {J.~M.}\ \bibnamefont {Schwarz}},\ and\
  \bibinfo {author} {\bibfnamefont {M.~L.}\ \bibnamefont {Manning}},\ }\href
  {https://doi.org/10.1038/nphys3471} {\bibfield  {journal} {\bibinfo
  {journal} {Nature Physics}\ }\textbf {\bibinfo {volume} {11}},\ \bibinfo
  {pages} {1074} (\bibinfo {year} {2015})}\BibitemShut {NoStop}%
\bibitem [{\citenamefont {Weaire}\ and\ \citenamefont
  {Rivier}(1984)}]{Weaire1984}%
  \BibitemOpen
  \bibfield  {author} {\bibinfo {author} {\bibfnamefont {D.}~\bibnamefont
  {Weaire}}\ and\ \bibinfo {author} {\bibfnamefont {N.}~\bibnamefont
  {Rivier}},\ }\href {https://doi.org/10.1080/00107518408210979} {\bibfield
  {journal} {\bibinfo  {journal} {Contemporary Physics}\ }\textbf {\bibinfo
  {volume} {25}},\ \bibinfo {pages} {59–99} (\bibinfo {year}
  {1984})}\BibitemShut {NoStop}%
\bibitem [{\citenamefont {Balasubramaniam}\ \emph {et~al.}(2022)\citenamefont
  {Balasubramaniam}, \citenamefont {Mège},\ and\ \citenamefont
  {Ladoux}}]{balasubramaniam2022active}%
  \BibitemOpen
  \bibfield  {author} {\bibinfo {author} {\bibfnamefont {L.}~\bibnamefont
  {Balasubramaniam}}, \bibinfo {author} {\bibfnamefont {R.-M.}\ \bibnamefont
  {Mège}},\ and\ \bibinfo {author} {\bibfnamefont {B.}~\bibnamefont
  {Ladoux}},\ }\href {https://doi.org/10.1016/j.gde.2021.101897} {\bibfield
  {journal} {\bibinfo  {journal} {Current Opinion in Genetics \& Development}\
  }\textbf {\bibinfo {volume} {73}},\ \bibinfo {pages} {101897} (\bibinfo
  {year} {2022})}\BibitemShut {NoStop}%
\bibitem [{\citenamefont {Alert}\ \emph {et~al.}(2022)\citenamefont {Alert},
  \citenamefont {Casademunt},\ and\ \citenamefont {Joanny}}]{alert2022active}%
  \BibitemOpen
  \bibfield  {author} {\bibinfo {author} {\bibfnamefont {R.}~\bibnamefont
  {Alert}}, \bibinfo {author} {\bibfnamefont {J.}~\bibnamefont {Casademunt}},\
  and\ \bibinfo {author} {\bibfnamefont {J.-F.}\ \bibnamefont {Joanny}},\
  }\href {https://doi.org/10.1146/annurev-conmatphys-082321-035957} {\bibfield
  {journal} {\bibinfo  {journal} {Annual Review of Condensed Matter Physics}\
  }\textbf {\bibinfo {volume} {13}},\ \bibinfo {pages} {143} (\bibinfo {year}
  {2022})}\BibitemShut {NoStop}%
\bibitem [{\citenamefont {Yashunsky}\ \emph {et~al.}(2022)\citenamefont
  {Yashunsky}, \citenamefont {Pearce}, \citenamefont {Blanch-Mercader},
  \citenamefont {Ascione}, \citenamefont {Silberzan},\ and\ \citenamefont
  {Giomi}}]{yashunsky2022chiral}%
  \BibitemOpen
  \bibfield  {author} {\bibinfo {author} {\bibfnamefont {V.}~\bibnamefont
  {Yashunsky}}, \bibinfo {author} {\bibfnamefont {D.~J.}\ \bibnamefont
  {Pearce}}, \bibinfo {author} {\bibfnamefont {C.}~\bibnamefont
  {Blanch-Mercader}}, \bibinfo {author} {\bibfnamefont {F.}~\bibnamefont
  {Ascione}}, \bibinfo {author} {\bibfnamefont {P.}~\bibnamefont {Silberzan}},\
  and\ \bibinfo {author} {\bibfnamefont {L.}~\bibnamefont {Giomi}},\ }\href
  {https://doi.org/10.1103/PhysRevX.12.041017} {\bibfield  {journal} {\bibinfo
  {journal} {Physical Review X}\ }\textbf {\bibinfo {volume} {12}},\ \bibinfo
  {pages} {041017} (\bibinfo {year} {2022})}\BibitemShut {NoStop}%
\bibitem [{\citenamefont {Voronoi}(1908{\natexlab{a}})}]{voronoi1908a}%
  \BibitemOpen
  \bibfield  {author} {\bibinfo {author} {\bibfnamefont {G.}~\bibnamefont
  {Voronoi}},\ }\href@noop {} {\bibfield  {journal} {\bibinfo  {journal}
  {Journal f{\"u}r die reine und angewandte Mathematik (Crelles Journal)}\
  }\textbf {\bibinfo {volume} {1908}},\ \bibinfo {pages} {97} (\bibinfo {year}
  {1908}{\natexlab{a}})}\BibitemShut {NoStop}%
\bibitem [{\citenamefont {Voronoi}(1908{\natexlab{b}})}]{voronoi1908b}%
  \BibitemOpen
  \bibfield  {author} {\bibinfo {author} {\bibfnamefont {G.}~\bibnamefont
  {Voronoi}},\ }\href@noop {} {\bibfield  {journal} {\bibinfo  {journal}
  {Journal f{\"u}r die reine und angewandte Mathematik (Crelles Journal)}\
  }\textbf {\bibinfo {volume} {1908}},\ \bibinfo {pages} {198} (\bibinfo {year}
  {1908}{\natexlab{b}})}\BibitemShut {NoStop}%
\bibitem [{\citenamefont {Farhadifar}\ \emph
  {et~al.}(2007{\natexlab{b}})\citenamefont {Farhadifar}, \citenamefont
  {Röper}, \citenamefont {Aigouy}, \citenamefont {Eaton},\ and\ \citenamefont
  {Jülicher}}]{Farhadifar2007}%
  \BibitemOpen
  \bibfield  {author} {\bibinfo {author} {\bibfnamefont {R.}~\bibnamefont
  {Farhadifar}}, \bibinfo {author} {\bibfnamefont {J.-C.}\ \bibnamefont
  {Röper}}, \bibinfo {author} {\bibfnamefont {B.}~\bibnamefont {Aigouy}},
  \bibinfo {author} {\bibfnamefont {S.}~\bibnamefont {Eaton}},\ and\ \bibinfo
  {author} {\bibfnamefont {F.}~\bibnamefont {Jülicher}},\ }\href
  {https://doi.org/10.1016/j.cub.2007.11.049} {\bibfield  {journal} {\bibinfo
  {journal} {Current Biology}\ }\textbf {\bibinfo {volume} {17}},\ \bibinfo
  {pages} {2095–2104} (\bibinfo {year} {2007}{\natexlab{b}})}\BibitemShut
  {NoStop}%
\bibitem [{\citenamefont {Marchetti}\ \emph
  {et~al.}(2013{\natexlab{a}})\citenamefont {Marchetti}, \citenamefont
  {Joanny}, \citenamefont {Ramaswamy}, \citenamefont {Liverpool}, \citenamefont
  {Prost}, \citenamefont {Rao},\ and\ \citenamefont {Simha}}]{Marchetti2013}%
  \BibitemOpen
  \bibfield  {author} {\bibinfo {author} {\bibfnamefont {M.~C.}\ \bibnamefont
  {Marchetti}}, \bibinfo {author} {\bibfnamefont {J.~F.}\ \bibnamefont
  {Joanny}}, \bibinfo {author} {\bibfnamefont {S.}~\bibnamefont {Ramaswamy}},
  \bibinfo {author} {\bibfnamefont {T.~B.}\ \bibnamefont {Liverpool}}, \bibinfo
  {author} {\bibfnamefont {J.}~\bibnamefont {Prost}}, \bibinfo {author}
  {\bibfnamefont {M.}~\bibnamefont {Rao}},\ and\ \bibinfo {author}
  {\bibfnamefont {R.~A.}\ \bibnamefont {Simha}},\ }\href
  {https://doi.org/10.1103/revmodphys.85.1143} {\bibfield  {journal} {\bibinfo
  {journal} {Reviews of Modern Physics}\ }\textbf {\bibinfo {volume} {85}},\
  \bibinfo {pages} {1143–1189} (\bibinfo {year}
  {2013}{\natexlab{a}})}\BibitemShut {NoStop}%
\bibitem [{\citenamefont {Hallatschek}\ \emph {et~al.}(2023)\citenamefont
  {Hallatschek}, \citenamefont {Datta}, \citenamefont {Drescher}, \citenamefont
  {Dunkel}, \citenamefont {Elgeti}, \citenamefont {Waclaw},\ and\ \citenamefont
  {Wingreen}}]{Hallatschek2023}%
  \BibitemOpen
  \bibfield  {author} {\bibinfo {author} {\bibfnamefont {O.}~\bibnamefont
  {Hallatschek}}, \bibinfo {author} {\bibfnamefont {S.~S.}\ \bibnamefont
  {Datta}}, \bibinfo {author} {\bibfnamefont {K.}~\bibnamefont {Drescher}},
  \bibinfo {author} {\bibfnamefont {J.}~\bibnamefont {Dunkel}}, \bibinfo
  {author} {\bibfnamefont {J.}~\bibnamefont {Elgeti}}, \bibinfo {author}
  {\bibfnamefont {B.}~\bibnamefont {Waclaw}},\ and\ \bibinfo {author}
  {\bibfnamefont {N.~S.}\ \bibnamefont {Wingreen}},\ }\href
  {https://doi.org/10.1038/s42254-023-00593-0} {\bibfield  {journal} {\bibinfo
  {journal} {Nature Reviews Physics}\ }\textbf {\bibinfo {volume} {5}},\
  \bibinfo {pages} {407–419} (\bibinfo {year} {2023})}\BibitemShut {NoStop}%
\bibitem [{\citenamefont {Petridou}\ and\ \citenamefont
  {Heisenberg}(2019)}]{Petridou2019}%
  \BibitemOpen
  \bibfield  {author} {\bibinfo {author} {\bibfnamefont {N.~I.}\ \bibnamefont
  {Petridou}}\ and\ \bibinfo {author} {\bibfnamefont {C.}~\bibnamefont
  {Heisenberg}},\ }\href {https://doi.org/10.15252/embj.2019102497} {\bibfield
  {journal} {\bibinfo  {journal} {The EMBO Journal}\ }\textbf {\bibinfo
  {volume} {38}},\ \bibinfo {pages} {e102497} (\bibinfo {year}
  {2019})}\BibitemShut {NoStop}%
\bibitem [{\citenamefont {Mongera}\ \emph {et~al.}(2018)\citenamefont
  {Mongera}, \citenamefont {Rowghanian}, \citenamefont {Gustafson},
  \citenamefont {Shelton}, \citenamefont {Kealhofer}, \citenamefont {Carn},
  \citenamefont {Serwane}, \citenamefont {Lucio}, \citenamefont {Giammona},\
  and\ \citenamefont {Campàs}}]{Mongera2018}%
  \BibitemOpen
  \bibfield  {author} {\bibinfo {author} {\bibfnamefont {A.}~\bibnamefont
  {Mongera}}, \bibinfo {author} {\bibfnamefont {P.}~\bibnamefont {Rowghanian}},
  \bibinfo {author} {\bibfnamefont {H.~J.}\ \bibnamefont {Gustafson}}, \bibinfo
  {author} {\bibfnamefont {E.}~\bibnamefont {Shelton}}, \bibinfo {author}
  {\bibfnamefont {D.~A.}\ \bibnamefont {Kealhofer}}, \bibinfo {author}
  {\bibfnamefont {E.~K.}\ \bibnamefont {Carn}}, \bibinfo {author}
  {\bibfnamefont {F.}~\bibnamefont {Serwane}}, \bibinfo {author} {\bibfnamefont
  {A.~A.}\ \bibnamefont {Lucio}}, \bibinfo {author} {\bibfnamefont
  {J.}~\bibnamefont {Giammona}},\ and\ \bibinfo {author} {\bibfnamefont
  {O.}~\bibnamefont {Campàs}},\ }\href
  {https://doi.org/10.1038/s41586-018-0479-2} {\bibfield  {journal} {\bibinfo
  {journal} {Nature}\ }\textbf {\bibinfo {volume} {561}},\ \bibinfo {pages}
  {401–405} (\bibinfo {year} {2018})}\BibitemShut {NoStop}%
\bibitem [{\citenamefont {Wu}\ \emph {et~al.}(2023)\citenamefont {Wu},
  \citenamefont {Yamada},\ and\ \citenamefont {Wang}}]{Wu2023}%
  \BibitemOpen
  \bibfield  {author} {\bibinfo {author} {\bibfnamefont {D.}~\bibnamefont
  {Wu}}, \bibinfo {author} {\bibfnamefont {K.~M.}\ \bibnamefont {Yamada}},\
  and\ \bibinfo {author} {\bibfnamefont {S.}~\bibnamefont {Wang}},\ }\href
  {https://doi.org/10.1146/annurev-cellbio-020223-031019} {\bibfield  {journal}
  {\bibinfo  {journal} {Annual Review of Cell and Developmental Biology}\
  }\textbf {\bibinfo {volume} {39}},\ \bibinfo {pages} {123–144} (\bibinfo
  {year} {2023})}\BibitemShut {NoStop}%
\bibitem [{\citenamefont {Lenne}\ and\ \citenamefont
  {Trivedi}(2022)}]{Lenne2022}%
  \BibitemOpen
  \bibfield  {author} {\bibinfo {author} {\bibfnamefont {P.-F.}\ \bibnamefont
  {Lenne}}\ and\ \bibinfo {author} {\bibfnamefont {V.}~\bibnamefont
  {Trivedi}},\ }\href {https://doi.org/10.1038/s41467-022-28151-9} {\bibfield
  {journal} {\bibinfo  {journal} {Nature Communications}\ }\textbf {\bibinfo
  {volume} {13}},\ \bibinfo {pages} {664} (\bibinfo {year} {2022})}\BibitemShut
  {NoStop}%
\bibitem [{\citenamefont {Petridou}\ \emph {et~al.}(2021)\citenamefont
  {Petridou}, \citenamefont {Corominas-Murtra}, \citenamefont {Heisenberg},\
  and\ \citenamefont {Hannezo}}]{Petridou2021}%
  \BibitemOpen
  \bibfield  {author} {\bibinfo {author} {\bibfnamefont {N.~I.}\ \bibnamefont
  {Petridou}}, \bibinfo {author} {\bibfnamefont {B.}~\bibnamefont
  {Corominas-Murtra}}, \bibinfo {author} {\bibfnamefont {C.-P.}\ \bibnamefont
  {Heisenberg}},\ and\ \bibinfo {author} {\bibfnamefont {E.}~\bibnamefont
  {Hannezo}},\ }\href {https://doi.org/10.1016/j.cell.2021.02.017} {\bibfield
  {journal} {\bibinfo  {journal} {Cell}\ }\textbf {\bibinfo {volume} {184}},\
  \bibinfo {pages} {1914} (\bibinfo {year} {2021})}\BibitemShut {NoStop}%
\bibitem [{\citenamefont {Mora}\ and\ \citenamefont {Bialek}(2011)}]{Mora2011}%
  \BibitemOpen
  \bibfield  {author} {\bibinfo {author} {\bibfnamefont {T.}~\bibnamefont
  {Mora}}\ and\ \bibinfo {author} {\bibfnamefont {W.}~\bibnamefont {Bialek}},\
  }\href {https://doi.org/10.1007/s10955-011-0229-4} {\bibfield  {journal}
  {\bibinfo  {journal} {Journal of Statistical Physics}\ }\textbf {\bibinfo
  {volume} {144}},\ \bibinfo {pages} {268–302} (\bibinfo {year}
  {2011})}\BibitemShut {NoStop}%
\bibitem [{\citenamefont {Hidalgo}\ \emph {et~al.}(2014)\citenamefont
  {Hidalgo}, \citenamefont {Grilli}, \citenamefont {Suweis}, \citenamefont
  {Muñoz}, \citenamefont {Banavar},\ and\ \citenamefont
  {Maritan}}]{Hidalgo2014}%
  \BibitemOpen
  \bibfield  {author} {\bibinfo {author} {\bibfnamefont {J.}~\bibnamefont
  {Hidalgo}}, \bibinfo {author} {\bibfnamefont {J.}~\bibnamefont {Grilli}},
  \bibinfo {author} {\bibfnamefont {S.}~\bibnamefont {Suweis}}, \bibinfo
  {author} {\bibfnamefont {M.~A.}\ \bibnamefont {Muñoz}}, \bibinfo {author}
  {\bibfnamefont {J.~R.}\ \bibnamefont {Banavar}},\ and\ \bibinfo {author}
  {\bibfnamefont {A.}~\bibnamefont {Maritan}},\ }\href
  {https://doi.org/10.1073/pnas.1319166111} {\bibfield  {journal} {\bibinfo
  {journal} {Proceedings of the National Academy of Sciences}\ }\textbf
  {\bibinfo {volume} {111}},\ \bibinfo {pages} {10095–10100} (\bibinfo {year}
  {2014})}\BibitemShut {NoStop}%
\bibitem [{\citenamefont {Krotov}\ \emph {et~al.}(2014)\citenamefont {Krotov},
  \citenamefont {Dubuis}, \citenamefont {Gregor},\ and\ \citenamefont
  {Bialek}}]{Krotov2014}%
  \BibitemOpen
  \bibfield  {author} {\bibinfo {author} {\bibfnamefont {D.}~\bibnamefont
  {Krotov}}, \bibinfo {author} {\bibfnamefont {J.~O.}\ \bibnamefont {Dubuis}},
  \bibinfo {author} {\bibfnamefont {T.}~\bibnamefont {Gregor}},\ and\ \bibinfo
  {author} {\bibfnamefont {W.}~\bibnamefont {Bialek}},\ }\href
  {https://doi.org/10.1073/pnas.1324186111} {\bibfield  {journal} {\bibinfo
  {journal} {Proceedings of the National Academy of Sciences}\ }\textbf
  {\bibinfo {volume} {111}},\ \bibinfo {pages} {3683–3688} (\bibinfo {year}
  {2014})}\BibitemShut {NoStop}%
\bibitem [{\citenamefont {Muñoz}(2018)}]{Munoz2018}%
  \BibitemOpen
  \bibfield  {author} {\bibinfo {author} {\bibfnamefont {M.~A.}\ \bibnamefont
  {Muñoz}},\ }\href {https://doi.org/10.1103/revmodphys.90.031001} {\bibfield
  {journal} {\bibinfo  {journal} {Reviews of Modern Physics}\ }\textbf
  {\bibinfo {volume} {90}},\ \bibinfo {pages} {031001} (\bibinfo {year}
  {2018})}\BibitemShut {NoStop}%
\bibitem [{\citenamefont {L{\'o}pez}\ \emph {et~al.}(2015)\citenamefont
  {L{\'o}pez}, \citenamefont {Gachelin}, \citenamefont {Douarche},
  \citenamefont {Auradou},\ and\ \citenamefont
  {Cl{\'e}ment}}]{lopez2015turning}%
  \BibitemOpen
  \bibfield  {author} {\bibinfo {author} {\bibfnamefont {H.~M.}\ \bibnamefont
  {L{\'o}pez}}, \bibinfo {author} {\bibfnamefont {J.}~\bibnamefont {Gachelin}},
  \bibinfo {author} {\bibfnamefont {C.}~\bibnamefont {Douarche}}, \bibinfo
  {author} {\bibfnamefont {H.}~\bibnamefont {Auradou}},\ and\ \bibinfo {author}
  {\bibfnamefont {E.}~\bibnamefont {Cl{\'e}ment}},\ }\href
  {https://doi.org/10.1103/PhysRevLett.115.028301} {\bibfield  {journal}
  {\bibinfo  {journal} {Physical Review Letters}\ }\textbf {\bibinfo {volume}
  {115}},\ \bibinfo {pages} {028301} (\bibinfo {year} {2015})}\BibitemShut
  {NoStop}%
\bibitem [{\citenamefont {Lubensky}\ \emph {et~al.}(2015)\citenamefont
  {Lubensky}, \citenamefont {Kane}, \citenamefont {Mao}, \citenamefont
  {Souslov},\ and\ \citenamefont {Sun}}]{Lubensky2015}%
  \BibitemOpen
  \bibfield  {author} {\bibinfo {author} {\bibfnamefont {T.~C.}\ \bibnamefont
  {Lubensky}}, \bibinfo {author} {\bibfnamefont {C.~L.}\ \bibnamefont {Kane}},
  \bibinfo {author} {\bibfnamefont {X.}~\bibnamefont {Mao}}, \bibinfo {author}
  {\bibfnamefont {A.}~\bibnamefont {Souslov}},\ and\ \bibinfo {author}
  {\bibfnamefont {K.}~\bibnamefont {Sun}},\ }\href
  {https://doi.org/10.1088/0034-4885/78/7/073901} {\bibfield  {journal}
  {\bibinfo  {journal} {Reports on Progress in Physics}\ }\textbf {\bibinfo
  {volume} {78}},\ \bibinfo {pages} {073901} (\bibinfo {year}
  {2015})}\BibitemShut {NoStop}%
\bibitem [{\citenamefont {Liu}\ and\ \citenamefont {Nagel}(2010)}]{Liu2010}%
  \BibitemOpen
  \bibfield  {author} {\bibinfo {author} {\bibfnamefont {A.~J.}\ \bibnamefont
  {Liu}}\ and\ \bibinfo {author} {\bibfnamefont {S.~R.}\ \bibnamefont
  {Nagel}},\ }\href {https://doi.org/10.1146/annurev-conmatphys-070909-104045}
  {\bibfield  {journal} {\bibinfo  {journal} {Annual Review of Condensed Matter
  Physics}\ }\textbf {\bibinfo {volume} {1}},\ \bibinfo {pages} {347–369}
  (\bibinfo {year} {2010})}\BibitemShut {NoStop}%
\bibitem [{\citenamefont {Nelson}\ \emph {et~al.}(2005)\citenamefont {Nelson},
  \citenamefont {Jean}, \citenamefont {Tan}, \citenamefont {Liu}, \citenamefont
  {Sniadecki}, \citenamefont {Spector},\ and\ \citenamefont
  {Chen}}]{nelson2005emergent}%
  \BibitemOpen
  \bibfield  {author} {\bibinfo {author} {\bibfnamefont {C.~M.}\ \bibnamefont
  {Nelson}}, \bibinfo {author} {\bibfnamefont {R.~P.}\ \bibnamefont {Jean}},
  \bibinfo {author} {\bibfnamefont {J.~L.}\ \bibnamefont {Tan}}, \bibinfo
  {author} {\bibfnamefont {W.~F.}\ \bibnamefont {Liu}}, \bibinfo {author}
  {\bibfnamefont {N.~J.}\ \bibnamefont {Sniadecki}}, \bibinfo {author}
  {\bibfnamefont {A.~A.}\ \bibnamefont {Spector}},\ and\ \bibinfo {author}
  {\bibfnamefont {C.~S.}\ \bibnamefont {Chen}},\ }\href
  {https://doi.org/10.1073/pnas.0502575102} {\bibfield  {journal} {\bibinfo
  {journal} {Proceedings of the National Academy of Sciences}\ }\textbf
  {\bibinfo {volume} {102}},\ \bibinfo {pages} {11594} (\bibinfo {year}
  {2005})}\BibitemShut {NoStop}%
\bibitem [{\citenamefont {Carpenter}\ \emph {et~al.}(2024)\citenamefont
  {Carpenter}, \citenamefont {Pérez-Verdugo},\ and\ \citenamefont
  {Banerjee}}]{Carpenter2024}%
  \BibitemOpen
  \bibfield  {author} {\bibinfo {author} {\bibfnamefont {L.~C.}\ \bibnamefont
  {Carpenter}}, \bibinfo {author} {\bibfnamefont {F.}~\bibnamefont
  {Pérez-Verdugo}},\ and\ \bibinfo {author} {\bibfnamefont {S.}~\bibnamefont
  {Banerjee}},\ }\href {https://doi.org/10.1016/j.bpj.2024.03.002} {\bibfield
  {journal} {\bibinfo  {journal} {Biophysical Journal}\ }\textbf {\bibinfo
  {volume} {123}},\ \bibinfo {pages} {909–919} (\bibinfo {year}
  {2024})}\BibitemShut {NoStop}%
\bibitem [{\citenamefont {Lu}\ \emph {et~al.}(2024)\citenamefont {Lu},
  \citenamefont {Guyomar}, \citenamefont {Vagne}, \citenamefont {Berthoz},
  \citenamefont {Torres-S{\'a}nchez}, \citenamefont {Lieb}, \citenamefont
  {Martin-Lemaitre}, \citenamefont {van Unen}, \citenamefont {Honigmann},
  \citenamefont {Pertz} \emph {et~al.}}]{lu2024polarity}%
  \BibitemOpen
  \bibfield  {author} {\bibinfo {author} {\bibfnamefont {L.}~\bibnamefont
  {Lu}}, \bibinfo {author} {\bibfnamefont {T.}~\bibnamefont {Guyomar}},
  \bibinfo {author} {\bibfnamefont {Q.}~\bibnamefont {Vagne}}, \bibinfo
  {author} {\bibfnamefont {R.}~\bibnamefont {Berthoz}}, \bibinfo {author}
  {\bibfnamefont {A.}~\bibnamefont {Torres-S{\'a}nchez}}, \bibinfo {author}
  {\bibfnamefont {M.}~\bibnamefont {Lieb}}, \bibinfo {author} {\bibfnamefont
  {C.}~\bibnamefont {Martin-Lemaitre}}, \bibinfo {author} {\bibfnamefont
  {K.}~\bibnamefont {van Unen}}, \bibinfo {author} {\bibfnamefont
  {A.}~\bibnamefont {Honigmann}}, \bibinfo {author} {\bibfnamefont
  {O.}~\bibnamefont {Pertz}}, \emph {et~al.},\ }\href
  {https://doi.org/10.1038/s41567-024-02460-w} {\bibfield  {journal} {\bibinfo
  {journal} {Nature Physics}\ }\textbf {\bibinfo {volume} {20}},\ \bibinfo
  {pages} {1194} (\bibinfo {year} {2024})}\BibitemShut {NoStop}%
\bibitem [{\citenamefont {Schmitt}\ \emph {et~al.}(2024)\citenamefont
  {Schmitt}, \citenamefont {Colen}, \citenamefont {Sala}, \citenamefont
  {Devany}, \citenamefont {Seetharaman}, \citenamefont {Caillier},
  \citenamefont {Gardel}, \citenamefont {Oakes},\ and\ \citenamefont
  {Vitelli}}]{Schmitt2024}%
  \BibitemOpen
  \bibfield  {author} {\bibinfo {author} {\bibfnamefont {M.~S.}\ \bibnamefont
  {Schmitt}}, \bibinfo {author} {\bibfnamefont {J.}~\bibnamefont {Colen}},
  \bibinfo {author} {\bibfnamefont {S.}~\bibnamefont {Sala}}, \bibinfo {author}
  {\bibfnamefont {J.}~\bibnamefont {Devany}}, \bibinfo {author} {\bibfnamefont
  {S.}~\bibnamefont {Seetharaman}}, \bibinfo {author} {\bibfnamefont
  {A.}~\bibnamefont {Caillier}}, \bibinfo {author} {\bibfnamefont {M.~L.}\
  \bibnamefont {Gardel}}, \bibinfo {author} {\bibfnamefont {P.~W.}\
  \bibnamefont {Oakes}},\ and\ \bibinfo {author} {\bibfnamefont
  {V.}~\bibnamefont {Vitelli}},\ }\href
  {https://doi.org/10.1016/j.cell.2023.11.041} {\bibfield  {journal} {\bibinfo
  {journal} {Cell}\ }\textbf {\bibinfo {volume} {187}},\ \bibinfo {pages} {481}
  (\bibinfo {year} {2024})}\BibitemShut {NoStop}%
\bibitem [{\citenamefont {Shumilin}\ \emph {et~al.}(2023)\citenamefont
  {Shumilin}, \citenamefont {Ryabov}, \citenamefont {Barannikov}, \citenamefont
  {Burnaev},\ and\ \citenamefont {Vanovskii}}]{shumilin2023method}%
  \BibitemOpen
  \bibfield  {author} {\bibinfo {author} {\bibfnamefont {S.}~\bibnamefont
  {Shumilin}}, \bibinfo {author} {\bibfnamefont {A.}~\bibnamefont {Ryabov}},
  \bibinfo {author} {\bibfnamefont {S.}~\bibnamefont {Barannikov}}, \bibinfo
  {author} {\bibfnamefont {E.}~\bibnamefont {Burnaev}},\ and\ \bibinfo {author}
  {\bibfnamefont {V.}~\bibnamefont {Vanovskii}},\ }\href@noop {} {\bibinfo
  {title} {A method for auto-differentiation of the voronoi tessellation}}
  (\bibinfo {year} {2023}),\ \Eprint {https://arxiv.org/abs/2312.16192}
  {arXiv:2312.16192 [cs.CG]} \BibitemShut {NoStop}%
\bibitem [{\citenamefont {Stringer}\ and\ \citenamefont
  {Pachitariu}(2025)}]{stringer2025cellpose3}%
  \BibitemOpen
  \bibfield  {author} {\bibinfo {author} {\bibfnamefont {C.}~\bibnamefont
  {Stringer}}\ and\ \bibinfo {author} {\bibfnamefont {M.}~\bibnamefont
  {Pachitariu}},\ }\href {https://doi.org/10.1038/s41592-025-02595-5}
  {\bibfield  {journal} {\bibinfo  {journal} {Nature Methods}\ }\textbf
  {\bibinfo {volume} {22}},\ \bibinfo {pages} {592} (\bibinfo {year}
  {2025})}\BibitemShut {NoStop}%
\bibitem [{\citenamefont {Alert}\ and\ \citenamefont
  {Trepat}(2020)}]{Alert2020}%
  \BibitemOpen
  \bibfield  {author} {\bibinfo {author} {\bibfnamefont {R.}~\bibnamefont
  {Alert}}\ and\ \bibinfo {author} {\bibfnamefont {X.}~\bibnamefont {Trepat}},\
  }\href {https://doi.org/10.1146/annurev-conmatphys-031218-013516} {\bibfield
  {journal} {\bibinfo  {journal} {Annual Review of Condensed Matter Physics}\
  }\textbf {\bibinfo {volume} {11}},\ \bibinfo {pages} {77–101} (\bibinfo
  {year} {2020})}\BibitemShut {NoStop}%
\bibitem [{\citenamefont {Cochet-Escartin}\ \emph {et~al.}(2014)\citenamefont
  {Cochet-Escartin}, \citenamefont {Ranft}, \citenamefont {Silberzan},\ and\
  \citenamefont {Marcq}}]{CochetEscartin2014}%
  \BibitemOpen
  \bibfield  {author} {\bibinfo {author} {\bibfnamefont {O.}~\bibnamefont
  {Cochet-Escartin}}, \bibinfo {author} {\bibfnamefont {J.}~\bibnamefont
  {Ranft}}, \bibinfo {author} {\bibfnamefont {P.}~\bibnamefont {Silberzan}},\
  and\ \bibinfo {author} {\bibfnamefont {P.}~\bibnamefont {Marcq}},\ }\href
  {https://doi.org/10.1016/j.bpj.2013.11.015} {\bibfield  {journal} {\bibinfo
  {journal} {Biophysical Journal}\ }\textbf {\bibinfo {volume} {106}},\
  \bibinfo {pages} {65–73} (\bibinfo {year} {2014})}\BibitemShut {NoStop}%
\bibitem [{\citenamefont {Recho}\ \emph {et~al.}(2016)\citenamefont {Recho},
  \citenamefont {Ranft},\ and\ \citenamefont {Marcq}}]{Recho2016}%
  \BibitemOpen
  \bibfield  {author} {\bibinfo {author} {\bibfnamefont {P.}~\bibnamefont
  {Recho}}, \bibinfo {author} {\bibfnamefont {J.}~\bibnamefont {Ranft}},\ and\
  \bibinfo {author} {\bibfnamefont {P.}~\bibnamefont {Marcq}},\ }\href
  {https://doi.org/10.1039/c5sm02857d} {\bibfield  {journal} {\bibinfo
  {journal} {Soft Matter}\ }\textbf {\bibinfo {volume} {12}},\ \bibinfo {pages}
  {2381–2391} (\bibinfo {year} {2016})}\BibitemShut {NoStop}%
\bibitem [{\citenamefont {Ranft}\ \emph {et~al.}(2010)\citenamefont {Ranft},
  \citenamefont {Basan}, \citenamefont {Elgeti}, \citenamefont {Joanny},
  \citenamefont {Prost},\ and\ \citenamefont {Jülicher}}]{Ranft2010}%
  \BibitemOpen
  \bibfield  {author} {\bibinfo {author} {\bibfnamefont {J.}~\bibnamefont
  {Ranft}}, \bibinfo {author} {\bibfnamefont {M.}~\bibnamefont {Basan}},
  \bibinfo {author} {\bibfnamefont {J.}~\bibnamefont {Elgeti}}, \bibinfo
  {author} {\bibfnamefont {J.-F.}\ \bibnamefont {Joanny}}, \bibinfo {author}
  {\bibfnamefont {J.}~\bibnamefont {Prost}},\ and\ \bibinfo {author}
  {\bibfnamefont {F.}~\bibnamefont {Jülicher}},\ }\href
  {https://doi.org/10.1073/pnas.1011086107} {\bibfield  {journal} {\bibinfo
  {journal} {Proceedings of the National Academy of Sciences}\ }\textbf
  {\bibinfo {volume} {107}},\ \bibinfo {pages} {20863–20868} (\bibinfo {year}
  {2010})}\BibitemShut {NoStop}%
\bibitem [{\citenamefont {Marchetti}\ \emph
  {et~al.}(2013{\natexlab{b}})\citenamefont {Marchetti}, \citenamefont
  {Joanny}, \citenamefont {Ramaswamy}, \citenamefont {Liverpool}, \citenamefont
  {Prost}, \citenamefont {Rao},\ and\ \citenamefont
  {Simha}}]{marchetti2013hydrodynamics}%
  \BibitemOpen
  \bibfield  {author} {\bibinfo {author} {\bibfnamefont {M.~C.}\ \bibnamefont
  {Marchetti}}, \bibinfo {author} {\bibfnamefont {J.-F.}\ \bibnamefont
  {Joanny}}, \bibinfo {author} {\bibfnamefont {S.}~\bibnamefont {Ramaswamy}},
  \bibinfo {author} {\bibfnamefont {T.~B.}\ \bibnamefont {Liverpool}}, \bibinfo
  {author} {\bibfnamefont {J.}~\bibnamefont {Prost}}, \bibinfo {author}
  {\bibfnamefont {M.}~\bibnamefont {Rao}},\ and\ \bibinfo {author}
  {\bibfnamefont {R.~A.}\ \bibnamefont {Simha}},\ }\href
  {https://doi.org/10.1103/RevModPhys.85.1143} {\bibfield  {journal} {\bibinfo
  {journal} {Reviews of Modern Physics}\ }\textbf {\bibinfo {volume} {85}},\
  \bibinfo {pages} {1143} (\bibinfo {year} {2013}{\natexlab{b}})}\BibitemShut
  {NoStop}%
\bibitem [{\citenamefont {JULICHER}\ \emph {et~al.}(2007)\citenamefont
  {JULICHER}, \citenamefont {KRUSE}, \citenamefont {PROST},\ and\ \citenamefont
  {JOANNY}}]{Julicher2007}%
  \BibitemOpen
  \bibfield  {author} {\bibinfo {author} {\bibfnamefont {F.}~\bibnamefont
  {JULICHER}}, \bibinfo {author} {\bibfnamefont {K.}~\bibnamefont {KRUSE}},
  \bibinfo {author} {\bibfnamefont {J.}~\bibnamefont {PROST}},\ and\ \bibinfo
  {author} {\bibfnamefont {J.}~\bibnamefont {JOANNY}},\ }\href
  {https://doi.org/10.1016/j.physrep.2007.02.018} {\bibfield  {journal}
  {\bibinfo  {journal} {Physics Reports}\ }\textbf {\bibinfo {volume} {449}},\
  \bibinfo {pages} {3–28} (\bibinfo {year} {2007})}\BibitemShut {NoStop}%
\bibitem [{\citenamefont {Prost}\ \emph {et~al.}(2015)\citenamefont {Prost},
  \citenamefont {Jülicher},\ and\ \citenamefont {Joanny}}]{Prost2015}%
  \BibitemOpen
  \bibfield  {author} {\bibinfo {author} {\bibfnamefont {J.}~\bibnamefont
  {Prost}}, \bibinfo {author} {\bibfnamefont {F.}~\bibnamefont {Jülicher}},\
  and\ \bibinfo {author} {\bibfnamefont {J.-F.}\ \bibnamefont {Joanny}},\
  }\href {https://doi.org/10.1038/nphys3224} {\bibfield  {journal} {\bibinfo
  {journal} {Nature Physics}\ }\textbf {\bibinfo {volume} {11}},\ \bibinfo
  {pages} {111–117} (\bibinfo {year} {2015})}\BibitemShut {NoStop}%
\bibitem [{\citenamefont {Avron}(1998)}]{avron1998odd}%
  \BibitemOpen
  \bibfield  {author} {\bibinfo {author} {\bibfnamefont {J.~E.}\ \bibnamefont
  {Avron}},\ }\href {https://doi.org/10.1023/a:1023084404080} {\bibfield
  {journal} {\bibinfo  {journal} {Journal of Statistical Physics}\ }\textbf
  {\bibinfo {volume} {92}},\ \bibinfo {pages} {543–557} (\bibinfo {year}
  {1998})}\BibitemShut {NoStop}%
\bibitem [{\citenamefont {Banerjee}\ \emph
  {et~al.}(2021{\natexlab{a}})\citenamefont {Banerjee}, \citenamefont
  {Vitelli}, \citenamefont {J{\"u}licher},\ and\ \citenamefont
  {Sur{\'o}wka}}]{banerjee2021active}%
  \BibitemOpen
  \bibfield  {author} {\bibinfo {author} {\bibfnamefont {D.}~\bibnamefont
  {Banerjee}}, \bibinfo {author} {\bibfnamefont {V.}~\bibnamefont {Vitelli}},
  \bibinfo {author} {\bibfnamefont {F.}~\bibnamefont {J{\"u}licher}},\ and\
  \bibinfo {author} {\bibfnamefont {P.}~\bibnamefont {Sur{\'o}wka}},\ }\href
  {https://doi.org/10.1103/PhysRevLett.126.138001} {\bibfield  {journal}
  {\bibinfo  {journal} {Physical Review Letters}\ }\textbf {\bibinfo {volume}
  {126}},\ \bibinfo {pages} {138001} (\bibinfo {year}
  {2021}{\natexlab{a}})}\BibitemShut {NoStop}%
\bibitem [{\citenamefont {Banerjee}\ \emph {et~al.}(2017)\citenamefont
  {Banerjee}, \citenamefont {Souslov}, \citenamefont {Abanov},\ and\
  \citenamefont {Vitelli}}]{banerjee2017odd}%
  \BibitemOpen
  \bibfield  {author} {\bibinfo {author} {\bibfnamefont {D.}~\bibnamefont
  {Banerjee}}, \bibinfo {author} {\bibfnamefont {A.}~\bibnamefont {Souslov}},
  \bibinfo {author} {\bibfnamefont {A.~G.}\ \bibnamefont {Abanov}},\ and\
  \bibinfo {author} {\bibfnamefont {V.}~\bibnamefont {Vitelli}},\ }\href
  {https://doi.org/10.1038/s41467-017-01378-7} {\bibfield  {journal} {\bibinfo
  {journal} {Nature Communications}\ }\textbf {\bibinfo {volume} {8}},\
  \bibinfo {pages} {1573} (\bibinfo {year} {2017})}\BibitemShut {NoStop}%
\bibitem [{\citenamefont {Chen}\ \emph {et~al.}(2021)\citenamefont {Chen},
  \citenamefont {Li}, \citenamefont {Scheibner}, \citenamefont {Vitelli},\ and\
  \citenamefont {Huang}}]{chen2021realization}%
  \BibitemOpen
  \bibfield  {author} {\bibinfo {author} {\bibfnamefont {Y.}~\bibnamefont
  {Chen}}, \bibinfo {author} {\bibfnamefont {X.}~\bibnamefont {Li}}, \bibinfo
  {author} {\bibfnamefont {C.}~\bibnamefont {Scheibner}}, \bibinfo {author}
  {\bibfnamefont {V.}~\bibnamefont {Vitelli}},\ and\ \bibinfo {author}
  {\bibfnamefont {G.}~\bibnamefont {Huang}},\ }\href
  {https://doi.org/10.1038/s41467-021-26034-z} {\bibfield  {journal} {\bibinfo
  {journal} {Nature Communications}\ }\textbf {\bibinfo {volume} {12}},\
  \bibinfo {pages} {5935} (\bibinfo {year} {2021})}\BibitemShut {NoStop}%
\bibitem [{\citenamefont {Banerjee}\ \emph
  {et~al.}(2021{\natexlab{b}})\citenamefont {Banerjee}, \citenamefont
  {Vitelli}, \citenamefont {Jülicher},\ and\ \citenamefont
  {Surówka}}]{Banerjee2021err}%
  \BibitemOpen
  \bibfield  {author} {\bibinfo {author} {\bibfnamefont {D.}~\bibnamefont
  {Banerjee}}, \bibinfo {author} {\bibfnamefont {V.}~\bibnamefont {Vitelli}},
  \bibinfo {author} {\bibfnamefont {F.}~\bibnamefont {Jülicher}},\ and\
  \bibinfo {author} {\bibfnamefont {P.}~\bibnamefont {Surówka}},\ }\href
  {https://doi.org/10.1103/PhysRevLett.127.189901} {\bibfield  {journal}
  {\bibinfo  {journal} {Physical Review Letters}\ }\textbf {\bibinfo {volume}
  {127}},\ \bibinfo {pages} {189901} (\bibinfo {year}
  {2021}{\natexlab{b}})}\BibitemShut {NoStop}%
\bibitem [{\citenamefont {Floyd}\ \emph {et~al.}(2024)\citenamefont {Floyd},
  \citenamefont {Dinner},\ and\ \citenamefont {Vaikuntanathan}}]{Floyd2024}%
  \BibitemOpen
  \bibfield  {author} {\bibinfo {author} {\bibfnamefont {C.}~\bibnamefont
  {Floyd}}, \bibinfo {author} {\bibfnamefont {A.~R.}\ \bibnamefont {Dinner}},\
  and\ \bibinfo {author} {\bibfnamefont {S.}~\bibnamefont {Vaikuntanathan}},\
  }\href {https://doi.org/10.1103/physrevresearch.6.033100} {\bibfield
  {journal} {\bibinfo  {journal} {Physical Review Research}\ }\textbf {\bibinfo
  {volume} {6}},\ \bibinfo {pages} {033100} (\bibinfo {year}
  {2024})}\BibitemShut {NoStop}%
\bibitem [{\citenamefont {Beris}\ and\ \citenamefont
  {Edwards}(1994)}]{Beris1994}%
  \BibitemOpen
  \bibfield  {author} {\bibinfo {author} {\bibfnamefont {A.}~\bibnamefont
  {Beris}}\ and\ \bibinfo {author} {\bibfnamefont {B.}~\bibnamefont
  {Edwards}},\ }\href@noop {} {\emph {\bibinfo {title} {Thermodynamics of
  Flowing Systems: with Internal Microstructure}}},\ Oxford Engineering Science
  Series\ (\bibinfo  {publisher} {Oxford University Press},\ \bibinfo {year}
  {1994})\BibitemShut {NoStop}%
\bibitem [{\citenamefont {Bodnár}\ and\ \citenamefont
  {Sequeira}(2022)}]{Bodnar2022}%
  \BibitemOpen
  \bibfield  {author} {\bibinfo {author} {\bibfnamefont {T.}~\bibnamefont
  {Bodnár}}\ and\ \bibinfo {author} {\bibfnamefont {A.}~\bibnamefont
  {Sequeira}},\ }\href {https://doi.org/10.1088/1742-6596/2367/1/012028}
  {\bibfield  {journal} {\bibinfo  {journal} {Journal of Physics: Conference
  Series}\ }\textbf {\bibinfo {volume} {2367}},\ \bibinfo {pages} {012028}
  (\bibinfo {year} {2022})}\BibitemShut {NoStop}%
\bibitem [{\citenamefont {Gordon}\ and\ \citenamefont
  {Schowalter}(1972)}]{Gordon1972}%
  \BibitemOpen
  \bibfield  {author} {\bibinfo {author} {\bibfnamefont {R.~J.}\ \bibnamefont
  {Gordon}}\ and\ \bibinfo {author} {\bibfnamefont {W.~R.}\ \bibnamefont
  {Schowalter}},\ }\href {https://doi.org/10.1122/1.549256} {\bibfield
  {journal} {\bibinfo  {journal} {Transactions of the Society of Rheology}\
  }\textbf {\bibinfo {volume} {16}},\ \bibinfo {pages} {79–97} (\bibinfo
  {year} {1972})}\BibitemShut {NoStop}%
\bibitem [{\citenamefont {Johnson}\ and\ \citenamefont
  {Segalman}(1977)}]{Johnson1977}%
  \BibitemOpen
  \bibfield  {author} {\bibinfo {author} {\bibfnamefont {M.}~\bibnamefont
  {Johnson}}\ and\ \bibinfo {author} {\bibfnamefont {D.}~\bibnamefont
  {Segalman}},\ }\href {https://doi.org/10.1016/0377-0257(77)80003-7}
  {\bibfield  {journal} {\bibinfo  {journal} {Journal of Non-Newtonian Fluid
  Mechanics}\ }\textbf {\bibinfo {volume} {2}},\ \bibinfo {pages} {255–270}
  (\bibinfo {year} {1977})}\BibitemShut {NoStop}%
\bibitem [{\citenamefont {Hinch}\ and\ \citenamefont
  {Harlen}(2021)}]{Hinch2021}%
  \BibitemOpen
  \bibfield  {author} {\bibinfo {author} {\bibfnamefont {J.}~\bibnamefont
  {Hinch}}\ and\ \bibinfo {author} {\bibfnamefont {O.}~\bibnamefont {Harlen}},\
  }\href {https://doi.org/10.1016/j.jnnfm.2021.104668} {\bibfield  {journal}
  {\bibinfo  {journal} {Journal of Non-Newtonian Fluid Mechanics}\ }\textbf
  {\bibinfo {volume} {298}},\ \bibinfo {pages} {104668} (\bibinfo {year}
  {2021})}\BibitemShut {NoStop}%
\bibitem [{\citenamefont {Castillo~Sánchez}\ \emph {et~al.}(2022)\citenamefont
  {Castillo~Sánchez}, \citenamefont {Jovanović}, \citenamefont {Kumar},
  \citenamefont {Morozov}, \citenamefont {Shankar}, \citenamefont
  {Subramanian},\ and\ \citenamefont {Wilson}}]{Castillo2022}%
  \BibitemOpen
  \bibfield  {author} {\bibinfo {author} {\bibfnamefont {H.~A.}\ \bibnamefont
  {Castillo~Sánchez}}, \bibinfo {author} {\bibfnamefont {M.~R.}\ \bibnamefont
  {Jovanović}}, \bibinfo {author} {\bibfnamefont {S.}~\bibnamefont {Kumar}},
  \bibinfo {author} {\bibfnamefont {A.}~\bibnamefont {Morozov}}, \bibinfo
  {author} {\bibfnamefont {V.}~\bibnamefont {Shankar}}, \bibinfo {author}
  {\bibfnamefont {G.}~\bibnamefont {Subramanian}},\ and\ \bibinfo {author}
  {\bibfnamefont {H.~J.}\ \bibnamefont {Wilson}},\ }\href
  {https://doi.org/10.1016/j.jnnfm.2022.104742} {\bibfield  {journal} {\bibinfo
   {journal} {Journal of Non-Newtonian Fluid Mechanics}\ }\textbf {\bibinfo
  {volume} {302}},\ \bibinfo {pages} {104742} (\bibinfo {year}
  {2022})}\BibitemShut {NoStop}%
\bibitem [{\citenamefont {Eggers}\ \emph {et~al.}(2023)\citenamefont {Eggers},
  \citenamefont {Liverpool},\ and\ \citenamefont {Mietke}}]{Eggers2023}%
  \BibitemOpen
  \bibfield  {author} {\bibinfo {author} {\bibfnamefont {J.}~\bibnamefont
  {Eggers}}, \bibinfo {author} {\bibfnamefont {T.~B.}\ \bibnamefont
  {Liverpool}},\ and\ \bibinfo {author} {\bibfnamefont {A.}~\bibnamefont
  {Mietke}},\ }\href {https://doi.org/10.1103/physrevlett.131.194002}
  {\bibfield  {journal} {\bibinfo  {journal} {Physical Review Letters}\
  }\textbf {\bibinfo {volume} {131}},\ \bibinfo {pages} {194002} (\bibinfo
  {year} {2023})}\BibitemShut {NoStop}%
\bibitem [{\citenamefont {Stone}\ \emph {et~al.}(2023)\citenamefont {Stone},
  \citenamefont {Shelley},\ and\ \citenamefont {Boyko}}]{Stone2023}%
  \BibitemOpen
  \bibfield  {author} {\bibinfo {author} {\bibfnamefont {H.~A.}\ \bibnamefont
  {Stone}}, \bibinfo {author} {\bibfnamefont {M.~J.}\ \bibnamefont {Shelley}},\
  and\ \bibinfo {author} {\bibfnamefont {E.}~\bibnamefont {Boyko}},\ }\href
  {https://doi.org/10.1039/d3sm00497j} {\bibfield  {journal} {\bibinfo
  {journal} {Soft Matter}\ }\textbf {\bibinfo {volume} {19}},\ \bibinfo {pages}
  {5353–5359} (\bibinfo {year} {2023})}\BibitemShut {NoStop}%
\bibitem [{\citenamefont {Tang}\ \emph {et~al.}(2024)\citenamefont {Tang},
  \citenamefont {Chen}, \citenamefont {Bowick},\ and\ \citenamefont
  {Bi}}]{tang2024cell}%
  \BibitemOpen
  \bibfield  {author} {\bibinfo {author} {\bibfnamefont {Y.}~\bibnamefont
  {Tang}}, \bibinfo {author} {\bibfnamefont {S.}~\bibnamefont {Chen}}, \bibinfo
  {author} {\bibfnamefont {M.~J.}\ \bibnamefont {Bowick}},\ and\ \bibinfo
  {author} {\bibfnamefont {D.}~\bibnamefont {Bi}},\ }\href
  {https://doi.org/10.1103/PhysRevLett.132.218402} {\bibfield  {journal}
  {\bibinfo  {journal} {Physical Review Letters}\ }\textbf {\bibinfo {volume}
  {132}},\ \bibinfo {pages} {218402} (\bibinfo {year} {2024})}\BibitemShut
  {NoStop}%
\bibitem [{Note1()}]{Note1}%
  \BibitemOpen
  \bibinfo {note} {For Voronoi model with $\alpha ^o=0$ it has been argued that
  $\mu $ can have a small but finite value in the liquid phase if $K_A>0$~\cite
  {sussman2018no}.}\BibitemShut {Stop}%
\bibitem [{\citenamefont {Hertaeg}\ \emph {et~al.}(2024)\citenamefont
  {Hertaeg}, \citenamefont {Fielding},\ and\ \citenamefont
  {Bi}}]{hertaeg2024discontinuous}%
  \BibitemOpen
  \bibfield  {author} {\bibinfo {author} {\bibfnamefont {M.~J.}\ \bibnamefont
  {Hertaeg}}, \bibinfo {author} {\bibfnamefont {S.~M.}\ \bibnamefont
  {Fielding}},\ and\ \bibinfo {author} {\bibfnamefont {D.}~\bibnamefont {Bi}},\
  }\href {https://doi.org/10.1103/PhysRevX.14.011027} {\bibfield  {journal}
  {\bibinfo  {journal} {Physical Review X}\ }\textbf {\bibinfo {volume} {14}},\
  \bibinfo {pages} {011027} (\bibinfo {year} {2024})}\BibitemShut {NoStop}%
\bibitem [{\citenamefont {Hargus}\ \emph {et~al.}(2021)\citenamefont {Hargus},
  \citenamefont {Epstein},\ and\ \citenamefont {Mandadapu}}]{hargus2021odd}%
  \BibitemOpen
  \bibfield  {author} {\bibinfo {author} {\bibfnamefont {C.}~\bibnamefont
  {Hargus}}, \bibinfo {author} {\bibfnamefont {J.~M.}\ \bibnamefont
  {Epstein}},\ and\ \bibinfo {author} {\bibfnamefont {K.~K.}\ \bibnamefont
  {Mandadapu}},\ }\href {https://doi.org/10.1103/PhysRevLett.127.178001}
  {\bibfield  {journal} {\bibinfo  {journal} {Physical Review Letters}\
  }\textbf {\bibinfo {volume} {127}},\ \bibinfo {pages} {178001} (\bibinfo
  {year} {2021})}\BibitemShut {NoStop}%
\bibitem [{\citenamefont {Fujita}\ \emph {et~al.}(2020)\citenamefont {Fujita},
  \citenamefont {Tasaka}, \citenamefont {Yanagisawa}, \citenamefont {Noto},\
  and\ \citenamefont {Murai}}]{Fujita2020}%
  \BibitemOpen
  \bibfield  {author} {\bibinfo {author} {\bibfnamefont {K.}~\bibnamefont
  {Fujita}}, \bibinfo {author} {\bibfnamefont {Y.}~\bibnamefont {Tasaka}},
  \bibinfo {author} {\bibfnamefont {T.}~\bibnamefont {Yanagisawa}}, \bibinfo
  {author} {\bibfnamefont {D.}~\bibnamefont {Noto}},\ and\ \bibinfo {author}
  {\bibfnamefont {Y.}~\bibnamefont {Murai}},\ }\href
  {https://doi.org/10.1007/s12650-020-00651-0} {\bibfield  {journal} {\bibinfo
  {journal} {Journal of Visualization}\ }\textbf {\bibinfo {volume} {23}},\
  \bibinfo {pages} {635–647} (\bibinfo {year} {2020})}\BibitemShut {NoStop}%
\bibitem [{\citenamefont {Pismen}(2010)}]{Pismen2010}%
  \BibitemOpen
  \bibfield  {author} {\bibinfo {author} {\bibfnamefont {L.}~\bibnamefont
  {Pismen}},\ }\href@noop {} {\emph {\bibinfo {title} {Patterns and Interfaces
  in Dissipative Dynamics}}},\ Springer Series in Synergetics\ (\bibinfo
  {publisher} {Springer Berlin Heidelberg},\ \bibinfo {year}
  {2010})\BibitemShut {NoStop}%
\bibitem [{\citenamefont {Cross}\ \emph {et~al.}(1994)\citenamefont {Cross},
  \citenamefont {Meiron},\ and\ \citenamefont {Tu}}]{Cross1994}%
  \BibitemOpen
  \bibfield  {author} {\bibinfo {author} {\bibfnamefont {M.~C.}\ \bibnamefont
  {Cross}}, \bibinfo {author} {\bibfnamefont {D.}~\bibnamefont {Meiron}},\ and\
  \bibinfo {author} {\bibfnamefont {Y.}~\bibnamefont {Tu}},\ }\href
  {https://doi.org/10.1063/1.166038} {\bibfield  {journal} {\bibinfo  {journal}
  {Chaos: An Interdisciplinary Journal of Nonlinear Science}\ }\textbf
  {\bibinfo {volume} {4}},\ \bibinfo {pages} {607–619} (\bibinfo {year}
  {1994})}\BibitemShut {NoStop}%
\bibitem [{\citenamefont {Cross}\ and\ \citenamefont
  {Greenside}(2009)}]{cross2009pattern}%
  \BibitemOpen
  \bibfield  {author} {\bibinfo {author} {\bibfnamefont {M.}~\bibnamefont
  {Cross}}\ and\ \bibinfo {author} {\bibfnamefont {H.}~\bibnamefont
  {Greenside}},\ }\href@noop {} {\emph {\bibinfo {title} {Pattern Formation and
  Dynamics in Nonequilibrium Systems}}}\ (\bibinfo  {publisher} {Cambridge
  University Press},\ \bibinfo {year} {2009})\BibitemShut {NoStop}%
\bibitem [{\citenamefont {Busse}\ and\ \citenamefont
  {Heikes}(1980)}]{Busse1980}%
  \BibitemOpen
  \bibfield  {author} {\bibinfo {author} {\bibfnamefont {F.~H.}\ \bibnamefont
  {Busse}}\ and\ \bibinfo {author} {\bibfnamefont {K.~E.}\ \bibnamefont
  {Heikes}},\ }\href {https://doi.org/10.1126/science.208.4440.173} {\bibfield
  {journal} {\bibinfo  {journal} {Science}\ }\textbf {\bibinfo {volume}
  {208}},\ \bibinfo {pages} {173–175} (\bibinfo {year} {1980})}\BibitemShut
  {NoStop}%
\bibitem [{\citenamefont {Bodenschatz}\ \emph {et~al.}(2000)\citenamefont
  {Bodenschatz}, \citenamefont {Pesch},\ and\ \citenamefont
  {Ahlers}}]{Bodenschatz2000recent}%
  \BibitemOpen
  \bibfield  {author} {\bibinfo {author} {\bibfnamefont {E.}~\bibnamefont
  {Bodenschatz}}, \bibinfo {author} {\bibfnamefont {W.}~\bibnamefont {Pesch}},\
  and\ \bibinfo {author} {\bibfnamefont {G.}~\bibnamefont {Ahlers}},\ }\href
  {https://doi.org/10.1146/annurev.fluid.32.1.709} {\bibfield  {journal}
  {\bibinfo  {journal} {Annual Review of Fluid Mechanics}\ }\textbf {\bibinfo
  {volume} {32}},\ \bibinfo {pages} {709} (\bibinfo {year} {2000})}\BibitemShut
  {NoStop}%
\bibitem [{\citenamefont {Ahlers}(2006)}]{Ahlers2006}%
  \BibitemOpen
  \bibfield  {author} {\bibinfo {author} {\bibfnamefont {G.}~\bibnamefont
  {Ahlers}},\ }\href {https://doi.org/10.1007/978-0-387-25111-0_4} {\bibfield
  {journal} {\bibinfo  {journal} {Springer Tracts in Modern Physics}\ ,\
  \bibinfo {pages} {67–94}} (\bibinfo {year} {2006})}\BibitemShut {NoStop}%
\bibitem [{\citenamefont {Guarino}\ and\ \citenamefont
  {Vidal}(2004)}]{Guarino2004}%
  \BibitemOpen
  \bibfield  {author} {\bibinfo {author} {\bibfnamefont {A.}~\bibnamefont
  {Guarino}}\ and\ \bibinfo {author} {\bibfnamefont {V.}~\bibnamefont
  {Vidal}},\ }\href {https://doi.org/10.1103/physreve.69.066311} {\bibfield
  {journal} {\bibinfo  {journal} {Physical Review E}\ }\textbf {\bibinfo
  {volume} {69}},\ \bibinfo {pages} {066311} (\bibinfo {year}
  {2004})}\BibitemShut {NoStop}%
\bibitem [{\citenamefont {Ecke}\ and\ \citenamefont
  {Shishkina}(2023)}]{Ecke2023}%
  \BibitemOpen
  \bibfield  {author} {\bibinfo {author} {\bibfnamefont {R.~E.}\ \bibnamefont
  {Ecke}}\ and\ \bibinfo {author} {\bibfnamefont {O.}~\bibnamefont
  {Shishkina}},\ }\href {https://doi.org/10.1146/annurev-fluid-120720-020446}
  {\bibfield  {journal} {\bibinfo  {journal} {Annual Review of Fluid
  Mechanics}\ }\textbf {\bibinfo {volume} {55}},\ \bibinfo {pages} {603–638}
  (\bibinfo {year} {2023})}\BibitemShut {NoStop}%
\bibitem [{\citenamefont {Echebarria}\ and\ \citenamefont
  {Pérez-García}(1998)}]{Echebarria1998}%
  \BibitemOpen
  \bibfield  {author} {\bibinfo {author} {\bibfnamefont {B.}~\bibnamefont
  {Echebarria}}\ and\ \bibinfo {author} {\bibfnamefont {C.}~\bibnamefont
  {Pérez-García}},\ }\href {https://doi.org/10.1209/epl/i1998-00315-2}
  {\bibfield  {journal} {\bibinfo  {journal} {Europhysics Letters (EPL)}\
  }\textbf {\bibinfo {volume} {43}},\ \bibinfo {pages} {35–40} (\bibinfo
  {year} {1998})}\BibitemShut {NoStop}%
\bibitem [{\citenamefont {Echebarria}\ and\ \citenamefont
  {Riecke}(2000{\natexlab{a}})}]{Echebarria2000}%
  \BibitemOpen
  \bibfield  {author} {\bibinfo {author} {\bibfnamefont {B.}~\bibnamefont
  {Echebarria}}\ and\ \bibinfo {author} {\bibfnamefont {H.}~\bibnamefont
  {Riecke}},\ }\href {https://doi.org/10.1016/s0167-2789(99)00212-2} {\bibfield
   {journal} {\bibinfo  {journal} {Physica D: Nonlinear Phenomena}\ }\textbf
  {\bibinfo {volume} {139}},\ \bibinfo {pages} {97–108} (\bibinfo {year}
  {2000}{\natexlab{a}})}\BibitemShut {NoStop}%
\bibitem [{\citenamefont {Echebarria}\ and\ \citenamefont
  {Riecke}(2000{\natexlab{b}})}]{Echebarria2000b}%
  \BibitemOpen
  \bibfield  {author} {\bibinfo {author} {\bibfnamefont {B.}~\bibnamefont
  {Echebarria}}\ and\ \bibinfo {author} {\bibfnamefont {H.}~\bibnamefont
  {Riecke}},\ }\href {https://doi.org/10.1016/s0167-2789(00)00101-9} {\bibfield
   {journal} {\bibinfo  {journal} {Physica D: Nonlinear Phenomena}\ }\textbf
  {\bibinfo {volume} {143}},\ \bibinfo {pages} {187–204} (\bibinfo {year}
  {2000}{\natexlab{b}})}\BibitemShut {NoStop}%
\bibitem [{\citenamefont {Rossby}(1969)}]{Rossby1969}%
  \BibitemOpen
  \bibfield  {author} {\bibinfo {author} {\bibfnamefont {H.~T.}\ \bibnamefont
  {Rossby}},\ }\href {https://doi.org/10.1017/s0022112069001674} {\bibfield
  {journal} {\bibinfo  {journal} {Journal of Fluid Mechanics}\ }\textbf
  {\bibinfo {volume} {36}},\ \bibinfo {pages} {309–335} (\bibinfo {year}
  {1969})}\BibitemShut {NoStop}%
\bibitem [{\citenamefont {Ecke}\ \emph {et~al.}(1992)\citenamefont {Ecke},
  \citenamefont {Zhong},\ and\ \citenamefont {Knobloch}}]{Ecke1992}%
  \BibitemOpen
  \bibfield  {author} {\bibinfo {author} {\bibfnamefont {R.~E.}\ \bibnamefont
  {Ecke}}, \bibinfo {author} {\bibfnamefont {F.}~\bibnamefont {Zhong}},\ and\
  \bibinfo {author} {\bibfnamefont {E.}~\bibnamefont {Knobloch}},\ }\href
  {https://doi.org/10.1209/0295-5075/19/3/005} {\bibfield  {journal} {\bibinfo
  {journal} {Europhysics Letters (EPL)}\ }\textbf {\bibinfo {volume} {19}},\
  \bibinfo {pages} {177–182} (\bibinfo {year} {1992})}\BibitemShut {NoStop}%
\bibitem [{\citenamefont {Kuo}\ and\ \citenamefont {Cross}(1993)}]{Kuo1993}%
  \BibitemOpen
  \bibfield  {author} {\bibinfo {author} {\bibfnamefont {E.~Y.}\ \bibnamefont
  {Kuo}}\ and\ \bibinfo {author} {\bibfnamefont {M.~C.}\ \bibnamefont
  {Cross}},\ }\href {https://doi.org/10.1103/physreve.47.r2245} {\bibfield
  {journal} {\bibinfo  {journal} {Physical Review E}\ }\textbf {\bibinfo
  {volume} {47}},\ \bibinfo {pages} {R2245–R2248} (\bibinfo {year}
  {1993})}\BibitemShut {NoStop}%
\bibitem [{\citenamefont {Favier}\ and\ \citenamefont
  {Knobloch}(2020)}]{Favier2020}%
  \BibitemOpen
  \bibfield  {author} {\bibinfo {author} {\bibfnamefont {B.}~\bibnamefont
  {Favier}}\ and\ \bibinfo {author} {\bibfnamefont {E.}~\bibnamefont
  {Knobloch}},\ }\href {https://doi.org/10.1017/jfm.2020.310} {\bibfield
  {journal} {\bibinfo  {journal} {Journal of Fluid Mechanics}\ }\textbf
  {\bibinfo {volume} {895}},\ \bibinfo {pages} {R1} (\bibinfo {year}
  {2020})}\BibitemShut {NoStop}%
\bibitem [{\citenamefont {Ecke}\ \emph {et~al.}(2022)\citenamefont {Ecke},
  \citenamefont {Zhang},\ and\ \citenamefont {Shishkina}}]{Ecke2022}%
  \BibitemOpen
  \bibfield  {author} {\bibinfo {author} {\bibfnamefont {R.~E.}\ \bibnamefont
  {Ecke}}, \bibinfo {author} {\bibfnamefont {X.}~\bibnamefont {Zhang}},\ and\
  \bibinfo {author} {\bibfnamefont {O.}~\bibnamefont {Shishkina}},\ }\href
  {https://doi.org/10.1103/physrevfluids.7.l011501} {\bibfield  {journal}
  {\bibinfo  {journal} {Physical Review Fluids}\ }\textbf {\bibinfo {volume}
  {7}},\ \bibinfo {pages} {l011501} (\bibinfo {year} {2022})}\BibitemShut
  {NoStop}%
\bibitem [{\citenamefont {de~Wit}\ \emph {et~al.}(2023)\citenamefont {de~Wit},
  \citenamefont {Boot}, \citenamefont {Madonia}, \citenamefont
  {Aguirre~Guzmán},\ and\ \citenamefont {Kunnen}}]{deWit2023}%
  \BibitemOpen
  \bibfield  {author} {\bibinfo {author} {\bibfnamefont {X.~M.}\ \bibnamefont
  {de~Wit}}, \bibinfo {author} {\bibfnamefont {W.~J.~M.}\ \bibnamefont {Boot}},
  \bibinfo {author} {\bibfnamefont {M.}~\bibnamefont {Madonia}}, \bibinfo
  {author} {\bibfnamefont {A.~J.}\ \bibnamefont {Aguirre~Guzmán}},\ and\
  \bibinfo {author} {\bibfnamefont {R.~P.~J.}\ \bibnamefont {Kunnen}},\ }\href
  {https://doi.org/10.1103/physrevfluids.8.073501} {\bibfield  {journal}
  {\bibinfo  {journal} {Physical Review Fluids}\ }\textbf {\bibinfo {volume}
  {8}},\ \bibinfo {pages} {073501} (\bibinfo {year} {2023})}\BibitemShut
  {NoStop}%
\bibitem [{\citenamefont {Wedi}\ \emph {et~al.}(2022)\citenamefont {Wedi},
  \citenamefont {Moturi}, \citenamefont {Funfschilling},\ and\ \citenamefont
  {Weiss}}]{Wedi2022}%
  \BibitemOpen
  \bibfield  {author} {\bibinfo {author} {\bibfnamefont {M.}~\bibnamefont
  {Wedi}}, \bibinfo {author} {\bibfnamefont {V.~M.}\ \bibnamefont {Moturi}},
  \bibinfo {author} {\bibfnamefont {D.}~\bibnamefont {Funfschilling}},\ and\
  \bibinfo {author} {\bibfnamefont {S.}~\bibnamefont {Weiss}},\ }\href
  {https://doi.org/10.1017/jfm.2022.195} {\bibfield  {journal} {\bibinfo
  {journal} {Journal of Fluid Mechanics}\ }\textbf {\bibinfo {volume} {939}},\
  \bibinfo {pages} {A14} (\bibinfo {year} {2022})}\BibitemShut {NoStop}%
\bibitem [{\citenamefont {Vasil}\ \emph {et~al.}(2024)\citenamefont {Vasil},
  \citenamefont {Burns}, \citenamefont {Lecoanet}, \citenamefont {Oishi},
  \citenamefont {Brown},\ and\ \citenamefont {Julien}}]{Vasil2024}%
  \BibitemOpen
  \bibfield  {author} {\bibinfo {author} {\bibfnamefont {G.~M.}\ \bibnamefont
  {Vasil}}, \bibinfo {author} {\bibfnamefont {K.~J.}\ \bibnamefont {Burns}},
  \bibinfo {author} {\bibfnamefont {D.}~\bibnamefont {Lecoanet}}, \bibinfo
  {author} {\bibfnamefont {J.~S.}\ \bibnamefont {Oishi}}, \bibinfo {author}
  {\bibfnamefont {B.~P.}\ \bibnamefont {Brown}},\ and\ \bibinfo {author}
  {\bibfnamefont {K.}~\bibnamefont {Julien}},\ }\href
  {https://doi.org/10.48550/arXiv.2409.20541} {\bibinfo {title} {Rapidly
  rotating wall-mode convection}} (\bibinfo {year} {2024}),\ \Eprint
  {https://arxiv.org/abs/2409.20541} {arXiv:2409.20541} \BibitemShut {NoStop}%
\end{thebibliography}%

\end{document}


\title{Supplementary Information: Chirality across scales in tissue dynamics}
\maketitle
\renewcommand{\thefigure}{S\arabic{figure}}
\section{Mean field theory}
\subsection{Dynamical matrix}

\begin{figure}[t]
\centering
\hspace{-2em}\includegraphics[width = 0.5\columnwidth]{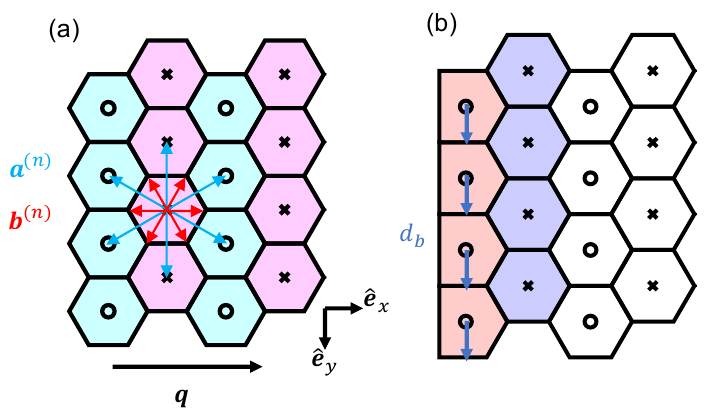}
\caption{{\bf  Mean-field model (MFT).} (a) Hexagonal structure of the MFT. (b) Illustration of the boundary cells. }
    \label{fig.m0}
\end{figure}

Here we detail the derivation of the mean-field theory (MFT). In the MFT, we assume a ground state of uniformly distributed cells which forms a honeycomb lattice-like voronoi tessellation (with side length $d$, see Fig. \ref{fig.m0}). All cell centers form a triangular lattice with discrete translational symmetry with three nearest neighbor bond vectors
	\begin{equation}
	\begin{aligned}
	\bm a^{(1)}&=\sqrt 3 d
	\begin{pmatrix}
	\sqrt 3 / 2 \\
	1/2
	\end{pmatrix},\\
	\bm a^{(2)}&=\sqrt 3 d
	\begin{pmatrix}
	0 \\
	1
	\end{pmatrix},\\
	\bm a^{(3)}&=\sqrt 3 d
	\begin{pmatrix}
	-\sqrt 3 / 2 \\
	1/2
	\end{pmatrix}
	\label{m3.1}
	\end{aligned}
	\end{equation}
	Each cell center has six nearest neighbors. The rest three neighbor bond vectors are $\bm a^{(4)}=-\bm a^{(1)}$, $\bm a^{(5)}=-\bm a^{(2)}$ and $\bm a^{(6)}=-\bm a^{(3)}$. Let the mean cell area $\ell_0^2 = (3\sqrt{3}/2)d^2 = 1$, we have $d/\ell_0 = 2^{1/2} /3^{3/4}\approx 0.62$. For simplicity we also assume $K_A=0$ because it does not qualitatively affect the fluid-solid phase transition in graph models. 
	
The cell locates at $\bm x$ has six nearest neighbors which locate at $\bm x^{(n)} = \bm x +\bm a^{(n)}$. Let $\bm b^{(n)}= [\bm a^{(n+1)}+\bm a^{(n+2)}]/3$, the cell at $\bm x$ also has six vertex at $ \bm x + \bm b^{(n)}$. Here, the vertex at $ \bm x + \bm b^{(n)}$ is formed by three cells at $\bm x$, $\bm x +  \bm a^{(n+1)}$ and $\bm x + \bm a^{(n+2)}$, and is the circumcentre of the three cell centers. Specifically, if $(x_v, y_v)$ is the circumcentre of three points $(x_i,y_i)$ where $i=1,2,3$, we have
\begin{equation}
\begin{aligned}
{x_v} &=\frac{\sum_{ij}\epsilon_{ij}(-y_i^2 y_j+y_i x_j^2)}{2 \sum_{ij}\epsilon_{ij}y_ix_j}, \\{y_v} &=\frac{\sum_{ij}\epsilon_{ij}(-x_i^2 x_j+x_i y_j^2)}{2 \sum_{ij}\epsilon_{ij}x_iy_j}
\label{s0.1}
\end{aligned}
\end{equation}

We assume displacement field $\bm u$ for all cell centers and $\bm v$ for all vertex. We define $\delta \bm u(\bm y) = \bm u(\bm x+\bm y)-\bm u(\bm x)$ and $\delta \bm v(\bm y) = \bm v(\bm x+\bm y)-\bm u(\bm x)$. In the linear regime, we have
\begin{equation}
\begin{aligned}
\bm v{(\bm x + \bm b^{(n)}})=\frac{2}{3d^2} [\bm b^{(n)} \cdot\bm u(\bm x )]\bm b^{(n)}+\frac{2}{3d^2} [\bm b^{(n+2)} \cdot\bm u(\bm x +\bm a^{(n+1)})]\bm b^{(n+2)}+\frac{2}{3d^2} [\bm b^{(n-2)} \cdot\bm u(\bm x +\bm a^{(n+2)})]\bm b^{(n-2)}
\label{s0.7}
\end{aligned}
\end{equation}

The force acting on cell at $\bm x$ is 
\begin{equation}
\begin{aligned}
\bm f(\bm x) = \sum_n\bm f_v(\bm x +\bm b^{(n)})\cdot \frac{\partial  \bm v(\bm x +\bm b^{(n)})}{\partial \bm u(\bm x)}.
\label{s0.2}
\end{aligned}
\end{equation}
where
\begin{equation}
\begin{aligned}
\bm f_v(\bm x +\bm b^{(n)}) = -\frac{\partial E}{\partial \bm v(\bm x +\bm b^{(n)})}  + \alpha^{\rm o} \hat{\bm e}_z \times  \frac{\partial E_C}{\partial \bm v(\bm x +\bm b^{(n)})}
\label{s0.6}
\end{aligned}
\end{equation}

In the reference state, $\bm f_v=0$ and ${\partial  \bm v(\bm x +\bm b^{(n)})}/{\partial \bm u(\bm x)}\neq 0$. Therefore, to obtain the linear dynamics, we need to expand $\bm f_v$ to linear order in $\bm u$, and evaluate ${\partial  \bm v(\bm x +\bm b^{(n)})}/{\partial \bm u(\bm x)}$ at $\bm u=0$, where ${\partial  \bm v(\bm x +\bm b^{(n)})}/{\partial \bm u(\bm x)}=(2/3)\bm b^{(n)}\bm b^{(n)}$.

The perimeter of cell at $\bm x$ is 
\begin{equation}
\begin{aligned}
P(\bm x) &= \sum_n|b^{(n)} + \bm v(\bm x +b^{(n)}) - \bm b^{(n+1)} - \bm v(\bm x +b^{(n+1)})|\\
&\approx 6d +\frac{1}{d}\sum_n \bm b^{(n+2)} \cdot[\bm v(\bm x +b^{(n+1)})-\bm v(\bm x +b^{(n)})] +  \frac{1}{6 d^3}\sum_n\left[\bm a^{(n+2)} \cdot[\bm v(\bm x +b^{(n+1)})-\bm v(\bm x +b^{(n)})]\right]^2
\label{s0.3}
\end{aligned}
\end{equation}

and 
\begin{equation}
\begin{aligned}
\frac {\partial P(\bm x)}{\partial \bm v(\bm x+\bm b^{(n)})} &=\frac{1}d \bm b^{(n)} + \frac{1}{3d^3}\left[\bm a^{(n+1)} \cdot[\bm v(\bm x +b^{(n)})-\bm v(\bm x +b^{(n-1)})]\right]\bm a^{(n+1)}\\&-\frac{1}{3d^3}\left[\bm a^{(n+2)} \cdot[\bm v(\bm x +b^{(n+1)})-\bm v(\bm x +b^{(n)})]\right]\bm a^{(n+2)}
\label{s0.4}
\end{aligned}
\end{equation}
Because the vertex at $ \bm x + \bm b^{(n)}$ is shared by three cells at $\bm x$, $\bm x +  \bm a^{(n+1)}$ and $\bm x + \bm a^{(n+2)}$. Note that 
$\bm b^{(n)} =\bm a^{(n+1)} + \bm b^{(n+2)} =\bm a^{(n+2)} + \bm b^{(n-2)}$. We then have
\begin{equation}
\begin{aligned}
\frac{\partial E}{\partial \bm v(\bm x +\bm b^{(n)})} &= K_P[P(\bm x)-P_0]\frac {\partial P(\bm x)}{\partial \bm v(\bm x+\bm b^{(n)})} + K_P[P(\bm x + \bm a^{(n+1)})-P_0]\frac {\partial P(\bm x+ \bm a^{(n+1)})}{\partial \bm v(\bm x+\bm b^{(n)})} \\&+ K_P[P(\bm x + \bm a^{(n+2)})-P_0]\frac {\partial P(\bm x+ \bm a^{(n+2)})}{\partial \bm v(\bm x+\bm b^{(n)})}
\label{s0.5}
\end{aligned}
\end{equation}

Substituting Eq.~(\ref{s0.7}) in Eq.~(\ref{s0.3}), we have (keep linear order)
\begin{equation}
\begin{aligned}
P(\bm x) &=6d + \frac{1}{3 d}\sum_{n=1}^6\bm a^{(n)} \cdot\delta \bm u (\bm a^{(n)})
\label{s0.8}
\end{aligned}
\end{equation}
and 
\begin{equation}
\begin{aligned}
P(\bm x+{\bm a}^{(m)}) &=6d + \frac{1}{3 d}\sum_{n\neq m+3}\bm a^{(n)} \cdot[\delta \bm u (\bm a^{(m)}+\bm a^{(n)})-\delta \bm u (\bm a^{(m)})] + \frac{1}{3 d} \bm a^{(m)} \cdot\delta \bm u (\bm a^{(m)})
\\&=6d + \frac{1}{3 d} [\bm a^{(m+1)}\cdot\delta \bm u (3\bm b^{(m-1)}) + \bm a^{(m+2)}\cdot\delta \bm u (\bm a^{(m+1)})+ \bm a^{(m)}\cdot\delta \bm u (2\bm a^{(m)}) + \bm a^{(m-1)}\cdot\delta \bm u (3\bm b^{(m-2)})]\\&+ \frac{1}{3d}[\bm a^{(m-2)}\cdot\delta \bm u (\bm a^{(m-1)})] 
\label{s0.9}
\end{aligned}
\end{equation}
Substituting Eq.~(\ref{s0.7}) in Eq.~(\ref{s0.4})gives
\begin{equation}
\begin{aligned}
\frac {\partial P(\bm x)}{\partial \bm v(\bm x+\bm b^{(n)})} &=\frac{1}d \bm b^{(n)} - \frac{1}{3d^3}\left[\bm b^{(n+1)} \cdot\delta\bm u(\bm a^{(n+1)}) \right]\bm a^{(n+1)}+ \frac{1}{3d^3}\left[\bm b^{(n+2)} \cdot\delta\bm u(\bm a^{(n+2)}) \right]\bm a^{(n+2)}
\label{s0.10}
\end{aligned}
\end{equation}
and 
\begin{equation}
\begin{aligned}
\frac {\partial P(\bm x+{\bm a}^{(n+1)})}{\partial \bm v(\bm x+\bm b^{(n)})} &=\frac{1}d \bm b^{(n+2)} - \frac{1}{3d^3}\left[\bm b^{(n)} \cdot[\delta\bm u(\bm a^{(n+2)})- \delta\bm u(\bm a^{(n+1)})]\right]\bm a^{(n)}- \frac{1}{3d^3}\left[\bm b^{(n+1)} \cdot\delta\bm u(\bm a^{(n+1)}) \right]\bm a^{(n+1)}\\\frac {\partial P(\bm x+{\bm a}^{(n+2)})}{\partial \bm v(\bm x+\bm b^{(n)})} &=\frac{1}d \bm b^{(n-2)} + \frac{1}{3d^3}\left[\bm b^{(n+2)} \cdot\delta\bm u(\bm a^{(n+2)})\right]\bm a^{(n+2)}+ \frac{1}{3d^3}\left[\bm b^{(n)} \cdot[\delta\bm u(\bm a^{(n+1)})- \delta\bm u(\bm a^{(n+2)})] \right]\bm a^{(n)}
\label{s0.11}
\end{aligned}
\end{equation}

We then have
\begin{align}
\frac{\partial E}{\partial \bm v(\bm x +\bm b^{(n)})} &= -\frac{K_P(6d-P_0)}{3d^3}\left[\bm b^{(n+1)} \cdot\delta\bm u(\bm a^{(n+1)}) \right]\bm a^{(n+1)}\\&+ \frac{K_P(6d-P_0)}{3d^3}\left[\bm b^{(n+2)} \cdot\delta\bm u(\bm a^{(n+2)}) \right]\bm a^{(n+2)}\\&+\frac{K_P}{3d^2}\left[\sum_{m=1}^6\bm a^{(m)} \cdot\delta \bm u (\bm a^{(m)})\right]\bm b^{(n)}\\&-\frac{K_P(6d-P_0)}{3d^3}\left[\bm b^{(n)} \cdot[\delta\bm u(\bm a^{(n+2)})- \delta\bm u(\bm a^{(n+1)})]\right]\bm a^{(n)}\\&- \frac{K_P(6d-P_0)}{3d^3}\left[\bm b^{(n+1)} \cdot\delta\bm u(\bm a^{(n+1)}) \right]\bm a^{(n+1)} \\ &+ \frac{K_P}{3d^2}\left[ \bm a^{(n+2)}\cdot\delta \bm u (3\bm b^{(n)})\right]\bm b^{(n+2)}\\&+ \frac{K_P}{3d^2}\left[ \bm a^{(n+3)}\cdot\delta \bm u (\bm a^{(n+2)})\right]\bm b^{(n+2)}\\&+ \frac{K_P}{3d^2}\left[ \bm a^{(n+1)}\cdot\delta \bm u (2\bm a^{(n+1)})\right]\bm b^{(n+2)}\\&+ \frac{K_P}{3d^2}\left[ \bm a^{(n)}\cdot\delta \bm u (3\bm b^{(n-1)})\right]\bm b^{(n+2)}\\&+ \frac{K_P}{3d^2}\left[ \bm a^{(n-1)}\cdot\delta \bm u (\bm a^{(n)})\right]\bm b^{(n+2)}\\&+\frac{K_P(6d-P_0)}{3d^3}\left[\bm b^{(n+2)} \cdot\delta\bm u(\bm a^{(n+2)})\right]\bm a^{(n+2)}\\&+\frac{K_P(6d-P_0)}{3d^3}\left[\bm b^{(n)} \cdot[\delta\bm u(\bm a^{(n+1)})- \delta\bm u(\bm a^{(n+2)})] \right]\bm a^{(n)}\\
&+ \frac{K_P}{3d^2}\left[ \bm a^{(n+3)}\cdot\delta \bm u (3\bm b^{(n+1)})\right]\bm b^{(n-2)}\\&+ \frac{K_P}{3d^2}\left[ \bm a^{(n+4)}\cdot\delta \bm u (\bm a^{(n+3)})\right]\bm b^{(n-2)}\\&+ \frac{K_P}{3d^2}\left[ \bm a^{(n+2)}\cdot\delta \bm u (2\bm a^{(n+2)})\right]\bm b^{(n-2)}\\&+ \frac{K_P}{3d^2}\left[ \bm a^{(n+1)}\cdot\delta \bm u (3\bm b^{(n)})\right]\bm b^{(n-2)}\\&+ \frac{K_P}{3d^2}\left[ \bm a^{(n)}\cdot\delta \bm u (\bm a^{(n+1)})\right]\bm b^{(n-2)}
\label{s0.12}
\end{align}
Similarly we calculate $\frac{\partial E_C}{\partial \bm v(\bm x +\bm b^{(n)})}$ and find
\begin{align}
\hat{\bm e}_z \times \frac{\partial E_C}{\partial \bm v(\bm x +\bm b^{(n)})} &= -\frac{2\sqrt 3 K_P}{d^2}\left[\bm b^{(n+1)} \cdot\delta\bm u(\bm a^{(n+1)}) \right]\bm b^{(n+1)}\\&+ \frac{2\sqrt 3 K_P}{d^2}\left[\bm b^{(n+2)} \cdot\delta\bm u(\bm a^{(n+2)}) \right]\bm b^{(n+2)}\\&-\frac{K_P}{3\sqrt 3d^2}\left[\sum_{m=1}^6\bm a^{(m)} \cdot\delta \bm u (\bm a^{(m)})\right]\bm a^{(n)}\\&-\frac{2\sqrt 3 K_P}{d^2}\left[\bm b^{(n)} \cdot[\delta\bm u(\bm a^{(n+2)})- \delta\bm u(\bm a^{(n+1)})]\right]\bm b^{(n)}\\&- \frac{2\sqrt 3 K_P}{d^2}\left[\bm b^{(n+1)} \cdot\delta\bm u(\bm a^{(n+1)}) \right]\bm b^{(n+1)} \\ &- \frac{K_P}{3\sqrt 3 d^2}\left[ \bm a^{(n+2)}\cdot\delta \bm u (3\bm b^{(n)})\right]\bm a^{(n+2)}\\&- \frac{K_P}{3\sqrt 3 d^2}\left[ \bm a^{(n+3)}\cdot\delta \bm u (\bm a^{(n+2)})\right]\bm a^{(n+2)}\\&- \frac{K_P}{3\sqrt 3 d^2}\left[ \bm a^{(n+1)}\cdot\delta \bm u (2\bm a^{(n+1)})\right]\bm a^{(n+2)}\\&- \frac{K_P}{3\sqrt 3 d^2}\left[ \bm a^{(n)}\cdot\delta \bm u (3\bm b^{(n-1)})\right]\bm a^{(n+2)}\\&- \frac{K_P}{3\sqrt 3 d^2}\left[ \bm a^{(n-1)}\cdot\delta \bm u (\bm a^{(n)})\right]\bm a^{(n+2)}\\&+\frac{2\sqrt 3 K_P}{d^2}\left[\bm b^{(n+2)} \cdot\delta\bm u(\bm a^{(n+2)})\right]\bm b^{(n+2)}\\&+\frac{2\sqrt 3 K_P}{d^2}\left[\bm b^{(n)} \cdot[\delta\bm u(\bm a^{(n+1)})- \delta\bm u(\bm a^{(n+2)})] \right]\bm b^{(n)}\\
&- \frac{K_P}{3\sqrt 3 d^2}\left[ \bm a^{(n+3)}\cdot\delta \bm u (3\bm b^{(n+1)})\right]\bm a^{(n-2)}\\&- \frac{K_P}{3\sqrt 3d^2}\left[ \bm a^{(n+4)}\cdot\delta \bm u (\bm a^{(n+3)})\right]\bm a^{(n-2)}\\&- \frac{K_P}{3\sqrt 3d^2}\left[ \bm a^{(n+2)}\cdot\delta \bm u (2\bm a^{(n+2)})\right]\bm a^{(n-2)}\\&- \frac{K_P}{3\sqrt 3 d^2}\left[ \bm a^{(n+1)}\cdot\delta \bm u (3\bm b^{(n)})\right]\bm a^{(n-2)}\\&- \frac{K_P}{3\sqrt 3 d^2}\left[ \bm a^{(n)}\cdot\delta \bm u (\bm a^{(n+1)})\right]\bm a^{(n-2)}
\label{s0.13}
\end{align}
We have
\begin{align}
\bm f(\bm x) &= \frac{2K_P(6d-P_0)}{3d^3}\sum_n\left[\bm b^{(n)} \cdot\delta\bm u(\bm a^{(n)}) \right]\bm b^{(n)} \\&- \frac{K_P}{9d^2}\sum_n\left[ \bm a^{(n+1)}\cdot\delta \bm u (3\bm b^{(n)})\right]\bm a^{(n-1)}\\&- \frac{K_P}{9d^2}\sum_n\left[ \bm a^{(n-1)}\cdot\delta \bm u (3\bm b^{(n)})\right]\bm a^{(n+1)}\\&+ \frac{K_P}{9d^2}\sum_n\left[ \bm a^{(n+1)}\cdot\delta \bm u (\bm a^{(n)})\right]\bm a^{(n-1)}\\&+ \frac{K_P}{9d^2}\sum_n\left[ \bm a^{(n-1)}\cdot\delta \bm u (\bm a^{(n)})\right]\bm a^{(n+1)}\\&+ \frac{K_P}{9d^2}\sum_n\left[ \bm a^{(n)}\cdot\delta \bm u (2\bm a^{(n)})\right]\bm a^{(n)}
\\&-\frac{4\sqrt 3 K_P \alpha^o}{3d^2}\sum_n\left[\bm b^{(n)} \cdot\delta\bm u(\bm a^{(n)}) \right]\bm a^{(n)}\\&-\frac{8\sqrt 3 K_P\alpha^o}{3d^2}\sum_n\left[\bm b^{(n+1)} \cdot\delta\bm u(\bm a^{(n)})\right]\bm b^{(n+1)}\\&+\frac{8\sqrt 3 K_P\alpha^o}{3d^2}\sum_n\left[\bm b^{(n-1)} \cdot\delta\bm u(\bm a^{(n)})\right]\bm b^{(n-1)}\\&- \frac{K_P\alpha^o}{3\sqrt 3 d^2}\sum_n\left[ \bm a^{(n+1)}\cdot\delta \bm u (\bm a^{(n)})\right]\bm b^{(n-1)}\\&- \frac{K_P\alpha^o}{3\sqrt 3 d^2}\sum_n\left[ \bm a^{(n-1)}\cdot\delta \bm u (\bm a^{(n)})\right]\bm b^{(n+1)}\\&-\frac{K_P\alpha^o}{3\sqrt 3 d^2}\sum_n\left[ \bm a^{(n)}\cdot\delta \bm u (2\bm a^{(n)})\right]\bm b^{(n)}\\&+ \frac{K_P\alpha^o}{3\sqrt 3 d^2}\sum_n\left[ \bm a^{(n-1)}\cdot\delta \bm u (3\bm b^{(n)})\right]\bm b^{(n+1)}\\
&+ \frac{K_P\alpha^o}{3\sqrt 3 d^2}\sum_n\left[ \bm a^{(n+1)}\cdot\delta \bm u (3\bm b^{(n)})\right]\bm b^{(n-1)}
\label{s0.14}
\end{align}

We then arrive at the dynamical matrix in Fourier space
\begin{align}
D(\bm q) &= \frac{2K_P(6d-P_0)}{3d^3}\sum_n\bm b^{(n)}\bm b^{(n)} \left[1- \cos(\bm a^{(n)}\cdot \bm q) \right] \\&- \frac{K_P}{9d^2}\sum_n \bm a^{(n-1)}\bm a^{(n+1)}\left[1- \cos(3\bm b^{(n)}\cdot \bm q)\right]\\&- \frac{K_P}{9d^2}\sum_n \bm a^{(n+1)}\bm a^{(n-1)}\left[1- \cos(3\bm b^{(n)}\cdot \bm q)\right]\\&+ \frac{K_P}{9d^2}\sum_n \bm a^{(n-1)}\bm a^{(n+1)}\left[1- \cos(\bm a^{(n)}\cdot \bm q) \right]\\&+ \frac{K_P}{9d^2}\sum_n \bm a^{(n+1)}\bm a^{(n-1)}\left[1- \cos(\bm a^{(n)}\cdot \bm q) \right]\\&+ \frac{K_P}{9d^2}\sum_n\bm a^{(n)}\bm a^{(n)}\left[1- \cos(2\bm a^{(n)}\cdot \bm q) \right]
\\&-\frac{4\sqrt 3 K_P \alpha^o}{3d^2}\sum_n\bm a^{(n)}\bm b^{(n)} \left[1- \cos(\bm a^{(n)}\cdot \bm q) \right]\\&-\frac{8\sqrt 3 K_P\alpha^o}{3d^2}\sum_n\bm b^{(n+1)}\bm b^{(n+1)} \left[1- \cos(\bm a^{(n)}\cdot \bm q) \right]\\&+\frac{8\sqrt 3 K_P\alpha^o}{3d^2}\sum_n\bm b^{(n-1)}\bm b^{(n-1)} \left[1- \cos(\bm a^{(n)}\cdot \bm q) \right]\\&- \frac{K_P\alpha^o}{3\sqrt 3 d^2}\sum_n\bm b^{(n-1)}\bm a^{(n+1)} \left[1- \cos(\bm a^{(n)}\cdot \bm q) \right]\\&- \frac{K_P\alpha^o}{3\sqrt 3 d^2}\sum_n\bm b^{(n+1)}\bm a^{(n-1)} \left[1- \cos(\bm a^{(n)}\cdot \bm q) \right]\\&- \frac{K_P\alpha^o}{3\sqrt 3 d^2}\sum_n\bm b^{(n)}\bm a^{(n)} \left[1- \cos(2\bm a^{(n)}\cdot \bm q) \right]\\&+ \frac{K_P\alpha^o}{3\sqrt 3 d^2}\sum_n\bm b^{(n-1)}\bm a^{(n+1)} \left[1- \cos(3\bm b^{(n)}\cdot \bm q) \right]\\
&+ \frac{K_P\alpha^o}{3\sqrt 3 d^2}\sum_n\bm b^{(n+1)}\bm a^{(n-1)} \left[1- \cos(3\bm b^{(n)}\cdot \bm q) \right]
\label{s0.15}
\end{align}

In the small $q=|\bm q|$ limit, we have
\begin{equation}
\begin{aligned}
D(\bm q) &= \frac{9K_Pdq^2}{4}\begin{pmatrix}
& 6d-P_0/3   & 2\alpha^od\\
&-2\alpha^od &  6d-P_0
\end{pmatrix},
\label{s0.16}
\end{aligned}
\end{equation}

For a continuum elastic medium, its dynamical matrix reads~\cite{scheibner2020odd,fruchart2022odd}
\begin{equation}
\begin{aligned}
D(\bm q) &= \bar A q^2\begin{pmatrix}
& B+\mu  & K^o\\
&A-K^o &  \mu
\end{pmatrix},
\label{s0.17}
\end{aligned}
\end{equation}
(note that $\bar A=(3\sqrt 3 /2)d^2=1$ is the average area and $A$ is the odd elasticity to bulk deformation.)  Comparing Eqs.~(\ref{s0.16},\ref{s0.17}), we find that the MFT can be coarse grained to a continuum medium with $A=0$, $B=3K_PP_0d/(2\bar A)\approx 0.93K_P P_0$, $\mu = 9K_P(6d-P_0)d/(4\bar A)\approx 1.4 K_P(3.72-P_0)$ and $K^o = 9\alpha^o K_Pd^2/(2\bar A)\approx 1.72 \alpha^oK_P$. These predictions have good quantitative agreement with the numerical results in the solid phase. 

We then consider $D(\bm q)$ for any $q$. When $q$ is not small, $D(\bm q)$ is general anisotropic and has a tedious mathematical form. The stability of $D(\bm q)$ is controlled by the real part of its eigenvalues, ${\rm Re}(\lambda_{\pm})$. Because of the next-nearest-neighbor interactions, ${\rm Re}(\lambda_{\pm})$ may have non-trivial dependence on $\bm q$, which is in stark contrast with continuum elastic materials. Below we show that there exists a most unstable mode $\bm q^*$ which determines the stability of the MFT: for arbitrary parameters $P_0$ and $\alpha^o$, as long as $\bm q^*$ is stable, all other modes must be stable. 

To show this, we consider the 12 special directions $\bm a^{(n)}$ and $\bm b^{(n)}$ in the MFT. Because of the discrete rotational symmetry, we only need to consider the case $\bm q=q\hat{\bm e}_x$ (parallel to $\bm b^{(2)}$) and $\bm q=q\hat{\bm e}_y$ (parallel to $\bm a^{(2)}$). We first consider $\bm q=q\hat{\bm e}_x$, where
\begin{equation}
\begin{aligned}
D(q\hat{\bm e}_x) = \frac{4K_P\sin^2(3dq/4)}{   d}\begin{pmatrix}
& [4+2\cos(3dq/2)]d-P_0/3   & 2\alpha^od\\
&[-4+2\cos(3dq/2)]\alpha^od &  (6d-P_0) 
\end{pmatrix},
\label{s0.18}
\end{aligned}
\end{equation}
The eigenvalues are
\begin{equation}
\begin{aligned}
\lambda_{\pm}=-\frac{4 K_P \sin^2(3dq/4)}{3 d}\left(\pm\sqrt{C_1^2  -C_2^2  }+2C_1 - C_3\right)
\label{s0.19}
\end{aligned}
\end{equation}
where $C_1 = P_0 + 3\cos(3dq/2)d-3d$, $C_2 = 6 |\alpha^o|\sqrt{2-\cos(3dq/2)}d\geq0$ and $C_3 =9[\cos(3dq/2)+1]d\geq0$. We find an exceptional point for each mode at $C_1=\pm C_2$. 

The stability is controlled by ${\rm Re}(\lambda_+)$: the mode is stable if ${\rm Re}(\lambda_+)>0$ and is unstable if ${\rm Re}(\lambda_+)<0$. For any given $q$ and $\alpha^o$, we have ${\rm Re}(\lambda_+)>0$ for small $P_0$ and ${\rm Re}(\lambda_+)<0$ for large $P_0$, and the stability changes at $P_0^*(q,\alpha^o)$ where ${\rm Re}(\lambda_+)=0$. We notice that $P_0^*(q=2\pi/(3d),\alpha^o)=6d$. Also, when $P_0=6d$, we always have ${\rm Re}(\lambda_+)\geq0$, suggesting that $P_0^*(q,\alpha^o)\leq 6d$. Therefore, the most unstable mode for $\bm q=q \hat{\bm e}_x$ is $q=2\pi/3$. The characteristic frequency can be found with $i\omega_\pm(q)=\lambda\pm$. 
Let $P_0^{\rm MF}=6d$, $s = 1-P_0/P_0^{\rm MF}$ and $\epsilon = \alpha^o/s$, we have 
\begin{equation}
\begin{aligned}
\omega_\pm(q)=i{2K_Ps}\left[2\pm\sqrt{1-3\epsilon^2}\right]
\label{s8}
\end{aligned}
\end{equation}
from which we identify exceptional points $|\epsilon|=1/\sqrt 3$. 

We repeat the calculation for $\bm q=q\hat{\bm e}_y$, and find that the most unstable mode is located at $q=2\pi/(\sqrt{3}d)$, where the characteristic frequencies are identical to Eq.~(\ref{s8}). 

\subsection{Prestress}
In this section we calculate the prestress in the MFT. The stress is defined using the generalized Irvine-Kirkwood formula for many-body interaction, see Sec.~\ref{IK} for details:
\begin{equation}
\begin{aligned}
\sigma = -\frac{1}{S} \sum_k \sum_{i\in N_k}\bm f_{k,i}\Delta \bm r_{k,i}
\label{m1.4}
\end{aligned}
\end{equation}

In the reference state of the MFT, the formula can be rewritten as
\begin{equation}
\begin{aligned}
\sigma = -\frac{1}{\bar A} \sum_{n}\bm f(\bm a^{(n)})\bm a^{(n)}
\label{s1.1}
\end{aligned}
\end{equation}
where 
\begin{equation}
\begin{aligned}
\bm f(\bm a^{(n)}) = \sum_m\left[ -\frac{\partial E(\bm x)}{\partial \bm v(\bm x + \bm b^{(m)})}+\alpha^o \hat{\bm e}_z \times \frac{\partial E_C(\bm x)}{\partial \bm v(\bm x + \bm b^{(m)})}\right]\cdot \frac{\partial \bm  v(\bm x + \bm b^{(m)})}{\partial \bm u(\bm x + \bm a^{(n)})}
\label{s1.2}
\end{aligned}
\end{equation}
in which 
\begin{equation}
\begin{aligned}
E(\bm x) &= \frac{K_P} 2 (P(\bm x)-P_0)^2\\
E_C(\bm x) &= \frac{K_P} 2 (P(\bm x))^2
\label{s1.3}
\end{aligned}
\end{equation}

When calculating the prestress the derivatives only need to be evaluated at $\bm u = \bm v = 0$. Substituting Eqs.~(\ref{s0.9}, \ref{s0.7}) into Eq.~(\ref{s1.3}), we have 
\begin{equation}
\begin{aligned}
\bm f(\bm a^{(n)}) &= \sum_m\left[ -\frac{K_P(6d-P_0)} d \bm b^{(m)}-\frac{6\alpha^o K_P} {\sqrt 3 } \bm a^{(m)}\right]\cdot \frac{\partial \bm  v(\bm x + \bm b^{(m)})}{\partial \bm u(\bm x + \bm a^{(n)})}\\&=\frac{2}{3}\left[ -\frac{K_P(6d-P_0)} d \bm b^{(n-1)}-\frac{6\alpha^o K_P} {\sqrt 3 } \bm a^{(n-1)}\right]\cdot \bm b^{(n+1)}\bm b^{(n+1)}\\&+\frac{2}{3}\left[ -\frac{K_P(6d-P_0)} d \bm b^{(n-2)}-\frac{6\alpha^o K_P} {\sqrt 3 } \bm a^{(n-2)}\right]\cdot \bm b^{(n-1)}\bm b^{(n-1)}\\&=-\frac{K_P(6d-P_0)} {3d} \bm a^{(n)}+\frac{6\alpha^o K_P} {\sqrt 3 } \bm b^{(n)}
\label{s1.4}
\end{aligned}
\end{equation}

We then have the stress tensor:
\begin{equation}
\begin{aligned}
\sigma&=\begin{pmatrix}
& \sigma_P   & -\tau_P\\
&\tau_P & \sigma_P
\end{pmatrix}
\\&=(K_P/2\bar A)(P_0^{\rm MF})^2\begin{pmatrix}
& s   & -\alpha^o\\
&\alpha^o & s
\end{pmatrix}
\label{s12}
\end{aligned}
\end{equation}
where $s=1-P_0/P_0^{\rm MF}$. 

\subsection{Chiral edge current}
The cortical chirality generates a pretorque $\tau_P$ in the tissue. When the tissue is in contact with a torque-free boundary, the cell in contact with the boundary would feel a net force. We consider a boundary in $\hat{\bm e}_y$. The boundary replaces the neighboring cells which are located at ${\bm a}^{(4)}$ and ${\bm a}^{(5)}$. We can estimate this force by turning off the chiral force of neighboring cells at ${\bm a}^{(4)}$ and ${\bm a}^{(5)}$, i.e.,
\begin{equation}
\begin{aligned}
\bm f_b &=-\frac{K_P(6d-P_0)} {3d} \sum_n\bm a^{(n)}+\frac{6\alpha^o K_P} {\sqrt 3 } \sum_{n\neq 4,5}\bm b^{(n)}=(0,-6\alpha^oK_P d)
\label{s1.4}
\end{aligned}
\end{equation}
The boundary cells feel a tangential force. This force would cause a displacement $d_b \hat{\bm e}_y$ of the boundary cells. $d_b$ can be solved from the force balance in $y$
\begin{equation}
\begin{aligned}
(\bm f_b+\bm f(\bm x))\cdot \hat{\bm e}_y &=0
\label{s1.5}
\end{aligned}
\end{equation}
where $\bm f(\bm x)$ is defined in Eq.~(\ref{s0.14}) and $\delta \bm u (\bm a^{(1)})=\delta \bm u (\bm a^{(2)})=\delta \bm u(3\bm b^{(1)})=\delta \bm u(3\bm b^{(5)})=\delta \bm u(3\bm b^{(6)})=-d_b \hat{\bm e}_y$, and other $\delta \bm u=0$. This gives $d_b = -12\alpha^o d^2/(6d-P_0)$. Edge current occurs when $|d_b|\geq \sqrt 3 d/2$ which triggers T1 transitions, i.e., 
\begin{equation}
\begin{aligned}
|\alpha^o|\geq \frac{\sqrt 3 (6d-P_0)}{24d}
\label{s1.6}
\end{aligned}
\end{equation}

\section{Details of numerical simulations}
In this section we detail the numerical simulation of chiral graph model. The simulation is performed in a rectangular box of size $L_x\times L_y$. The cell centers are initialized at random locations with average cell area $\ell_0^2=1$ and minimum cell-cell distance $d_{\rm min}=0.8\ell_0$. The dynamics are calculated by a numerical integration of Eq.~(2) of the main text with the fourth-order Runge-Kutta method with $\Delta t=0.01\gamma/K_P$ or $\Delta t=0.02\gamma/K_P$.  
We adopt two different boundary conditions: 

(1). Periodic boundary condition

\begin{figure}[t]
\centering
\hspace{0em}\includegraphics[width = 0.6\columnwidth]{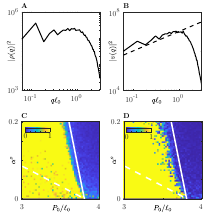}
\caption{{\bf Additional simulation results of the graph model.} (A) Density fluctuations in the Fourier space. (B) Velocity fluctuations in the Fourier space. Dashed line indicates an exponent $1$.  (C) Relative fluctuations of the steady-state cell velocity for 10 different random initial conditions. (D) Relative difference of the steady-state cell velocity for system sizes $W=10\ell_0$ and $W=20\ell_0$. In (A) and (B) $P_0/\ell_0=4.0$, $\alpha^o=0.2$ and $W=100\ell_0$.    In all panels $A_0=\ell_0^2=1$, $K_P=1$ and $K_A=0.1$. 
    }
    \label{fig.m2}
\end{figure}

Periodic boudnary condition with $L_x = L_y=W$ is used throughout the main text and Methods, except in Extended Data Fig. 5B of the Methods. $W=10\ell_0$ is used throughout the paper except in Extended Data Fig. 3EF of Methods ($W=100\ell_0$) and in Fig. \ref{fig.m2}CD ($W=20\ell_0$). 

(2). Torque-free boundary

In Extended Data Fig. 5B of Methods we assume torque-free boundary at the edges of the rectangular box. Specifically, we create mirror images with respect to each edge of all cells that are in contact with that edge. Because the boundary is frictionless, all edge vertices are not included in Eq.~(2) of the main text. We use $L_x=L_y=W=10\ell_0$. 

The Jacobian matrix is calculated by automatic differentiation. Specifically, it is defined by 
$J=\partial \bm f/\partial r$. Here ${\bm f}$ stands for the $2N_c$ components of the forces and ${\bm r}$ stands for the $2N_c$ components of the positions of all $N_c$ cells ($N_c$ is the total number of cells).

In Fig.~\ref{fig.m2}AB we analyze the chaotic flow structure in the steady state for $\alpha=0.2$ and $P_0/\ell_0=4.0$. We find no large-scale structure in both the density field and the velocity field. The only peak in the Fourier space is located at $q\ell_0\approx 1$, i.e., the scale of a single cell. This is in stark contrast with the active turbulence in which vortices form at large lengthscales.

In Fig.~\ref{fig.m2}C we study the dependence of the steady-state cell velocity on initial conditions. The fact that $P_0^*(\alpha^o)<P_0^{\rm MF}$ suggests that the steady-state behavior must have some dependence on the initial conditions. For example, for $P_0^*(\alpha^o)<P_0<P_0^{\rm MF}$, if the initial condition is the hexagonal lattice studied in MFT, the model is in solid phase and the velocity vanishes. However, the simulation results indicate that the model is in the liquid phase for disordered initial conditions. In this case, the model may have two absorbing states at the same time: one fluid state with continuous chaotic motion, and one solid state in which the model freezes. On the other hand, the probability that the initial condition is a lattice can be extremely small. To study whether the steady-state property depends on the disordered initial conditions used in the simulation, we calculate the relative fluctuations: the ratio between the variance and the mean of $\langle \bm v^2\rangle$ obtained for 10 different initial conditions. We find that the ratio is very small in the liquid phase. It is large in the solid phase because $\langle \bm v^2\rangle$ vanishes and the ratio becomes meaningless. Therefore, the random initial condition does not have a considerable impact on the phase diagram. 

In Fig.~\ref{fig.m2}D we study the dependence of the steady-state cell velocity on system size $W$. We compare the relative difference between the velocity for $W=20$ and that for $W=10$ and find negligible difference in the liquid phase (again the yellow region in the solid phase is simply because $\langle \bm v^2\rangle$ vanishes). There might be a finite-size effect on the critical exponents, whose identification requires a finite-size scaling, which is beyond our current numerical capability. 

\section{Computation of the stress in the chiral graph model}
\label{IK}
As our graph model is active, the stress cannot be derived from a derivative of the free energy. To calculate the stress, we first define 
\begin{equation}
\begin{aligned}
E_k = \frac{K_P}2 (P_k-P_0)^2 + \frac{K_A}2 (A_k-A_0)^2 
\label{m1.1}
\end{aligned}
\end{equation}
to be the Hamiltonian component of the $j$-th cell. Similarly 
\begin{equation}
\begin{aligned}
E^C_k = \frac{K_P}2 (P_k)^2 
\label{m1.2}
\end{aligned}
\end{equation}
is the contractile part of $E_k$. With that, the force acting on cell $i$ is rewritten as
\begin{equation}
\begin{aligned}
\bm f_i = \sum_{k\in N_i} \bm f_{k,i}
\label{m1.3}
\end{aligned}
\end{equation}
where 
\begin{equation}
\begin{aligned}
 \bm f_{k,i}=\sum_j\left(-\frac {\partial E_k}{\partial \bm r^v_j} + \alpha^o \hat{\bm e}_z \times  \frac{\partial E^C_k}{\partial \bm r^v_j}\right)\cdot\frac{\partial \bm r^v_j}{\partial \bm r_i}\,,
\label{m1.3-2}
\end{aligned}
\end{equation}
and $N_i$ is the set formed by cell $i$ and its neighbors. $\bm f_{k,i}$ is the force acting on cell $i$ that originates from the Hamiltonian of cell $k$. Hence, for the Hamiltonian of a given cell $k$, we can define a many-body interaction within $N_k$. One can show that this interaction conserves momentum, i.e., $\sum_{i\in N_k} \bm f_{k,i}=0$. This allows us to calculate the stress tensor using a generalized Irving-Kirkwood formula for many-body interaction~\cite{heinz2005calculation} (mass is neglected in the overdamped limit):
\begin{equation}
\begin{aligned}
\sigma = -\frac{1}{S} \sum_k \sum_{i\in N_k}\bm f_{k,i}\Delta \bm r_{k,i}
\label{m1.4}
\end{aligned}
\end{equation}
 where $S$ is the total area and $\Delta \bm r_{k,i}=\bm r_{i}-\bm r_{k}$ with $\bm r_{i}$ being the position of cell $i$. 

Equation~(\ref{m1.4}) is used to calculate the stress tensor numerically in the simulation with Lees-Edwards periodic boundary
conditions. For the elasticity tensor, we apply small bulk expansion/simple shear strain, and calculate the steady-state stress before and after the deformation. For the viscosity tensor, we impose a macroscopic time-dependent simple shear strain $e(t)$, which shears the $xy$ plane in the $x$ direction. We adopt the "affine solvent model": we assume that the model is attached to an affine substrate fluid, which produces a frictional force $\gamma(\partial_t \bm r_i - \bm v_f(\bm r_i))$. Here $\bm v_f(\bm r_i) = y_i \dot{e}(t)$ is the fluid velocity at $\bm r_i=(x_i, y_i)$. We use an oscillatory simple shear strain rate,
\begin{equation}
\begin{aligned}
\dot{e} (t)=\left\{ 
\begin{aligned}
&\epsilon\qquad&(nT<t\leq(n+\frac 1 2) T\\
&-\epsilon  \qquad&(n+\frac 1 2) T<t\leq (n+1)T\\
\end{aligned}
\right.
\end{aligned} \, .
\label{m1.5}
\end{equation}
In the simulation we use $\epsilon=0.01$ and $T=200$. The average stress tensor is defined by $\sigma = \langle \sigma(t) {\rm sgn}[\dot{e}(t)]\rangle_t$~\cite{han2021fluctuating}.

\section{Comparison to active nematohydrodynamics theories} 
\label{comparison_nematohydrodynamics}

\begin{figure}[t]
\centering
\hspace{-2em}\includegraphics[width = 0.6\columnwidth]{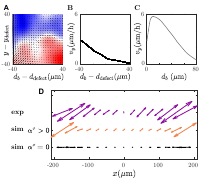}
\caption{{\bf  Tilt of topological defects around boundaries.} (A) Average flow field around boundary defects (defect center within $90{\rm \mu m}$ to the boundaries. Colors indicate positive (red) and negative (blue) vorticity. $d_b$ is the distance from the data point to the boundaries. $d_{\rm defect}$ is the distance from the defect center to the boundaries.  (B) Average chiral flow $v_y$ around boundary defects. (C) Average chiral flow $v_y$ for all regions around boundaries. (D) Mean shear flow as function of $x$. Arrows indicate the compressive direction of the shear flow and the sizes indicate the magnitude of the shear flow. 
    }
    \label{fig.m6}
\end{figure}

For monolayers composed of anisotropic cells, a nematic phase with topological defects can also emerge. While
$+1/2$ defects are typically distributed isotropically within the bulk, they tend to align perpendicularly to the boundary when located nearby. In Ref.~\cite{yashunsky2022chiral} nematic order is indeed observed in HT1080 cells. Interestingly, the defects near the boundary are tilted by an average angle of $16^{\circ}$, influenced by chirality. The two vortices produced by each defect are also tilted, see Extended Data Fig.~\ref{fig.m6}A for the average flow field around boundary defects (distance to boundary $d_{\rm defect}<90{\rm \mu m}$). Although the tilt of the boundary defects also affects the chiral flow $v_y$ around them~\cite{yashunsky2022chiral}, we believe that these defects do not have a major influence on the average flow field: We find that the chiral flow around the defects has magnitude similar to the overall chiral flow (Fig.~\ref{fig.m6}BC). Given that the defects are sparsely distributed in the sample, their contribution to the overall flow field is negligible.

The tilting of the defects can be explained by the tilted shear flows around the boundaries, with which the defects tend to align (as they are anisotropic). To show this, we calculate the shear flow produced by Eq.~(33) of Methods, $\dot{e}_s=(\nabla \bm v + (\nabla \bm v)^T)/2-(\nabla \cdot \bm v){\bm I}/3$:
\begin{equation}
\begin{aligned}
\dot{e}_s&=
-\frac{\ell_0^2\partial^2_x \rho}{\gamma}\begin{pmatrix}
&(B+\mu)/4  &  (A-K^o)/2 \\
&(A-K^o)/2 & -(B+\mu)/4
\end{pmatrix}
\end{aligned}
\end{equation}
The shear flow is proportional to $\partial_x \rho$, which is large near the boundary and diminishes in the bulk. When 
$\alpha^o=0$, the orientation vector $\hat{\bm n}_s=(1,0)$, causing the defects to align perpendicular to the boundary. Nonzero $\alpha^o$ tilts the  shear flows, leading to a corresponding tilt in the alignment of the defects. 

The chiral graph model is composed of isotropic cells, hence there is no intrinsic alignment interaction which is required for nematic order. Nevertheless, we may still use the direction of the shear flow to indicate the alignment of potential defects. The shear flow direction found in the simulation qualitatively agrees with both the continuum theory and the experiment, see Extended Data Fig. \ref{fig.m6}D. Nonzero $\alpha^o$ leads to nonzero $A$ which tilts the shear flow. Near the boundary, the strong shear flow tilts the defects. In the bulk, the shear flow is weak and we expect nearly isotropic distribution of defect orientations, in alignment with the experimental results~\cite{yashunsky2022chiral}. To reproduce the tilting behavior in simulations, a nematic extension of the graph model, such as that described in Ref.~\cite{lin2023structure}, would be required. 

While upon coarse-graining the chiral graph model is compatible with most of the qualitative predictions of active nematohydrodynamics, its domain of appplicability extends to narrow strips
just a few cells wide (where homogeneization cannot be performed). It also provides more accurate predictions of the chiral flow $v_y$ even for wide tracks, and explains the observed correlation between the converging flow $v_x$, the chiral flow $v_y$ and the density field.

\bibliography{citation}